\documentclass[%
 reprint,
superscriptaddress,
 amsmath,amssymb,
 aps,
prd,
floatfix,
nofootinbib
]{revtex4-2}
\usepackage{array}[=2016-10-06]
\usepackage{aas_macros}
\usepackage{tablefootnote}
\usepackage{upgreek}
\usepackage{graphicx}
\usepackage{bm}
\usepackage{enumitem}
\usepackage{mathtools}
\usepackage[colorlinks=true]{hyperref}
\usepackage[mathlines]{lineno} 
\usepackage[table]{xcolor}
\usepackage[normalem]{ulem}
\usepackage{multirow}
\usepackage{orcidlink}
\usepackage{empheq}
\usepackage{amsmath}
\usepackage{color,booktabs}
\definecolor{grey}{gray}{0.9}
\usepackage{float}
\usepackage{tikz}
\definecolor{sptcol}{HTML}{386A7A}
\hypersetup{
  linkcolor=sptcol,
  citecolor=sptcol,
  urlcolor=sptcol
}

\newcommand{\agora}{\textsc{Agora}}

\newcommand{\nside}{N_{\rm side}}
\newcommand{\planck}{{\it Planck}}
\newcommand{\sqdeg}{{\rm deg}^{2}}
\newcommand{\ukam}{\mu {\rm K}\textnormal{-}{\rm arcmin}}

\newcommand{\som}{\sigma_{8}\Omega_{\rm m}^{0.25}}
\newcommand{\omk}{\Omega_{\rm k}}
\newcommand{\summnu}{\sum m_{\nu}}

\newcommand{\baodesi}{{\rm BAO}_{\rm DESI}}
\newcommand{\baopredesi}{{\rm BAO}_{\rm pre\textnormal{-}DESI}}
\newcommand{\kgmv}{\kappa_{\rm GMV}}
\newcommand{\kgmvprof}{\kappa_{\rm GMVprof}}
\newcommand{\kpp}{\kappa_{\rm PP}}
\newcommand{\kmuse}{\kappa_{\rm MUSE}}

\newcommand{\kgpa}{\kappa_{\rm GPA}}
\newcommand{\kmpa}{\kappa_{\rm MPA}}
\newcommand{\kact}{\kappa_{\rm ACT}}
\newcommand{\kplk}{\kappa_{\rm Planck}}
\newcommand{\kpa}{\kappa_{\rm PA}}

\newcommand{\cmbspa}{{\rm CMB}_{\rm SPA}}
\newcommand{\cmbspt}{{\rm CMB}_{\rm SPT\textnormal{-}3G}}

\newcommand{\rdNzero}{N_{L}^{(0),\rm RD}}
\newcommand{\None}{N_{L}^{(1)}}

\newcommand{\gmvprof}{{\rm GMV}_{\rm prof} }
\newcommand{\gmvtsz}{{\rm GMV}_{\rm tSZdpj} }
\newcommand{\gmvxilc}{{\rm GMV}_{\rm xILC} }
\newcommand{\gmvsys}{{\rm GMV}_{\rm sys.corr}}

\newcommand{\checked}[1]{\textcolor{red}{(checked)}}

\begin{document}
\preprint{APS/123-QED}

\title{SPT-3G D1: Quadratic-Estimator CMB Lensing Reconstruction and Cosmology}


\author{Y.~Omori\,\orcidlink{0000-0002-0963-7310}}
\email{yomori@uchicago.edu}
\affiliation{Department of Astronomy and Astrophysics, University of Chicago, 5640 South Ellis Avenue, Chicago, IL, 60637, USA}
\affiliation{Kavli Institute for Cosmological Physics, University of Chicago, 5640 South Ellis Avenue, Chicago, IL, 60637, USA}
\affiliation{NSF-Simons AI Institute for the Sky (SkAI), 172 E. Chestnut St., Chicago, IL 60611, USA}
\author{W.~L.~K.~Wu\,\orcidlink{0000-0001-5411-6920}}
\email{wlw@caltech.edu}
\affiliation{California Institute of Technology, 1200 East California Boulevard., Pasadena, CA, 91125, USA}
\affiliation{Kavli Institute for Particle Astrophysics and Cosmology, Stanford University, 452 Lomita Mall, Stanford, CA, 94305, USA}
\affiliation{SLAC National Accelerator Laboratory, 2575 Sand Hill Road, Menlo Park, CA, 94025, USA}
\author{Y.~Nakato}
\affiliation{Department of Physics, Stanford University, 382 Via Pueblo Mall, Stanford, CA, 94305, USA}
\author{F.~Bianchini\,\orcidlink{0000-0003-4847-3483}}
\affiliation{Kavli Institute for Particle Astrophysics and Cosmology, Stanford University, 452 Lomita Mall, Stanford, CA, 94305, USA}
\affiliation{Department of Physics, Stanford University, 382 Via Pueblo Mall, Stanford, CA, 94305, USA}
\affiliation{SLAC National Accelerator Laboratory, 2575 Sand Hill Road, Menlo Park, CA, 94025, USA}
\author{L.~Balkenhol\,\orcidlink{0000-0001-6899-1873}}
\affiliation{Sorbonne Universit\'e, CNRS, UMR 7095, Institut d'Astrophysique de Paris, 98 bis bd Arago, 75014 Paris, France}
\author{C.~Daley\,\orcidlink{0000-0002-3760-2086}}
\affiliation{Universit\'e Paris-Saclay, Universit\'e Paris Cit\'e, CEA, CNRS, AIM, 91191, Gif-sur-Yvette, France}
\affiliation{Department of Astronomy, University of Illinois Urbana-Champaign, 1002 West Green Street, Urbana, IL, 61801, USA}
\author{W.~Quan\,\orcidlink{0009-0002-2589-5501}}
\affiliation{High-Energy Physics Division, Argonne National Laboratory, 9700 South Cass Avenue, Lemont, IL, 60439, USA}
\affiliation{Department of Physics, University of Chicago, 5640 South Ellis Avenue, Chicago, IL, 60637, USA}
\affiliation{Kavli Institute for Cosmological Physics, University of Chicago, 5640 South Ellis Avenue, Chicago, IL, 60637, USA}
\author{E.~Anderes\,\orcidlink{0009-0003-3245-3979}}
\affiliation{Department of Statistics, University of California, One Shields Avenue, Davis, CA 95616, USA}
\author{A.~J.~Anderson\,\orcidlink{0000-0002-4435-4623}}
\affiliation{Fermi National Accelerator Laboratory, MS209, P.O. Box 500, Batavia, IL, 60510, USA}
\affiliation{Kavli Institute for Cosmological Physics, University of Chicago, 5640 South Ellis Avenue, Chicago, IL, 60637, USA}
\affiliation{Department of Astronomy and Astrophysics, University of Chicago, 5640 South Ellis Avenue, Chicago, IL, 60637, USA}
\author{B.~Ansarinejad}
\affiliation{School of Physics, University of Melbourne, Parkville, VIC 3010, Australia}
\author{M.~Archipley\,\orcidlink{0000-0002-0517-9842}}
\affiliation{Department of Astronomy and Astrophysics, University of Chicago, 5640 South Ellis Avenue, Chicago, IL, 60637, USA}
\affiliation{Kavli Institute for Cosmological Physics, University of Chicago, 5640 South Ellis Avenue, Chicago, IL, 60637, USA}
\author{D.~R.~Barron\,\orcidlink{0000-0002-1623-5651}}
\affiliation{Department of Physics and Astronomy, University of New Mexico, Albuquerque, NM, 87131, USA}
\author{P.~S.~Barry\,\orcidlink{0000-0001-9103-9354}}
\affiliation{School of Physics and Astronomy, Cardiff University, Cardiff, CF24 3AA, UK}
\author{K.~Benabed}
\affiliation{Sorbonne Universit\'e, CNRS, UMR 7095, Institut d'Astrophysique de Paris, 98 bis bd Arago, 75014 Paris, France}
\author{A.~N.~Bender\,\orcidlink{0000-0001-5868-0748}}
\affiliation{High-Energy Physics Division, Argonne National Laboratory, 9700 South Cass Avenue, Lemont, IL, 60439, USA}
\affiliation{Kavli Institute for Cosmological Physics, University of Chicago, 5640 South Ellis Avenue, Chicago, IL, 60637, USA}
\affiliation{Department of Astronomy and Astrophysics, University of Chicago, 5640 South Ellis Avenue, Chicago, IL, 60637, USA}
\author{B.~A.~Benson\,\orcidlink{0000-0002-5108-6823}}
\affiliation{Fermi National Accelerator Laboratory, MS209, P.O. Box 500, Batavia, IL, 60510, USA}
\affiliation{Kavli Institute for Cosmological Physics, University of Chicago, 5640 South Ellis Avenue, Chicago, IL, 60637, USA}
\affiliation{Department of Astronomy and Astrophysics, University of Chicago, 5640 South Ellis Avenue, Chicago, IL, 60637, USA}
\author{L.~E.~Bleem\,\orcidlink{0000-0001-7665-5079}}
\affiliation{High-Energy Physics Division, Argonne National Laboratory, 9700 South Cass Avenue, Lemont, IL, 60439, USA}
\affiliation{Kavli Institute for Cosmological Physics, University of Chicago, 5640 South Ellis Avenue, Chicago, IL, 60637, USA}
\affiliation{Department of Astronomy and Astrophysics, University of Chicago, 5640 South Ellis Avenue, Chicago, IL, 60637, USA}
\author{S.~Bocquet\,\orcidlink{0000-0002-4900-805X}}
\affiliation{University Observatory, Faculty of Physics, LMU Munich, Scheinerstr.~1, 81679 Munich, Germany}
\author{F.~R.~Bouchet\,\orcidlink{0000-0002-8051-2924}}
\affiliation{Sorbonne Universit\'e, CNRS, UMR 7095, Institut d'Astrophysique de Paris, 98 bis bd Arago, 75014 Paris, France}
\author{E.~Camphuis\,\orcidlink{0000-0003-3483-8461}}
\affiliation{Sorbonne Universit\'e, CNRS, UMR 7095, Institut d'Astrophysique de Paris, 98 bis bd Arago, 75014 Paris, France}
\author{M.~G.~Campitiello}
\affiliation{High-Energy Physics Division, Argonne National Laboratory, 9700 South Cass Avenue, Lemont, IL, 60439, USA}
\author{J.~E.~Carlstrom\,\orcidlink{0000-0002-2044-7665}}
\affiliation{Kavli Institute for Cosmological Physics, University of Chicago, 5640 South Ellis Avenue, Chicago, IL, 60637, USA}
\affiliation{Enrico Fermi Institute, University of Chicago, 5640 South Ellis Avenue, Chicago, IL, 60637, USA}
\affiliation{Department of Physics, University of Chicago, 5640 South Ellis Avenue, Chicago, IL, 60637, USA}
\affiliation{High-Energy Physics Division, Argonne National Laboratory, 9700 South Cass Avenue, Lemont, IL, 60439, USA}
\affiliation{Department of Astronomy and Astrophysics, University of Chicago, 5640 South Ellis Avenue, Chicago, IL, 60637, USA}
\author{J.~Carron\,\orcidlink{0000-0002-5751-1392}}
\affiliation{Istituto ricerche solari Aldo e Cele Dacc\`o (IRSOL), Faculty of Informatics, Universit\`a della Svizzera italiana, 6605 Locarno, Switzerland}
\affiliation{Universit\'e de Gen\`eve, D\'epartement de Physique Th\'eorique, 24 Quai Ansermet, CH-1211 Gen\`eve 4, Switzerland}
\author{C.~L.~Chang}
\affiliation{High-Energy Physics Division, Argonne National Laboratory, 9700 South Cass Avenue, Lemont, IL, 60439, USA}
\affiliation{Kavli Institute for Cosmological Physics, University of Chicago, 5640 South Ellis Avenue, Chicago, IL, 60637, USA}
\affiliation{Department of Astronomy and Astrophysics, University of Chicago, 5640 South Ellis Avenue, Chicago, IL, 60637, USA}
\author{P.~M.~Chichura\,\orcidlink{0000-0002-5397-9035}}
\affiliation{Department of Physics, University of Chicago, 5640 South Ellis Avenue, Chicago, IL, 60637, USA}
\affiliation{Kavli Institute for Cosmological Physics, University of Chicago, 5640 South Ellis Avenue, Chicago, IL, 60637, USA}
\author{A.~Chokshi}
\affiliation{Department of Astronomy and Astrophysics, University of Chicago, 5640 South Ellis Avenue, Chicago, IL, 60637, USA}
\author{T.-L.~Chou\,\orcidlink{0000-0002-3091-8790}}
\affiliation{Department of Astronomy and Astrophysics, University of Chicago, 5640 South Ellis Avenue, Chicago, IL, 60637, USA}
\affiliation{Kavli Institute for Cosmological Physics, University of Chicago, 5640 South Ellis Avenue, Chicago, IL, 60637, USA}
\affiliation{National Taiwan University, No. 1, Sec. 4, Roosevelt Road, Taipei 106319, Taiwan}
\author{A.~Coerver\,\orcidlink{0000-0002-2707-1672}}
\affiliation{Department of Physics, University of California, Berkeley, CA, 94720, USA}
\author{T.~M.~Crawford\,\orcidlink{0000-0001-9000-5013}}
\affiliation{Department of Astronomy and Astrophysics, University of Chicago, 5640 South Ellis Avenue, Chicago, IL, 60637, USA}
\affiliation{Kavli Institute for Cosmological Physics, University of Chicago, 5640 South Ellis Avenue, Chicago, IL, 60637, USA}
\author{T.~de~Haan\,\orcidlink{0000-0001-5105-9473}}
\affiliation{High Energy Accelerator Research Organization (KEK), Tsukuba, Ibaraki 305-0801, Japan}
\author{K.~R.~Dibert}
\affiliation{Department of Astronomy and Astrophysics, University of Chicago, 5640 South Ellis Avenue, Chicago, IL, 60637, USA}
\affiliation{Kavli Institute for Cosmological Physics, University of Chicago, 5640 South Ellis Avenue, Chicago, IL, 60637, USA}
\author{M.~A.~Dobbs}
\affiliation{Department of Physics and McGill Space Institute, McGill University, 3600 Rue University, Montreal, Quebec H3A 2T8, Canada}
\affiliation{Canadian Institute for Advanced Research, CIFAR Program in Gravity and the Extreme Universe, Toronto, ON, M5G 1Z8, Canada}
\author{M.~Doohan}
\affiliation{School of Physics, University of Melbourne, Parkville, VIC 3010, Australia}
\author{D.~Dutcher\,\orcidlink{0000-0002-9962-2058}}
\affiliation{Joseph Henry Laboratories of Physics, Jadwin Hall, Princeton University, Princeton, NJ 08544, USA}
\author{C.~Feng}
\affiliation{Department of Astronomy, University of Science and Technology of China, Hefei 230026, China}
\affiliation{School of Astronomy and Space Science, University of Science and Technology of China, Hefei 230026}
\affiliation{Department of Physics, University of Illinois Urbana-Champaign, 1110 West Green Street, Urbana, IL, 61801, USA}
\author{K.~R.~Ferguson\,\orcidlink{0000-0002-4928-8813}}
\affiliation{Department of Physics and Astronomy, University of California, Los Angeles, CA, 90095, USA}
\affiliation{Department of Physics and Astronomy, Michigan State University, East Lansing, MI 48824, USA}
\author{N.~C.~Ferree\,\orcidlink{0000-0002-7130-7099}}
\affiliation{California Institute of Technology, 1200 East California Boulevard., Pasadena, CA, 91125, USA}
\affiliation{Kavli Institute for Particle Astrophysics and Cosmology, Stanford University, 452 Lomita Mall, Stanford, CA, 94305, USA}
\affiliation{Department of Physics, Stanford University, 382 Via Pueblo Mall, Stanford, CA, 94305, USA}
\author{K.~Fichman}
\affiliation{Department of Physics, University of Chicago, 5640 South Ellis Avenue, Chicago, IL, 60637, USA}
\affiliation{Kavli Institute for Cosmological Physics, University of Chicago, 5640 South Ellis Avenue, Chicago, IL, 60637, USA}
\author{A.~Foster\,\orcidlink{0000-0002-7145-1824}}
\affiliation{Joseph Henry Laboratories of Physics, Jadwin Hall, Princeton University, Princeton, NJ 08544, USA}
\author{S.~Galli}
\affiliation{Sorbonne Universit\'e, CNRS, UMR 7095, Institut d'Astrophysique de Paris, 98 bis bd Arago, 75014 Paris, France}
\author{A.~E.~Gambrel}
\affiliation{Kavli Institute for Cosmological Physics, University of Chicago, 5640 South Ellis Avenue, Chicago, IL, 60637, USA}
\author{A.~K.~Gao}
\affiliation{Department of Physics, University of Illinois Urbana-Champaign, 1110 West Green Street, Urbana, IL, 61801, USA}
\author{F.~Ge}
\affiliation{California Institute of Technology, 1200 East California Boulevard., Pasadena, CA, 91125, USA}
\affiliation{Kavli Institute for Particle Astrophysics and Cosmology, Stanford University, 452 Lomita Mall, Stanford, CA, 94305, USA}
\affiliation{Department of Physics, Stanford University, 382 Via Pueblo Mall, Stanford, CA, 94305, USA}
\affiliation{Department of Physics \& Astronomy, University of California, One Shields Avenue, Davis, CA 95616, USA}
\author{F.~Guidi\,\orcidlink{0000-0001-7593-3962}}
\affiliation{Department of Physics \& Astronomy, University of California, One Shields Avenue, Davis, CA 95616, USA}
\affiliation{Sorbonne Universit\'e, CNRS, UMR 7095, Institut d'Astrophysique de Paris, 98 bis bd Arago, 75014 Paris, France}
\author{S.~Guns}
\affiliation{Department of Physics, University of California, Berkeley, CA, 94720, USA}
\author{N.~W.~Halverson}
\affiliation{CASA, Department of Astrophysical and Planetary Sciences, University of Colorado, Boulder, CO, 80309, USA }
\affiliation{Department of Physics, University of Colorado, Boulder, CO, 80309, USA}
\author{E.~Hivon\,\orcidlink{0000-0003-1880-2733}}
\affiliation{Sorbonne Universit\'e, CNRS, UMR 7095, Institut d'Astrophysique de Paris, 98 bis bd Arago, 75014 Paris, France}
\author{G.~P.~Holder\,\orcidlink{0000-0002-0463-6394}}
\affiliation{Department of Physics, University of Illinois Urbana-Champaign, 1110 West Green Street, Urbana, IL, 61801, USA}
\author{W.~L.~Holzapfel}
\affiliation{Department of Physics, University of California, Berkeley, CA, 94720, USA}
\author{J.~C.~Hood}
\affiliation{Kavli Institute for Cosmological Physics, University of Chicago, 5640 South Ellis Avenue, Chicago, IL, 60637, USA}
\author{A.~Hryciuk}
\affiliation{Department of Physics, University of Chicago, 5640 South Ellis Avenue, Chicago, IL, 60637, USA}
\affiliation{Kavli Institute for Cosmological Physics, University of Chicago, 5640 South Ellis Avenue, Chicago, IL, 60637, USA}
\author{N.~Huang\,\orcidlink{0000-0003-3595-0359}}
\affiliation{Department of Physics, University of California, Berkeley, CA, 94720, USA}
\author{T.~Jhaveri}
\affiliation{Department of Astronomy and Astrophysics, University of Chicago, 5640 South Ellis Avenue, Chicago, IL, 60637, USA}
\affiliation{Kavli Institute for Cosmological Physics, University of Chicago, 5640 South Ellis Avenue, Chicago, IL, 60637, USA}
\author{F.~K\'eruzor\'e}
\affiliation{High-Energy Physics Division, Argonne National Laboratory, 9700 South Cass Avenue, Lemont, IL, 60439, USA}
\author{A.~R.~Khalife\,\orcidlink{0000-0002-8388-4950}}
\affiliation{Sorbonne Universit\'e, CNRS, UMR 7095, Institut d'Astrophysique de Paris, 98 bis bd Arago, 75014 Paris, France}
\author{L.~Knox}
\affiliation{Department of Physics \& Astronomy, University of California, One Shields Avenue, Davis, CA 95616, USA}
\author{K.~Kornoelje}
\affiliation{Department of Astronomy and Astrophysics, University of Chicago, 5640 South Ellis Avenue, Chicago, IL, 60637, USA}
\affiliation{Kavli Institute for Cosmological Physics, University of Chicago, 5640 South Ellis Avenue, Chicago, IL, 60637, USA}
\affiliation{High-Energy Physics Division, Argonne National Laboratory, 9700 South Cass Avenue, Lemont, IL, 60439, USA}
\author{C.-L.~Kuo}
\affiliation{Kavli Institute for Particle Astrophysics and Cosmology, Stanford University, 452 Lomita Mall, Stanford, CA, 94305, USA}
\affiliation{Department of Physics, Stanford University, 382 Via Pueblo Mall, Stanford, CA, 94305, USA}
\affiliation{SLAC National Accelerator Laboratory, 2575 Sand Hill Road, Menlo Park, CA, 94025, USA}
\author{K.~Levy}
\affiliation{School of Physics, University of Melbourne, Parkville, VIC 3010, Australia}
\author{Y.~Li\,\orcidlink{0000-0002-4820-1122}}
\affiliation{Kavli Institute for Cosmological Physics, University of Chicago, 5640 South Ellis Avenue, Chicago, IL, 60637, USA}
\author{A.~E.~Lowitz\,\orcidlink{0000-0002-4747-4276}}
\affiliation{Kavli Institute for Cosmological Physics, University of Chicago, 5640 South Ellis Avenue, Chicago, IL, 60637, USA}
\author{C.~Lu}
\affiliation{Department of Physics, University of Illinois Urbana-Champaign, 1110 West Green Street, Urbana, IL, 61801, USA}
\author{G.~P.~Lynch\,\orcidlink{0009-0004-3143-1708}}
\affiliation{Department of Physics \& Astronomy, University of California, One Shields Avenue, Davis, CA 95616, USA}
\author{T.~J.~Maccarone\,\orcidlink{0000-0003-0976-4755}}
\affiliation{Department of Physics \& Astronomy, Box 41051, Texas Tech University, Lubbock TX 79409-1051, USA}
\author{A.~S.~Maniyar\,\orcidlink{0000-0002-4617-9320}}
\affiliation{Kavli Institute for Particle Astrophysics and Cosmology, Stanford University, 452 Lomita Mall, Stanford, CA, 94305, USA}
\affiliation{Department of Physics, Stanford University, 382 Via Pueblo Mall, Stanford, CA, 94305, USA}
\affiliation{SLAC National Accelerator Laboratory, 2575 Sand Hill Road, Menlo Park, CA, 94025, USA}
\author{E.~S.~Martsen}
\affiliation{Department of Astronomy and Astrophysics, University of Chicago, 5640 South Ellis Avenue, Chicago, IL, 60637, USA}
\affiliation{Kavli Institute for Cosmological Physics, University of Chicago, 5640 South Ellis Avenue, Chicago, IL, 60637, USA}
\author{F.~Menanteau}
\affiliation{Department of Astronomy, University of Illinois Urbana-Champaign, 1002 West Green Street, Urbana, IL, 61801, USA}
\affiliation{Center for AstroPhysical Surveys, National Center for Supercomputing Applications, Urbana, IL, 61801, USA}
\author{M.~Millea\,\orcidlink{0000-0001-7317-0551}}
\affiliation{Department of Physics, University of California, Berkeley, CA, 94720, USA}
\author{J.~Montgomery}
\affiliation{Department of Physics and McGill Space Institute, McGill University, 3600 Rue University, Montreal, Quebec H3A 2T8, Canada}
\author{T.~Natoli}
\affiliation{Kavli Institute for Cosmological Physics, University of Chicago, 5640 South Ellis Avenue, Chicago, IL, 60637, USA}
\author{A.~Ouellette\,\orcidlink{0000-0003-0170-5638}}
\affiliation{Department of Physics, University of Illinois Urbana-Champaign, 1110 West Green Street, Urbana, IL, 61801, USA}
\author{Z.~Pan\,\orcidlink{0000-0002-6164-9861}}
\affiliation{High-Energy Physics Division, Argonne National Laboratory, 9700 South Cass Avenue, Lemont, IL, 60439, USA}
\affiliation{Kavli Institute for Cosmological Physics, University of Chicago, 5640 South Ellis Avenue, Chicago, IL, 60637, USA}
\affiliation{Department of Physics, University of Chicago, 5640 South Ellis Avenue, Chicago, IL, 60637, USA}
\author{P.~Paschos}
\affiliation{Enrico Fermi Institute, University of Chicago, 5640 South Ellis Avenue, Chicago, IL, 60637, USA}
\author{K.~A.~Phadke\,\orcidlink{0000-0001-7946-557X}}
\affiliation{Department of Astronomy, University of Illinois Urbana-Champaign, 1002 West Green Street, Urbana, IL, 61801, USA}
\affiliation{Center for AstroPhysical Surveys, National Center for Supercomputing Applications, Urbana, IL, 61801, USA}
\affiliation{NSF-Simons AI Institute for the Sky (SkAI), 172 E. Chestnut St., Chicago, IL 60611, USA}
\author{A.~W.~Pollak}
\affiliation{Department of Astronomy and Astrophysics, University of Chicago, 5640 South Ellis Avenue, Chicago, IL, 60637, USA}
\author{K.~Prabhu}
\affiliation{Department of Physics \& Astronomy, University of California, One Shields Avenue, Davis, CA 95616, USA}
\author{M.~Rahimi}
\affiliation{School of Physics, University of Melbourne, Parkville, VIC 3010, Australia}
\author{A.~Rahlin\,\orcidlink{0000-0003-3953-1776}}
\affiliation{Department of Astronomy and Astrophysics, University of Chicago, 5640 South Ellis Avenue, Chicago, IL, 60637, USA}
\affiliation{Kavli Institute for Cosmological Physics, University of Chicago, 5640 South Ellis Avenue, Chicago, IL, 60637, USA}
\author{C.~L.~Reichardt\,\orcidlink{0000-0003-2226-9169}}
\affiliation{School of Physics, University of Melbourne, Parkville, VIC 3010, Australia}
\author{M.~Rouble}
\affiliation{Department of Physics and McGill Space Institute, McGill University, 3600 Rue University, Montreal, Quebec H3A 2T8, Canada}
\author{J.~E.~Ruhl}
\affiliation{Department of Physics, Case Western Reserve University, Cleveland, OH, 44106, USA}
\author{A.~C.~Silva~Oliveira\,\orcidlink{0000-0001-5755-5865}}
\affiliation{California Institute of Technology, 1200 East California Boulevard., Pasadena, CA, 91125, USA}
\affiliation{Kavli Institute for Particle Astrophysics and Cosmology, Stanford University, 452 Lomita Mall, Stanford, CA, 94305, USA}
\affiliation{Department of Physics, Stanford University, 382 Via Pueblo Mall, Stanford, CA, 94305, USA}
\author{A.~Simpson}
\affiliation{Department of Astronomy and Astrophysics, University of Chicago, 5640 South Ellis Avenue, Chicago, IL, 60637, USA}
\affiliation{Kavli Institute for Cosmological Physics, University of Chicago, 5640 South Ellis Avenue, Chicago, IL, 60637, USA}
\author{J.~A.~Sobrin\,\orcidlink{0000-0001-6155-5315}}
\affiliation{Department of Physics, Villanova University, 800 E Lancaster Ave, Villanova, PA 19085, USA}
\affiliation{Fermi National Accelerator Laboratory, MS209, P.O. Box 500, Batavia, IL, 60510, USA}
\author{A.~A.~Stark}
\affiliation{Center for Astrophysics \textbar{} Harvard \& Smithsonian, 60 Garden Street, Cambridge, MA, 02138, USA}
\author{J.~Stephen}
\affiliation{Enrico Fermi Institute, University of Chicago, 5640 South Ellis Avenue, Chicago, IL, 60637, USA}
\author{C.~Tandoi\,\orcidlink{0000-0002-2077-6004}}
\affiliation{Department of Astronomy, University of Illinois Urbana-Champaign, 1002 West Green Street, Urbana, IL, 61801, USA}
\author{C.~Trendafilova}
\affiliation{CERCA/ISO, Department of Physics, Case Western Reserve University, Cleveland, OH, 44106, USA}
\affiliation{Center for AstroPhysical Surveys, National Center for Supercomputing Applications, Urbana, IL, 61801, USA}
\author{J.~D.~Vieira\,\orcidlink{0000-0001-7192-3871}}
\affiliation{Department of Astronomy, University of Illinois Urbana-Champaign, 1002 West Green Street, Urbana, IL, 61801, USA}
\affiliation{Department of Physics, University of Illinois Urbana-Champaign, 1110 West Green Street, Urbana, IL, 61801, USA}
\affiliation{Center for AstroPhysical Surveys, National Center for Supercomputing Applications, Urbana, IL, 61801, USA}
\author{A.~G.~Vieregg\,\orcidlink{0000-0002-4528-9886}}
\affiliation{Kavli Institute for Cosmological Physics, University of Chicago, 5640 South Ellis Avenue, Chicago, IL, 60637, USA}
\affiliation{Department of Astronomy and Astrophysics, University of Chicago, 5640 South Ellis Avenue, Chicago, IL, 60637, USA}
\affiliation{Enrico Fermi Institute, University of Chicago, 5640 South Ellis Avenue, Chicago, IL, 60637, USA}
\affiliation{Department of Physics, University of Chicago, 5640 South Ellis Avenue, Chicago, IL, 60637, USA}
\author{A.~Vitrier\,\orcidlink{0009-0009-3168-092X}}
\affiliation{Sorbonne Universit\'e, CNRS, UMR 7095, Institut d'Astrophysique de Paris, 98 bis bd Arago, 75014 Paris, France}
\author{Y.~Wan}
\affiliation{Department of Astronomy, University of Illinois Urbana-Champaign, 1002 West Green Street, Urbana, IL, 61801, USA}
\affiliation{Center for AstroPhysical Surveys, National Center for Supercomputing Applications, Urbana, IL, 61801, USA}
\author{N.~Whitehorn\,\orcidlink{0000-0002-3157-0407}}
\affiliation{Department of Physics and Astronomy, Michigan State University, East Lansing, MI 48824, USA}
\author{M.~R.~Young}
\affiliation{Fermi National Accelerator Laboratory, MS209, P.O. Box 500, Batavia, IL, 60510, USA}
\affiliation{Kavli Institute for Cosmological Physics, University of Chicago, 5640 South Ellis Avenue, Chicago, IL, 60637, USA}
\author{J.~A.~Zebrowski}
\affiliation{Kavli Institute for Cosmological Physics, University of Chicago, 5640 South Ellis Avenue, Chicago, IL, 60637, USA}
\affiliation{Department of Astronomy and Astrophysics, University of Chicago, 5640 South Ellis Avenue, Chicago, IL, 60637, USA}
\affiliation{Fermi National Accelerator Laboratory, MS209, P.O. Box 500, Batavia, IL, 60510, USA}
\collaboration{SPT-3G Collaboration}
\noaffiliation

\date{\today}

\begin{abstract}
We present a map of the cosmic microwave background (CMB) lensing potential reconstructed from observations taken during the 2019 and 2020 seasons with the third-generation camera on the South Pole Telescope (SPT), covering the $1500\,\sqdeg$ SPT-3G Main field, referred to as the SPT-3G D1 dataset. From the multi-frequency temperature and polarization data, we reconstruct the CMB lensing field using a quadratic estimator that jointly accounts for the $T$, $E$, and $B$ fields and their covariance. The resulting lensing map is dominated by polarization information for $L \lesssim 600$ and provides the highest signal-to-noise measurement per mode reported to date. At this precision, residual foreground and instrumental systematic effects must be accounted for explicitly, and we propagate their impact through the lensing likelihood using a simulation-based emulator. With these nuisance parameters fixed to their best-fit values, we measure a lensing amplitude consistent with unity at $2\%$ precision relative to the $\Lambda$CDM model that best fits the combined \planck{}, ACT DR6, and SPT-3G D1 $TT$/$TE$/$EE$ likelihoods ($\cmbspa{}$). We further measure the structure-growth parameter $\som$ to be $0.6046\pm0.0096$ from the SPT-3G D1 lensing spectrum alone and $0.6020\pm0.0084$ when combined with ACT DR6 and $\planck{}$ PR4 CMB lensing, providing the tightest constraint on this parameter from CMB lensing to date. By further combining this with $\cmbspa{}$ and the latest Dark Energy Spectroscopic Instrument (DESI) DR2 baryon acoustic oscillation (BAO) data, we obtain $\summnu < 0.072\,\mathrm{eV}$ (95\% C.L.) when allowing the neutrino mass to vary within $\Lambda$CDM. Compared with previous work, the better agreement of our measurement with DESI DR2 BAO yields both this relaxed upper bound and reduced ($\mathord{\sim}2\sigma$) preferences for nonzero spatial curvature and for deviations of $(w_0,w_a)$ from the $\Lambda$CDM expectation. When we combine CMB lensing with the Dark Energy Survey (DES)\,Y3 3$\times$2pt analysis, we obtain $S_{8}=0.811\pm0.011$, corresponding to a $1.4\%$ constraint on the late-time clustering amplitude. This precision is competitive with that obtained from the primary CMB within $\Lambda$CDM, providing a sensitive test of the consistency of structure growth across cosmic time.
\end{abstract}

\maketitle

\tableofcontents

\section{Introduction}\label{sec:intro}
As photons travel from the cosmic microwave background (CMB) last-scattering surface to the observer, they pass through gravitational potentials sourced by the intervening large-scale structure. These potentials deflect the paths of the photons, an effect known as gravitational lensing~\citep{blanchard1987}. These deflections, typically a few arcminutes in magnitude and coherent over degree scales, remap the primordial CMB anisotropies and introduce characteristic distortions in the observed CMB. Measurements of these distortions can be used to reconstruct the projected gravitational potential integrated along the line of sight, providing a direct probe of the intervening matter distribution and large-scale structure in the late-time Universe~\citep{lewis2006}.

The physics of the CMB is well understood; the CMB is emitted at effectively a single redshift, and that redshift is very well known. These features make CMB lensing an exceptionally clean probe of the matter distribution in the Universe. The sensitivity of CMB lensing to structure along the line of sight peaks at $z \sim 2$, making it particularly well suited to probing the intermediate-redshift regime of structure formation and complementary to low-redshift large-scale structure surveys. Moreover, its broad redshift sensitivity allows CMB lensing to be cross-correlated with a wide range of probes, including galaxy clustering, galaxy weak lensing, the thermal and kinetic Sunyaev-Zel’dovich effects, the cosmic infrared background, and upcoming line-intensity mapping surveys (including 21\,cm measurements). These cross-correlation measurements enable robust tests of theoretical predictions for structure formation and provide new insights into the interplay between dark matter, baryons, and large-scale structure across cosmic time.

The amplitude and shape of the CMB lensing power spectrum are sensitive to both the growth and the geometry of the Universe, providing complementary information that helps break degeneracies and improve joint constraints on cosmological parameters such as $\Omega_{\rm m}$, $\sigma_{8}$, $H_{0}$, and the sum of the neutrino masses $\sum m_{\nu}$ when combined with primary CMB and Baryon acoustic oscillation (BAO) measurements. Indeed, the CMB lensing power spectrum has been used extensively to constrain cosmology, using data from the South Pole Telescope (SPT; \cite{simard2018,wu2019,pan2023}), \planck{} \citep{planck2013lens,planck2015lens,planck2018lens,carron2022}, and the Atacama Cosmology Telescope (ACT; \cite{madhavacheril2024}).
Differences in observing strategy, angular resolution, and noise level lead to lensing reconstructions with distinct characteristics. The SPT-3G Main survey employs an observing strategy focused on deep integration over a 1,500\,$\mathrm{deg}^2$ patch, designed to achieve a lensing map with the highest signal-to-noise ratio per mode. In this low-noise regime, polarization becomes the dominant channel for lensing reconstruction, allowing us to produce lensing maps less susceptible to biases from astrophysical foregrounds. This strategy is motivated by the goal of delensing the degree-scale $B$-mode measurements from the BICEP experiment, which requires a very low-noise lensing map to perform effective delensing, a crucial step towards detecting primordial gravitational waves generated during inflation~\cite{bk18, bkdelens}.

In this work, we reconstruct CMB lensing maps using data over the $1500\,{\rm deg}^{2}$ Main survey field taken during the 2019 and 2020 observing seasons (hereafter referred to as D1), present the lensing power spectrum, and derive cosmological parameters using this measurement. Compared with the previous SPT-3G lensing study~\cite{pan2023}, this analysis uses substantially deeper temperature maps and additionally incorporates polarization data, both of which improve the lensing signal-to-noise ratio. We additionally employ the global minimum-variance (GMV) estimator~\cite{maniyar2021}, which optimally combines the temperature and polarization information in the lensing reconstruction. We find that residual foreground-induced shifts are small relative to the statistical uncertainties by comparing the baseline reconstruction with variants with reduced foreground sensitivity. Finally, we also introduce an improved cosmological inference pipeline, in which systematic effects are modeled through an emulator-based forward-modeling approach and their associated parameters are jointly sampled with the cosmological parameters.

A complementary CMB lensing analysis of the SPT-3G D1 data was presented by~\cite{ge2025}, in which CMB lensing bandpowers were inferred from polarization data using the Marginal Unbiased Score Expansion (MUSE; \cite{millea2022}) framework. This approach is methodologically distinct from the quadratic-estimator approach used in our analysis and uses polarization data alone, whereas our reconstruction combines both temperature and polarization. We compare the lensing reconstructions and cosmological constraints from these two analyses at several points throughout this paper.\footnote{For the MUSE comparisons presented in this work, we perform cosmological inference using the publicly available MUSE CMB lensing-only likelihood. This isolates the lensing information from the SPT-3G $EE$ likelihood, allowing the MUSE lensing measurement to be combined with different external datasets.}

The key results of this paper are
\begin{enumerate}
\item {\bf Highest signal-to-noise measurement of individual lensing modes to date, with polarization dominating the lensing reconstruction for $L \lesssim 600$.} With the exceptional depth of the SPT-3G D1 dataset, polarization contributes more signal-to-noise than temperature to the lensing reconstruction on these scales, enabling particularly clean measurements with reduced contamination from extragalactic foregrounds.
\item {\bf A new emulator-based framework for joint cosmological and systematic inference.} We build an emulator that characterizes the change in the theoretical CMB lensing power spectrum as the extragalactic foreground and instrumental systematic parameters are varied. This enables joint sampling of the cosmological, foreground, instrumental systematic, and other nuisance parameters, consistently capturing their degeneracies and propagating both statistical and systematic uncertainties into the final cosmological constraints, for the first time in a QE-based CMB lensing analysis.

\item {\bf State-of-the-art constraints on $\som$.} 
Our lensing power spectrum yields the tightest constraint on $\som$ from CMB lensing alone to date, at 1.6\% precision. When combining our lensing result with \planck{} and ACT DR6 lensing measurements, we constrain $\som$ at 1.4\%, a precision comparable to that of \planck{} primary CMB measurements.

\item \textbf{An $S_{8}$ constraint surpassing the precision of the primary CMB.}
We combine our baseline CMB lensing measurement with galaxy clustering, galaxy-galaxy lensing, and cosmic shear measurements (3$\times$2pt) from DES\,Y3 and obtain a $1.4\%$ constraint on $S_{8} \equiv \sigma_8(\Omega_{\rm m}/0.3)^{0.5}$, with an uncertainty smaller than that obtained from the primary CMB.

\item {\bf Consistency of $S_8$-type parameters between CMB lensing, CMB lensing + DES\,Y3 3$\times$2pt, and primary CMB.}
We find good agreement between the $S_8$ constraints from CMB lensing + DES\,Y3 3$\times$2pt and the primary CMB, which differ by $1.6\sigma$.

\end{enumerate}

This paper is organized as follows: In Section \ref{sec:data}, we provide a brief overview of the SPT-3G instrument and describe the dataset used in this analysis. We outline the mapmaking procedure and summarize the data processing steps applied to the input $T/Q/U$ maps, including calibration, masking, and filtering. In Section \ref{sec:simulations}, we describe the two classes of simulations used throughout this work. We detail how the simulated skies are generated, how instrumental effects are incorporated, and how these simulations are used to characterize biases, uncertainties, and covariances. In Section \ref{sec:lensing}, we introduce the quadratic estimators (QE) employed for CMB lensing reconstruction. We describe the estimator formalism, the implementation choices specific to this analysis, and the procedures used to obtain unbiased estimates of the lensing power spectrum. In Section \ref{sec:validation}, we present a series of validation tests applied to the reconstructed lensing maps and power spectra. These include null tests, consistency checks across estimators, and tests using simulations to verify the robustness of our reconstruction pipeline. In Section \ref{sec:inference}, we describe our cosmological parameter inference pipeline. We outline the likelihood framework, covariance estimation, and modeling assumptions, and present validation tests performed at the parameter inference level to ensure unbiased and stable constraints. In Section \ref{sec:results}, we present our main cosmological results, including constraints on parameters such as the amplitude of the reconstructed CMB lensing power spectrum and $\som$. We compare these constraints to previous measurements and discuss their implications. We then combine the lensing measurement with primary CMB and/or BAO measurements to constrain cosmological parameters such as $\Omega_{\rm m}$, $\sigma_8$, $H_0$, the sum of neutrino masses, spatial curvature, and evolving dark energy equation of state.
Finally, in Section \ref{sec:summary}, we summarize our findings and highlight prospects for future improvements and analyses. The data and likelihood code used in this paper are publicly available.\footnote{\url{https://pole.uchicago.edu/public/data/omori26/}} \\

\section{SPT-3G data}\label{sec:data}
The SPT~\cite{carlstrom2011} is a 10-meter telescope located at the Amundsen-Scott South Pole Station in Antarctica. The SPT-3G instrument \citep{benson2014}, the third-generation camera on the SPT, has been in operation since 2018 and delivers substantially improved performance compared with its predecessors, SPT-SZ~\citep{carlstrom2011} and SPTpol~\citep{austermann2012}. It contains over 16000 superconducting transition-edge sensor bolometers with trichroic, dual-polarization pixels that observe simultaneously in three frequency bands centered approximately at 95\,GHz, 150\,GHz, and 220\,GHz. The combination of a large detector count and $\mathord{\sim}1'$ angular resolution provides the low-noise and high-resolution data required to generate high-signal-to-noise CMB lensing maps.

\begin{figure*}
	\includegraphics[width=0.98\textwidth]{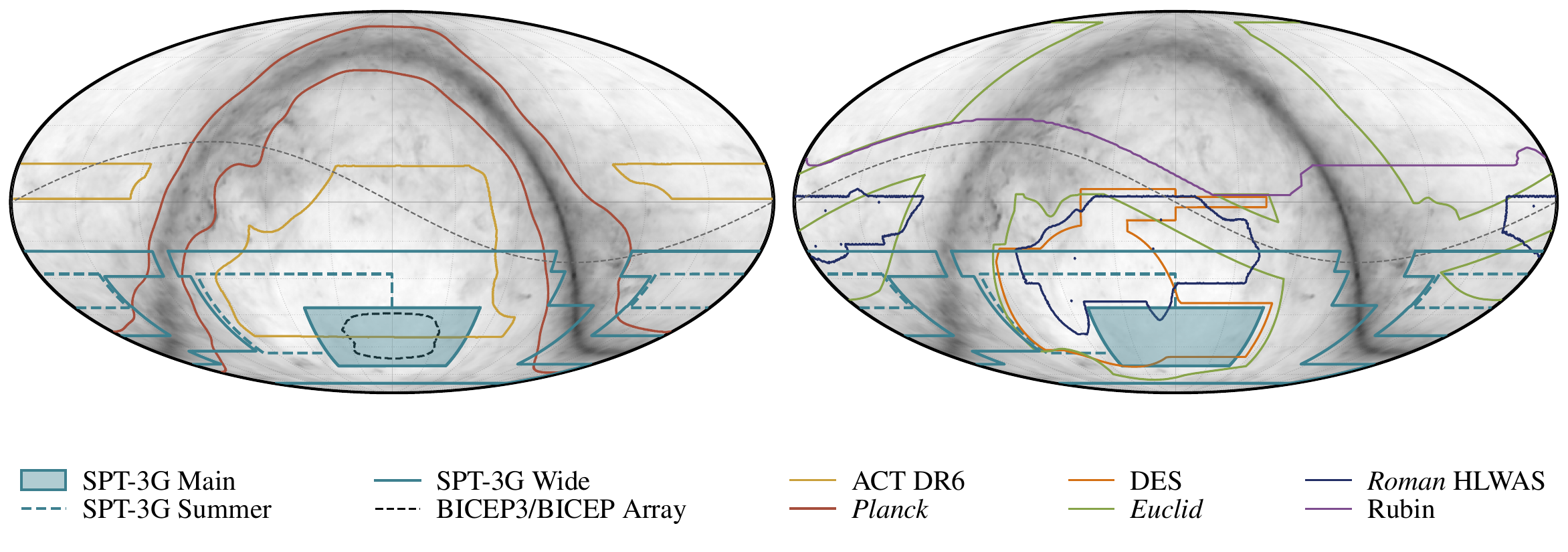}
    \makeatletter\long\def\@ifdim#1#2#3{#2}\makeatother
    \caption{Footprints of the SPT-3G Main survey (teal filled;~\cite{quan2026}), SPT-3G Summer survey (teal dashed line), Wide survey (teal solid line; excluding Main and Summer), ACT DR6 (gold solid line;~\cite{act_inst,naess2025}), \planck{} (red solid line;~\cite{planck_mission,planck2018_overview}), BICEP3/BICEP Array (black dashed line;~\cite{bicep3,biceparray,bk18}), DES (dark orange line;~\cite{flaugher2015,des2016}), {\it Euclid} (olive solid line;~\cite{euclid}), Rubin-LSST (below the purple line;~\cite{rubin}), and the {\it Roman} High-Latitude-Wide-Area Survey (HLWAS; navy solid line;~\cite{roman,roman_rotac2025}), shown in equatorial coordinates. The sinusoidal curve indicates the ecliptic plane.}
  
    \label{fig:footprint}
\end{figure*}

\begin{figure}
	\includegraphics[width=\columnwidth]{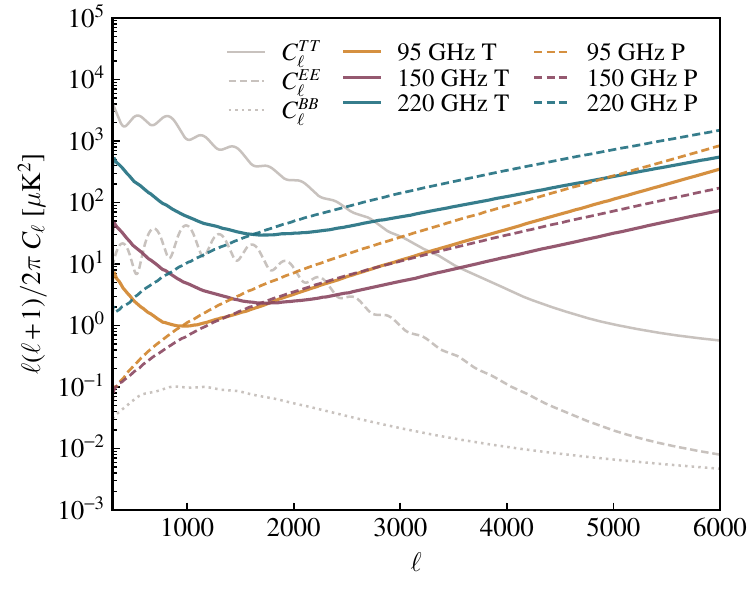}
    \caption{Noise levels computed from sign-flip realizations in temperature (solid) and polarization (dashed) for the 95 (orange), 150 (purple), and 220 (teal) GHz channels. The $C_{\ell}^{TT}$, $C_{\ell}^{EE}$, and $C_{\ell}^{BB}$ theory spectra from \planck{} 2018 best-fit cosmology are shown in the background (gray solid, dashed, and dotted lines). }
    \label{fig:noiselevels}
\end{figure}

In this analysis, we use data collected during the 2019 and 2020 observing seasons over the SPT-3G Main field as shown in  Figure~\ref{fig:footprint}. The Main SPT-3G survey covers approximately $1500\,\sqdeg$ of sky, spanning in right ascension from $20^\textrm{h}40^\textrm{m}0^\textrm{s}$ to $3^\textrm{h}20^\textrm{m}0^\textrm{s}$ and in declination from $-42^\circ$ to $-70^\circ$. The survey footprint is divided into four subfields centered at declinations of $-44.75^\circ$, $-52.25^\circ$, $-59.75^\circ$, and $-67.25^\circ$.  Relative to the 2018 dataset, these maps achieve substantially lower noise levels, reaching 5, 4, and 16\,$\ukam$ at 95, 150, and 220\,GHz (in temperature), respectively (see Figure~\ref{fig:noiselevels}). This improved depth places the analysis in a new regime for CMB lensing, where polarization data play a dominant role in the reconstruction, particularly on large angular scales. A detailed description of the mapmaking and data-processing pipeline, including the data products, filtering, calibration, and associated validation and null tests, is provided in \cite{quan2026}. We also refer the reader to the primary power-spectrum analysis of \cite[][hereafter C26]{camphuis2026}, which used the same $T/Q/U$ maps and performed extensive null tests. We additionally refer the reader to the MUSE analysis~\cite{ge2025}, in which several calibration approaches were compared. We summarize below the main elements adopted from these works, along with the additional calibration procedures and simulations specific to this lensing analysis.

\subsection{Beam}\label{sec:beams}
A detailed description of the SPT-3G beam characterization procedure will be presented in Huang et al.\ (in preparation); here, we summarize only the key aspects. The angular resolution of SPT is fundamentally limited by diffraction from its 10-meter primary mirror. The telescope beam (or point-spread function) acts to smooth the observed sky signal, suppressing power on small angular scales. Accurate knowledge of the beam profile is therefore essential for recovering the true sky power spectra and ensuring unbiased measurements of small-scale anisotropies. In this analysis, the beams for the temperature and polarization maps are derived separately.

The fiducial temperature beam is characterized using observations of both planets and bright compact sources, the majority of which are active galactic nuclei (AGNs). These two classes of sources are used to measure the beam over different angular scales. Due to their brightness, planets enable precise measurements of the outer beam profile over large angular scales (tens of arcminutes), but their cores saturate the detectors, preventing accurate characterization of the inner regions. Conversely, AGNs are much fainter and effectively unresolved, allowing detailed measurement of the inner beam core. The two measurements are then combined to construct a composite beam profile spanning from the central core to the outer sidelobes.

For the polarization channels, the beam is modeled as \cite{ge2025,camphuis2026}:
\begin{equation}\label{eq:betapol}
B_{\ell,\nu}^{P}=B_{\ell,\nu}^{\rm main} +\beta_{\rm pol,\nu} (B_{\ell,\nu}^{T}-B_{\ell,\nu}^{\rm main}),
\end{equation}
normalized to 1 at $\ell=800$. Here, $B_{\ell}^{\rm main}$ is the main-lobe beam, a physically motivated beam profile based on the telescope optics that is fit to the temperature-derived beam profile at $R<0.75'$, and $\beta_{\rm pol, \nu}$ are free scaling parameters fitted in \cite{ge2025,camphuis2026}.  
The $\beta_{\rm pol, \nu}$ parameters are taken from marginalized posteriors obtained by jointly fitting the $\Lambda$CDM model, beam, and other systematic parameters to the measured CMB power spectra.

\subsection{Calibration}
The calibration of the SPT-3G D1 maps and the corrections for instrumental effects, including temperature-to-polarization leakage, are described in detail in \cite{quan2026}. Briefly, the absolute calibration of the 150\,GHz maps is determined by cross-correlating the SPT-3G maps with the \planck{} 143\,GHz maps, while the 95 and 220\,GHz maps are calibrated internally relative to the 150\,GHz maps using cross-frequency spectra. A monopole temperature-to-polarization leakage correction is applied by subtracting scaled temperature templates from the $Q/U$ maps, and a global polarization-angle offset $\Delta\psi$ is determined from the observed $EB$ correlation and removed by rotating the $Q/U$ maps. We denote the residual absolute temperature and polarization calibration factors at 150\,GHz by $T_{\rm cal}$ and $P_{\rm cal}$, respectively, and vary these parameters in the cosmological likelihood while keeping the inter-frequency calibration fixed to the values adopted in \cite{quan2026}.

\subsection{Extragalactic foregrounds and mitigation strategy}\label{sec:extragalacticfg}
CMB lensing maps are reconstructed from distortions imprinted on the primary CMB fluctuations. However, the observed millimeter-wave sky contains not only the CMB but also emission and spectral distortions from astrophysical sources such as infrared galaxies, radio galaxies, and galaxy clusters. Infrared sources are predominantly dusty star-forming galaxies whose emission traces obscured star formation~\cite{lagache2005}, particularly during the peak of cosmic star formation at redshifts $z\sim1$--$2$~\cite{madau2014}. The brightest sources are detected individually, while the combined emission from unresolved infrared galaxies forms the cosmic infrared background (CIB). Radio emission is primarily associated with active galactic nuclei, where relativistic electrons in magnetized jets produce synchrotron radiation~\cite{dezotti2010}. In galaxy clusters, hot ionized gas causes inverse Compton scattering of CMB photons, distorting the CMB spectral energy distribution and producing the thermal Sunyaev-Zel'dovich (tSZ) effect~\cite{sunyaev1970,sunyaev1972,carlstrom2002}. The tSZ signal is one of the dominant foreground contaminants in temperature-based CMB lensing reconstruction.

Because these foreground signals arise from astrophysical objects that trace the nonlinear late-time large-scale structure, the resulting foreground fluctuations are intrinsically non-Gaussian. When decomposed into Fourier modes, this non-Gaussianity manifests as correlations between modes in the observed maps. As we will describe in Section~\ref{sec:lensing}, lensing reconstruction exploits analogous mode coupling induced by gravitational lensing and can therefore misinterpret foreground-induced correlations as a true lensing signal. Moreover, galaxies and galaxy clusters trace the underlying matter density field and are therefore correlated with the same large-scale structure that produces the true CMB lensing signal. Foregrounds can thus bias the reconstructed lensing auto-spectrum both through their own non-Gaussian mode coupling (trispectrum) and through their correlation with the true lensing field (bispectrum)~\cite{vanengelen2014,osborne2014,sailer2020}. 

We use several complementary approaches to reduce and characterize these biases. We mask the brightest infrared and radio sources and the galaxy clusters detected at the highest significance. To assess contamination from the remaining unmasked populations, we construct alternative temperature maps in which the tSZ or CIB signals are suppressed through frequency-based component separation. We also apply a profile-hardened lensing estimator~\cite{sailer2020} designed to reduce the response of the lensing estimator to compact and extended foreground sources. These foreground-mitigated reconstructions are used as robustness tests. We adopt the minimum-variance reconstruction as our baseline to retain the full statistical sensitivity and account for residual foreground contamination through modeling and marginalization in the likelihood. The implementation of each of these steps is described in the following sections.

\subsection{Model-informed linear combination}\label{sec:lc}
We combine the 95/150/220\,GHz maps linearly using the model-informed linear-combination (LC) approach of \cite{bleem2022}, in which the combined map is constructed as a weighted sum of the frequency channels, with the weights determined from simulations, instrument noise models, and analytic fits. Because these weights do not depend on the specific realization of the sky, the realization-dependent cancellation that impacts \emph{internal} linear-combination (ILC)~\cite{delabrouille2009} approach is avoided, and the resulting maps are unbiased with respect to the target signal under correct spectral and calibration assumptions. This leads to slightly suboptimal component separation because the model-informed LC method does not use the exact information from the data map, but ensures that the resulting maps recover unbiased power.

\begin{figure}
\includegraphics[width=\columnwidth]{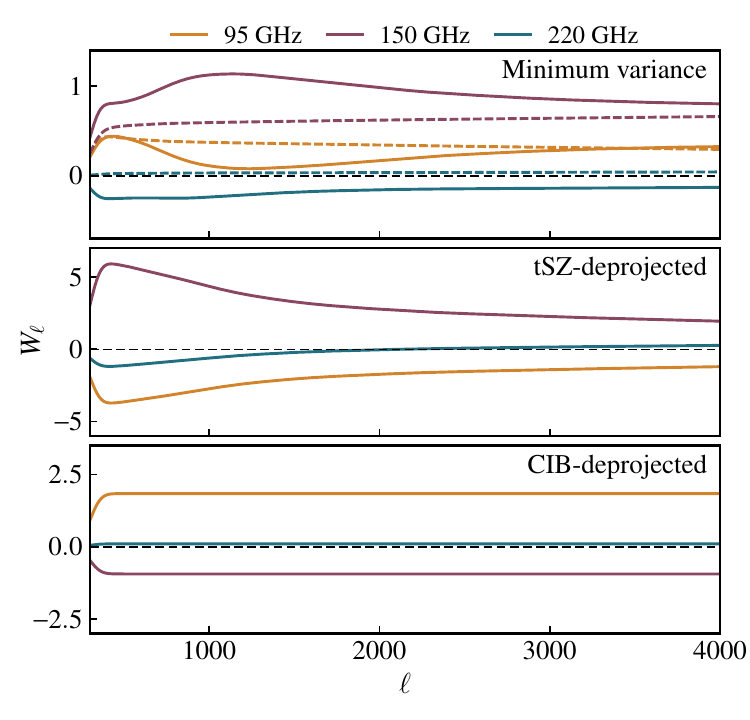}
\caption{Weights for the 95/150/220\,GHz channels in forming the minimum-variance, tSZ-deprojected, and CIB-deprojected combinations. The solid lines represent the weights for the temperature maps, whereas the dashed lines represent the weights for the polarization maps.}
\label{fig:ilcweights}
\end{figure}

We combine the frequency channels in three different ways: 
\begin{itemize}[leftmargin=8pt,labelsep=0.6em]
\item[-]{\it Minimum variance}:
We form a map by linearly combining the frequency channels with weights that enforce unit response to the CMB while minimizing the variance from noise and foregrounds. For each mode $(\ell,m)$, we estimate the inter-frequency covariance and compute the corresponding minimum-variance weights.

\item[-]{\it tSZ-deprojected}: We construct a constrained linear-combination (cLC) map following \cite{remazeilles2011}, choosing the weights so that the combination has unit response to the CMB and vanishing response to the thermal SZ spectral energy distribution (SED). This preserves the CMB signal while projecting out the tSZ signal, thereby reducing a potentially important source of foreground bias in the maps entering the lensing estimator. Imposing this additional nulling condition reduces the freedom available to optimize the weights for noise minimization. As a result, the cLC map is noisier than the minimum-variance linear combination.

\item[-]{\it CIB-deprojected}:
Unlike the tSZ effect, the emission from the infrared galaxies that make up the CIB cannot be described by a single SED, because the CIB is the integrated emission from many galaxies over a broad range of redshifts. To mitigate CIB contamination in our maps, we jointly deproject two effective\footnote{The spectral indices and dust temperatures of the individual effective SED components are not intended to have a direct physical interpretation on their own. Instead, the sum of the two components is constructed to reproduce a realistic CIB SED.} modified blackbody SED templates, each described by
\begin{equation}
I_{\nu}(\beta_{\rm d},T_{\rm d})\propto\frac{2h\nu^3}{c^2}\frac{(\nu/\nu_{0})^{\beta_{\rm d}}}{\exp(h\nu/k_{\rm B}T_{\rm d})-1}.
\end{equation}
The two templates have $(\beta_{\rm d},T_{\rm d})=(3.00,32\,{\rm K})$ and $(2.20,10\,{\rm K})$. These values are selected through a four-parameter grid search using \textsc{Agora} simulations~\cite{omori2024}, choosing the pair that produces the smallest residual CIB contamination. For each trial pair, the SEDs are evaluated in each frequency channel and imposed as separate nulling constraints, with the constrained linear-combination weights computed from the simulated foreground covariance and the measured auto- and cross-noise spectra between frequency channels (see Section~\ref{sec:Agora}).
\end{itemize}
For visualization, Figure~\ref{fig:ilcweights} shows the frequency-channel weights for the different LC variants, computed from the $m$-averaged inter-band covariance rather than the per-$(\ell,m)$ covariance used in the analysis.

Before applying these weights to the data maps in harmonic space, bright point sources are inpainted using a simple iterative averaging scheme to avoid ringing artifacts due to band-limited spherical harmonic transforms. The LC weighting is performed without deconvolving the transfer function to preserve some modes that are otherwise challenging to characterize; we refer the reader to \cite{quan2026} for details on the transfer function. The resulting linear-combination map is given by
\begin{equation}
X_{\rm LC}=\sum_{\nu}w_{\nu}X^{\rm calib}_{\nu},
\end{equation}
where $X^{\rm calib}_{\nu}$ are the calibrated $T/Q/U$ maps from~\cite{quan2026}. Once the frequency maps are combined, we convolve the resulting map with a common beam, which we choose to be the 150\,GHz beam.  Pairs of these linearly combined maps are then fed into the lensing estimator after $C^{-1}$ filtering (described in Section~\ref{sec:cinv}), yielding a suite of reconstructed lensing maps that trade off foreground sensitivity in a controlled way, including combinations that are optimized for minimum variance as well as combinations that suppress specific contaminants such as tSZ or CIB.

\subsection{Source masking}
To remove the objects expected to produce the largest localized foreground-induced biases in the lensing reconstruction, we construct a source mask from our fiducial catalog\footnote{This list was constructed from a preliminary version of the source catalog. Because we restrict the mask to high-signal-to-noise detections, it is effectively identical to the one from the final catalog. The official point source catalog is described in Archipley et al.\ (in preparation).} we construct a source mask from our fiducial catalog of infrared and radio sources detected above 6\,mJy at 150\,GHz and galaxy clusters detected above a significance threshold of $\xi = 10$, where $\xi$ denotes the signal-to-noise ratio of the multi-frequency matched-filter detection~\cite{bleem2015}.
For infrared and radio sources, we choose the masking radius as the radius at which the product of the source detection significance and the radial beam profile falls below unity. For galaxy clusters, we apply the same criterion using the cluster $\beta$-profile convolved with the telescope beam.

The resulting binary mask is then apodized with a Gaussian kernel of width $3'$. In total, this masks 2116 sources and 537 clusters, corresponding to a fractional loss of $4\%$ of the survey area. We refer to this as the analysis mask and use this mask to compute the final lensing power spectrum.

Additionally, we construct a less conservative mask for map processing, specifically for the $C^{-1}$ filtering step described in the following section. The nominal masking criterion is intentionally conservative and extends far into the low-amplitude tails of the source and cluster profiles. In these outskirts, the residual foreground signal is both subdominant to the CMB fluctuations and much more slowly varying than the central signal and is therefore expected to produce a substantially smaller response in the lensing reconstruction. Using the nominal radii during $C^{-1}$ filtering would consequently discard predominantly CMB information for little additional suppression of foreground contamination. During map processing, we therefore use a mask that removes the foreground-dominated cores while preserving as much of the surrounding area for the lensing reconstruction as possible. For temperature, we determine these radii from the difference between the original and Gaussian-constrained-inpainted maps~\cite{hoffman1991,benoitlevy2013}. The constrained inpainting fills the masked region with a CMB realization conditioned on the surrounding pixels, such that taking the difference largely removes the underlying CMB fluctuations and reveals the source or cluster signal. We measure the radial profiles in these difference maps and select the radius at which the profile falls below $30\,\mu{\rm K}$.

In polarization, the number of detectable point sources is much smaller because extragalactic sources are typically only weakly polarized, at the level of a few percent. Applying a mask derived from temperature detections would therefore remove an unnecessarily large fraction of the sky. Instead, we begin with our fiducial source catalog containing sources brighter than $6\,\mathrm{mJy}$ in total intensity and measure the peak $Q$ and $U$ values in the $95\,\mathrm{GHz}$ map, where polarized point sources are expected to be most detectable because the polarized source population is dominated by radio galaxies, whose emission generally decreases with increasing frequency. In parallel, we generate $Q/U$ maps in which these sources have been inpainted. A source is classified as polarized if the difference between the original and inpainted maps exceeds $50\,\mu\mathrm{K}$ in either $Q$ or $U$. This procedure identifies 137 sources, substantially fewer than are masked in temperature, allowing us to mask only those sources most likely to contaminate the polarization-based lensing reconstruction. Adopting a source mask with fewer holes for the polarization maps allows us to reconstruct each lensing mode using more CMB mode pairs than when using the temperature source mask. The final lensing power spectrum is nevertheless computed using the analysis mask, which includes the locations of the temperature sources.

\subsection{$C^{-1}$ filtering}\label{sec:cinv}
Before running lensing reconstruction, the linearly combined map $X_{\rm LC}$ is inverse-variance filtered to downweight modes dominated by noise or residual foregrounds. In the context of CMB lensing, this procedure is commonly referred to as $C^{-1}$ filtering. It implements the inverse-variance weighting of sky modes based on their total variance, which includes contributions from the CMB signal, instrumental and atmospheric noise, and astrophysical foregrounds.

The observed map can be modeled as
\begin{equation}
X_{\rm LC}(\hat{n}) = \mathcal{U} \otimes 
\left[ X_{\rm CMB}(\hat{n}) + F(\hat{n}) \right] + n(\hat{n}),
\end{equation}
where $\mathcal{U}$ denotes the combined response of the mapmaking and filtering pipeline, including the common beam, $F$ represents residual astrophysical foregrounds and colored noise components modeled in harmonic space, and $n$ is the instrumental noise. For SPT-3G, the dominant filtering effect is the suppression of modes slowly varying along the scan direction, which is modeled as a declination-dependent cut in azimuthal frequency, $m_{\rm cut}(\delta)$. The inverse-variance-filtered map is given by $\overline{X} = C^{-1} X$, where $C = \mathcal{U} S \mathcal{U}^{\dagger} + N_{\rm flat}$ is the total covariance. Here, $S$ and $N_{\rm flat}$ denote the covariance of map components modeled in harmonic and pixel space, respectively. The harmonic component $S$ includes contributions from the CMB, astrophysical foregrounds, and residual noise $C_{\ell}^{\mathrm{fg+n_{res}}}$,
\begin{equation}
S = C_{\ell}^{\mathrm{CMB}} + C_{\ell}^{\mathrm{fg+n_{res}}},
\end{equation}
where $C_{\ell}^{\mathrm{fg+n_{res}}}$ is modeled by subtracting off a white-noise level from the total noise and foreground power spectrum, with the white noise component being modeled in the pixel-space covariance. The inverse pixel-space covariance $N_{\rm flat}^{-1}$ is constructed by multiplying the inverse pixel variance by the apodized boundary mask $M^{\rm b}$ and binary point source mask $M^{\rm ps}$. We adopt map depths of $5.0$ and $4.2\,\ukam$ for the temperature and polarization channels, respectively.\footnote{The $N_{\rm flat}$ level for the temperature maps is higher than that of the polarization maps due to contributions from foregrounds.} We additionally generate a filtered map {\it without} the point source mask for the purpose of computing an unbiased response function. 

In practice, directly evaluating $C^{-1}X$ is computationally prohibitive, as it inverts a large covariance matrix that couples all pixels and harmonic modes through the beam and noise properties of the experiment. Rearranging $C^{-1}$ in a computationally convenient form, we have:
\begin{align}\label{eq:cinvcore}
\overline{X} &=C^{-1}X\nonumber\\
&=S^{-1}\!\left[S^{-1}+\mathcal{U}^{\dagger}N_{\rm flat}^{-1}\mathcal{U}\right]^{-1}
\mathcal{U}^{\dagger}N_{\rm flat}^{-1}X,
\end{align}
following the implementation described in \cite{story2015}. Equation~\eqref{eq:cinvcore} is solved iteratively using preconditioned conjugate-gradient descent, which requires only the ability to apply the operators $S^{-1}$ and $N_{\rm flat}^{-1}$ to a map. In each iteration, multiplications by $S^{-1}$ are performed in harmonic space, where each mode $(\ell,m)$ is weighted by the inverse of its expected CMB, foreground, and colored-noise variance. Multiplications by $N_{\rm flat}^{-1}$ are carried out in pixel space, where the noise is approximated to be uncorrelated between pixels and $N_{\rm flat}^{-1}$ is zero for masked pixels. This alternating procedure yields an efficient realization of the full inverse-variance weighting given by the formal expression above, without explicitly constructing or inverting the full covariance matrix.

The filtered maps are then used as input to the quadratic estimator, ensuring that the lensing reconstruction attains close to optimal performance under realistic noise and foreground modeling.

\section{Simulations}\label{sec:simulations}
We use two types of simulations in this work. One captures the non-Gaussian nature of the lensing potential and foregrounds, including the correlations between them, while the other consists of 500 Gaussian realizations constructed to match the relevant signal and noise properties. These simulations are used for different parts of the analysis, which we describe in turn below.

\subsection{Non-Gaussian simulation}\label{sec:Agora}

As discussed in Section~\ref{sec:extragalacticfg}, astrophysical foregrounds can bias the reconstructed lensing signal through both their non-Gaussian structure and their correlation with the underlying large-scale structure. Quantifying these effects requires simulations with realistic foreground populations, including their clustering and range of amplitudes, while preserving their correlations with the true lensing field. This is particularly important for the tSZ signal from galaxy clusters. Because clusters are highly biased tracers of the matter distribution, they preferentially reside in regions of high lensing signal~\citep{madhavacheril2018}. Spurious lensing reconstructed from their tSZ signal is therefore correlated with the true lensing field, producing a bias in the reconstructed lensing power spectrum~\cite{osborne2014,omori2023}. In addition, masking bright foreground sources and clusters can preferentially remove high-lensing-signal regions, further biasing the inferred amplitude of matter fluctuations~\cite{osborne2014,lembo2022}.

For this purpose, we use the \textsc{Agora} simulation suite, which provides extragalactic foreground maps including tSZ, kSZ, CIB, and radio emission. These components are generated on a common lightcone from the same underlying large-scale structure as the lensing field, thereby preserving the expected correlations between the foregrounds and CMB lensing.

Although \textsc{Agora} is designed to produce realistic maps that resemble the observed sky, small discrepancies may remain. We therefore calibrate the foreground maps to match observational constraints from SPT-3G at high $\ell$ (see Appendix~\ref{sec:sim_calibration} for details). In addition, we rescale the amplitude of the lensing potential to correct for the difference between the \planck{} 2013 best-fit cosmology assumed by the underlying MultiDark Planck II (MDPL2) simulation used in \textsc{Agora} and the \planck{} 2018 best-fit cosmology adopted in this analysis. This correction corresponds to a $\mathord{\sim}0.4\%$ adjustment in the amplitude of the lensing potential map.

Once the individual components are calibrated, we measure the auto- and cross-power spectra of the calibrated foreground maps. These spectra are subsequently used to generate correlated Gaussian realizations of the foreground components, as described below. We also extract ten $1500\,{\rm deg}^2$ patches from the single full-sky realization of \textsc{Agora} available to construct simulated SPT-3G fields.

Before running the mock-observation pipeline on these maps, we apply a simple masking and inpainting procedure so that the masking in the simulations is roughly equivalent to that in the data. For clusters, we use an abundance-matching prescription: we match the number of clusters detected above approximately $\xi=10$, rescaled to the expected number over the full sky using an $f_{\rm sky}$ factor, and select clusters that fall within each patch. At the corresponding halo locations, we apply an iterative smoothing procedure in the Compton-$y$ map to suppress the cluster signal. For point sources, we mask pixels for which the total flux from all (CIB+radio) sources falling in a pixel exceeds 6 mJy. The CMB realization is added afterward and is left unmodified. This allows us to pass the maps through the mock-observation pipeline as if the most massive clusters and brightest point sources were absent from the foreground sky. We then provide the true point-source locations to the filtering stage of the mock observations, so that the masking and transfer function treatment remains consistent with that applied to the data.

\subsection{Gaussian foreground simulation}\label{sec:Gsim}
Because the \textsc{Agora} foreground maps are constructed from a high-resolution $N$-body simulation, only one independent realization is available. For applications that require large ensembles of simulations, we instead construct Gaussian foreground realizations whose power spectra match those measured from the available non-Gaussian realization.\footnote{After masking the brightest sources, the remaining foreground maps are reasonably close to Gaussian, making this approximation adequate for applications that require large ensembles of simulations.} These Gaussian simulations are used to estimate the mean-field bias and response function of the reconstructed lensing maps (see Section~\ref{sec:lensing}), both of which require a large number of realizations for accurate estimation.

The Gaussian-foreground mock skies used in this work consist of three components: the primary CMB, Gaussian realizations of the extragalactic foregrounds, and instrumental noise. We describe the generation of the CMB and foreground components below and the noise generation, which is common to both the Gaussian- and non-Gaussian-foreground simulations, in the next subsection.\\[0.05cm]

\begin{itemize}[leftmargin=1em]
\item[-] {\it Primary CMB}: A total of 250 full-sky realizations of the unlensed primary CMB are generated by first computing the unlensed CMB power spectra for the \planck{} 2018 best-fit cosmology using \texttt{CAMB}, and then drawing Gaussian random fields with the corresponding power spectra. Gaussian realizations of the lensing potential are generated in the same manner.\footnote{In generating these $T/Q/U$ and $\phi$ maps, we also include the correlations between them using $C_{L}^{T\phi}$ and $C_{L}^{E\phi}$ computed from \texttt{CAMB}. } The unlensed CMB maps are then deflected using the package \texttt{Lenspix}. This operation is carried out at $\nside{}=8192$, up to $\ell_{\rm max}=17000$. We extract two non-overlapping $1500\,\sqdeg{}$ patches located on opposite hemispheres from each full-sky realization to reach 500 realizations (of which we use 498) and label this set of simulations \texttt{cmb1phi1}. We additionally generate 250 independent realizations of the unlensed CMB map but deflect them using the same lensing potential as the first set for the purpose of computing the $N_{L}^{(1)}$ bias (described in Section \ref{sec:N0N1}), which we label as \texttt{cmb2phi1}. 
\item[-] {\it Foregrounds}: We generate 250 full-sky Gaussian foreground realizations correlated with the lensing potential. For each Gaussian realization of the lensing potential, we construct foreground maps using the foreground-lensing cross-correlations measured from \textsc{Agora}. In this way, each set of Gaussian foreground maps preserves its cross-correlation with the corresponding lensing realization at the two-point level, matched to those measured in \textsc{Agora}.

In temperature, kSZ, CIB, and radio sources are bundled, whereas Compton-$y$ is treated separately and is converted into a temperature map using
\begin{equation}
\Delta T = T_{\rm CMB}\, g(\nu)\, y,
\end{equation}
where $T_{\rm CMB}=2.726\ {\rm K}$ and 
\begin{equation}
g(\nu)=x\, {\rm coth}(x/2)-4,
\end{equation}
with $x=h\nu/k_{\rm B}T_{\rm CMB}$. In polarization, we include the contribution from radio sources, although their impact on the maps used in this analysis is minimal due to the applied masking threshold and the multipole range used for lensing reconstruction~\cite{gupta2019}.
These foreground components are then added to the lensed CMB maps. For \texttt{cmb2phi1}, the foreground realizations are extracted from different regions of the full-sky foreground maps so that they differ from those used in the first simulation set.
\end{itemize}

For these Gaussian foreground simulations, the maps are already based on power spectra computed after masking $\xi>10$ clusters and sources above 6 mJy, so we can directly pass the maps through the mock observation pipeline.

\subsection{Noise}
As described in \cite{quan2026}, noise realizations are generated from sign-flip coadds. The full coadd is subtracted from each individual-observation map in order to remove the astrophysical sky signal as much as possible. Full-depth noise coadds are then formed by assigning random $\pm 1$ signs to the individual-observation maps before coadding, with the assignments chosen such that the two halves have nearly equal total weights. For a given observation, the same sign is applied across all three frequency bands in order to preserve the inter-frequency noise correlation structure. In practice, residual sky signal can remain due to effects such as time-varying astrophysical sources. We therefore identify the brightest residual sources and replace those pixels in the noise maps by randomly assigning neighboring pixel values. The resulting noise spectra for the $95$, $150$, and $220\,\mathrm{GHz}$ channels are shown in Figure~\ref{fig:noiselevels}. These noise realizations are added to both the $\texttt{cmb1phi1}$ and $\texttt{cmb2phi1}$ simulation sets, as well as to the non-Gaussian simulations. For $\texttt{cmb2phi1}$, the same set of noise realizations is reused with the realization indices rearranged, so that a given CMB realization is paired with a different noise realization than in $\texttt{cmb1phi1}$.
\section{Quadratic estimator}\label{sec:lensing}
\subsection{Suboptimal quadratic estimator (SQE)}
The general derivation of the curved-sky quadratic estimator is given in \cite{okamoto2003}, and the same formalism has been employed in several CMB lensing analyses \citep{planck2013lens,planck2015lens,omori2017,omori2023}. Briefly, the unnormalized estimator for the lensing potential can be written as
\begin{multline}\label{eq:QE_grad}
\overline{\phi}^{\alpha}_{L M}=\frac{(-1)^{M}}{2}\sum_{\ell_{1} m_{1}\ell_{2} m_{2}}\begin{pmatrix}
\ell_{1} & \ell_{2} & L \\
m_{1} & m_{2} & -M \\
\end{pmatrix} \times \\
W_{\ell_{1}\ell_{2}L}^{\alpha,\phi}\overline{X}_{\ell_{1}m_{1}}\overline{Y}_{\ell_{2}m_{2}}.
\end{multline}
Here $W_{\ell_{1}\ell_{2}L}^{\alpha,\phi}$ is the standard curved-sky lensing weight for the estimator pair $\alpha=XY$, constructed using fiducial lensed-gradient CMB spectra in the lensing response. Its functional form differs for the individual estimators $\alpha=\{\text{TT, EE, TE, TB, EB, ET, BT, BE}\}$.\footnote{Throughout this work, italicized field pairs, e.g., $TT$, denote power spectra, while upright field pairs, e.g., $\mathrm{TT}$, denote quadratic estimators.} $\overline{X}$ and $\overline{Y}$ are two inverse-variance filtered maps where $X,Y\!\in\left\{ T,E,B \right\}$. We use ($\ell$, $m$) and ($L$, $M$) to denote temperature/polarization and lensing modes, respectively throughout this work. 
In evaluating Equation \eqref{eq:QE_grad} we make use of the separability of the estimator introduced in \cite{dvorkin2009}, following an approach similar to that of  \cite{planck2015lens}, by performing the harmonic-space convolution as a map-space multiplication.

Once the filtered $\overline{\phi}$ estimates are computed, we form a minimum-variance lensing estimate by taking a weighted sum of the individual estimators~\cite{planck2013lens,planck2015lens}
\begin{equation}
\hat{\phi}_{LM}^{\rm MV} = \frac{\sum_{\alpha} \overline{\phi}_{LM}^{\alpha}}{\sum_{\alpha}\mathcal{R}_{L}^{\alpha\phi}},
\end{equation}
where 
$\overline{\phi}_{LM}^{\alpha}=\hat{\phi}_{LM}^{\alpha}\mathcal{R}_{L}^{\alpha\phi}$ is the response-weighted estimate of $\phi$, and $\mathcal{R}_{L}^{\alpha\phi}$ is the analytic response function for the given estimator $\alpha$ to the lensing potential $\phi$:

\begin{align}
\mathcal{R}^{\alpha\phi}_{L}
&=
\frac{1}{2(2L+1)}
\sum_{\ell_1\ell_2}
W^{\alpha}_{\ell_1\ell_2L}
\mathcal{W}^{\alpha,\phi}_{\ell_1\ell_2L}
F^{X}_{\ell_1}
F^{Y}_{\ell_2},
\end{align}
where $\mathcal{W}^{\alpha,\phi}_{\ell_1\ell_2L}$ denotes the physical covariance response of the $\alpha$ fields to lensing\footnote{$\mathcal{W}^{\alpha,\phi}_{\ell_1\ell_2L}$ and ${W}^{\alpha,\phi}_{\ell_1\ell_2L}$ are the same for a standard matched quadratic estimator, but need not be identical for modified estimators such as profile-hardened estimators.} and $F_{\ell}$ represents the diagonal filtering function $1/(C_{\ell}+N_{\ell})$.

We refer to this as the suboptimal minimum-variance estimator~\cite{maniyar2021} and denote it as ``MV.'' In practice, we replace the analytic response used for the normalization with a response measured from simulations. For each estimator, we compute the cross-spectrum between the reconstructed lensing map $\overline{\phi}$ and the corresponding input lensing potential for each simulation realization, average these cross-spectra over all realizations, and divide by the input lensing power spectrum:\footnote{We denote simulation-based and analytical response functions with and without a tilde, respectively.}

\begin{equation}\label{eqn:simresp}
\widetilde{\mathcal{R}}_{L}^{\alpha\phi}=\left\langle\frac{C_{L} ( \phi^{{\rm in}},\overline{\phi}^{\alpha}) } {C_{L} ( \phi^{{\rm in}},\phi^{ {\rm in}}) }\right\rangle.
\end{equation}
The response used here can be understood as the analytic response combined with a simulation-based MC correction, analogous to the MC-corrected analytic response adopted in~\cite{wu2019, pan2023}. This correction accounts for realistic effects that are not captured by the analytic calculation, including masking and filtering. For SPT-3G, the dominant contribution to this correction, at the level of roughly $20\%$ for the MV combination (and at a similar level for the global minimum variance estimator introduced next), arises from mode loss induced by the filtering.

\subsection{Global minimum variance}
As noted in~\cite{maniyar2021}, the ``minimum-variance" combination discussed above, formed by taking a weighted sum of the individual estimators, does not yield the lensing estimator with the lowest possible variance. A lower-variance estimator can instead be obtained by jointly filtering the $T$, $E$, and $B$ fields, accounting for their correlations, and optimally combining the resulting quadratic mode pairs. This is referred to as the ``global minimum-variance" estimator (GMV hereafter), which we adopt as our baseline choice in this work. Mathematically the reconstruction can be written as
\begin{align}\label{eq:gmvint}
\hat{\phi}_{LM}&=(\widetilde{\mathcal{R}}^{\rm GMV\, \phi})^{-1}\frac{(-1)^{M}}{2}\nonumber\\
&\times\sum_{\ell_{1} m_{1}\ell_{2} m_{2}} \begin{pmatrix}
\ell_{1} & \ell_{2} & L \\
m_{1} & m_{2} & -M \\
\end{pmatrix}  \overline{\mathbf{ X}}_{\ell_{1} m_{1}}^{\top} \mathbf{ W}_{\ell_{1}\ell_{2}L}^{{\rm GMV},\phi} \overline{\mathbf{ X}}_{\ell_{2}m_{2}},
\end{align}
where
\begin{equation}\label{eq:xbar}
	\overline{\mathbf{X}}_{\ell m} \equiv \bm{C}_\ell^{-1}\mathbf{X}_{\ell m},
\end{equation}
with $\mathbf{X}_{\ell m} = [T_{\ell m}, E_{\ell m}, B_{\ell m}]$ and
\begin{align}\label{eq:invcl}
	\bm{C}_\ell^{-1}
	&=
	\begin{bmatrix}
		C_\ell^{TT} & C_\ell^{TE} & 0\\
		C_\ell^{TE} & C_\ell^{EE} & 0\\
		0 & 0 & C_\ell^{BB}\\
	\end{bmatrix}^{-1}\nonumber\\
	&=
	\begin{bmatrix}
		C_\ell^{EE}/D_\ell & -C_\ell^{TE}/D_\ell & 0\\
		-C_\ell^{TE}/D_\ell &C_\ell^{TT}/D_\ell & 0\\
		0 & 0 & 1/C_\ell^{BB}\\
	\end{bmatrix},
\end{align}
with $D_\ell \equiv C_\ell^{TT}C_\ell^{EE} - \left[C_\ell^{TE}\right]^2$, where $C_{\ell}$ includes the CMB signal, noise, and foreground residuals. The weight function also consists of all the combinations
\begin{equation}
\mathbf{W}_{\ell_{1}\ell_{2}L}^{{\rm GMV},\phi} =\begin{bmatrix}
f_{\rm TT} & f_{\rm TE} & f_{\rm TB}\\
f_{\rm ET} & f_{\rm EE} & f_{\rm EB}\\
f_{\rm BT} & f_{\rm BE} & f_{\rm BB}\\
\end{bmatrix},
\end{equation}
where the functional forms of $f_{ij}$ are identical to those used in constructing the suboptimal quadratic estimator (see e.g., \cite{okamoto2003}).

\subsection{tSZ deprojection and cross-ILC}
As described in Section~\ref{sec:lc}, we use the LC procedure to produce both minimum-variance and foreground-deprojected temperature maps. Here, we use different combinations of these maps in the temperature block of the lensing estimator to reduce foreground biases while retaining as much statistical sensitivity as possible.

In this work, we consider two variants of this approach: tSZ deprojection~\cite{madhavacheril2018} and cross-ILC~\cite{raghunathan2023}. We incorporate both directly into the full GMV reconstruction following~\cite{nakato2026}, which allows different temperature maps to be used in the two estimator legs while consistently combining the temperature and polarization information within the GMV framework.

The first variant combines a tSZ-deprojected temperature map with a minimum-variance temperature map. Since the tSZ signal is nulled in one of the two input maps, tSZ-induced correlations between the two legs cannot generate the leading spurious mode coupling that would otherwise be interpreted as lensing. At the same time, using a minimum-variance map in the other leg avoids the substantial noise penalty that would result from deprojecting both input maps. We adopt the shorthand notation ``tSZdpj" for this type of reconstruction.
The second variant follows a similar approach but uses two differently foreground-deprojected maps: a tSZ-deprojected and a CIB-deprojected temperature map. This allows the reconstruction to suppress contamination from both tSZ and CIB simultaneously. We refer to this type of reconstruction as ``xILC" throughout this work.

\subsection{Profile hardening}
One extension to the global minimum-variance framework is the profile-hardening technique~\cite{osborne2014,sailer2020}, a generalization of the ``bias hardening" technique described in~\cite{namikawa2013}, which modifies the lensing estimator so that it has zero linear response to a specified source profile. These profiles are chosen to match those of contaminants such as galaxy clusters, infrared galaxies, and radio galaxies, thereby suppressing their contribution to the reconstructed lensing signal. Hereafter, we use ``prof" to denote the profile-hardened GMV reconstruction.

The weight function of this estimator can be written as
\begin{equation}\label{eq:profhard}
W^{\alpha,\phi-s}_{\ell_{1}\ell_{2}L} = W^{\alpha,\phi}_{\ell_{1}\ell_{2}L} - \frac{\mathcal{R}_{L}^{{\rm TT},s\phi} }{\mathcal{R}_{L}^{{\rm TT},ss}} W^{{\rm TT},s}_{\ell_{1}\ell_{2}L},
\end{equation}
where
\begin{equation}
W^{{\rm TT},s}_{\ell_{1}\ell_{2}L}=u_{\ell_{1}}u_{\ell_{2}}\sqrt{\frac{(2\ell_{1}+1)(2\ell_{2}+1)(2L+1)}{4\pi} }\begin{pmatrix}
\ell_{1} & \ell_{2} & L\\
0 & 0 & 0
\end{pmatrix},
\end{equation}
$u_{\ell_{1}}$ and $u_{\ell_{2}}$ denote the source profiles to be hardened against (with point-source hardening corresponding to the special case $u_{\ell_1}=u_{\ell_2}=1$), and the response functions are given by:
\begin{align}
\mathcal{R}_{L}^{{\rm TT},s\phi}&=\frac{1}{2(2L+1)}\sum_{\ell_{1}\ell_{2}}W^{{\rm TT},s}_{\ell_{1}\ell_{2}L} (M^{-1}_{\ell_{1}\ell_{2}})_{11}W^{{\rm TT},\phi}_{\ell_{1}\ell_{2}L}\\
\mathcal{R}_{L}^{{\rm TT},ss}&=\frac{1}{2(2L+1)}\sum_{\ell_{1}\ell_{2}}W^{{\rm TT},s}_{\ell_{1}\ell_{2}L} (M^{-1}_{\ell_{1}\ell_{2}})_{11}W^{{\rm TT},s}_{\ell_{1}\ell_{2}L}
\end{align}
with
\begin{align}
(M^{-1}_{\ell_{1}\ell_{2}})_{11}= \frac{ C^{EE}_{\ell_{1}} C^{EE}_{\ell_{2}}  }{2D_{\ell_{1}}D_{\ell_{2}}},
\end{align}
which is the inverse pair covariance for the TT-block. In practice, we compute these two response functions analytically using the flat-sky approximation. The difference between the flat- and curved-sky responses is small, with the largest difference seen on the largest scales, and much of this geometric difference cancels further when taking the ratio of the two flat-sky responses.

Plugging this into Equation~\eqref{eq:gmvint} gives us:
\begin{align}
&\hat{\phi}_{LM}^{\rm GMV,BH}=(\widetilde{\mathcal{R}}^{\rm GMV\, \phi})^{-1}\frac{(-1)^{M}}{2}
\sum_{\ell_{1} m_{1}\ell_{2} m_{2}} \begin{pmatrix}
\ell_{1} & \ell_{2} & L \\
m_{1} & m_{2} & -M \\
\end{pmatrix}\nonumber\\
&\times\Biggl[\overline{\bm{ X}}_{\ell_{1} m_{1}}^{\top} \bm{ W}_{\ell_{1}\ell_{2}L}^{{\rm GMV},\phi} \overline{\bm{ X}}_{\ell_{2}m_{2}}-\frac{\mathcal{R}_{L}^{{\rm TT},s\phi}}{\mathcal{R}_{L}^{{\rm TT},ss}}\overline{T}_{\ell_{1}m_{1}}W^{{\rm TT},s}_{\ell_{1}\ell_{2}L}\overline{T}_{\ell_{2}m_{2}}\biggl].
\end{align}
We note that in this GMV implementation, the hardening is applied only to source contributions entering through $\overline{T}$ with weight $f_{\rm TT}$. Consequently, it does not remove source contamination that enters other GMV terms through the $T$ contribution to the inverse-filtered field $\overline{E}$, which arises from the nonzero $TE$ covariance.

Previous studies have shown that, although profile-hardening estimators are constructed for Poisson-distributed sources with a specified profile, they also substantially suppress contamination from sources with different profiles and from the clustering of these sources~\cite{qu2024,maccrann2023}.
In this work, we choose our baseline profile to match that of a modified tSZ power-spectrum profile from \textsc{Agora}, $(C_{\ell}^{\rm tSZ})^{\gamma}$,\footnote{The overall normalization of the profile is arbitrary, as it cancels in the profile-hardening construction.} which results in a cleaner lensing map as tested by stacking at the locations of galaxy clusters and sources in our data, as discussed in the following section.

\begin{figure*}
\includegraphics[width=0.9\linewidth]{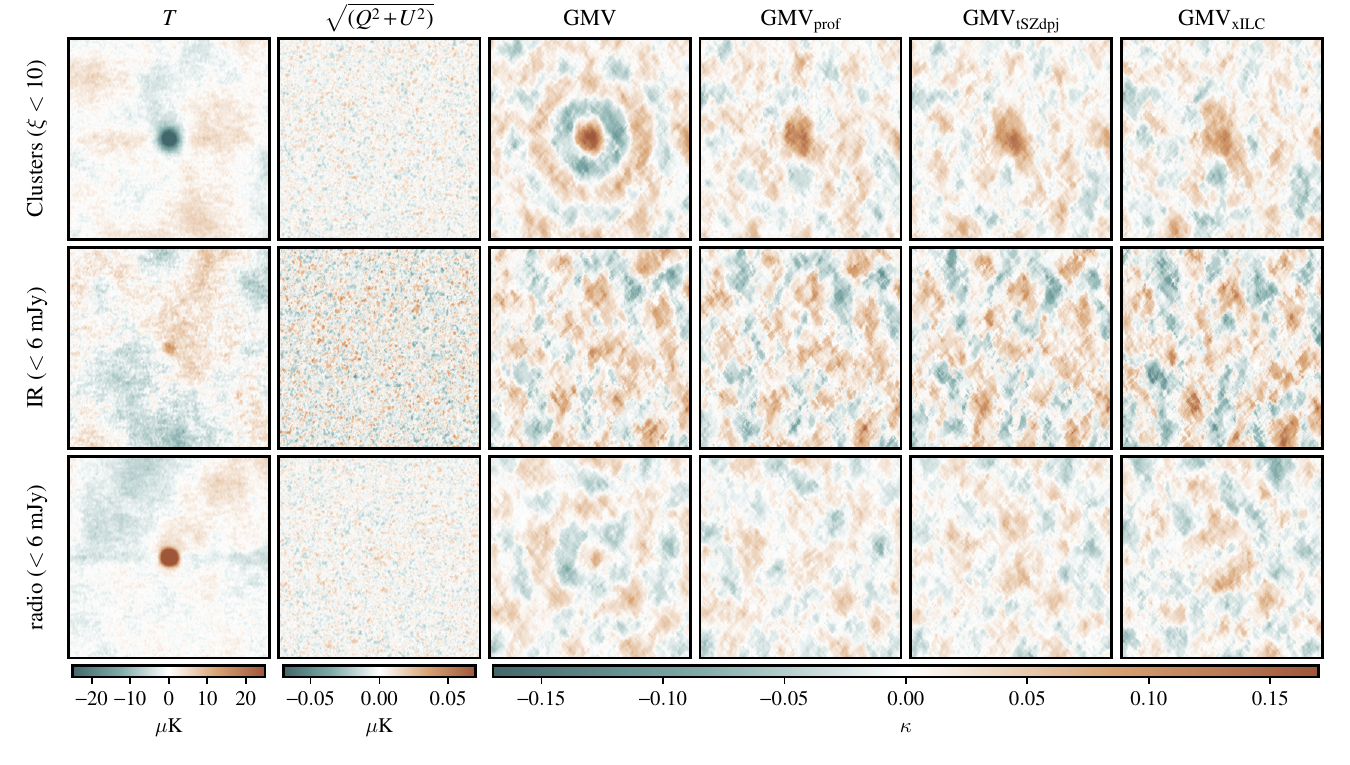} 
\caption{Images of maps stacked at the location of galaxy clusters (top), infrared sources (middle), and radio sources (bottom) just below our masking threshold ($\xi=10$ for clusters and 6\,mJy for infrared/radio sources). The first two columns show stacks from the linearly combined minimum-variance temperature map (first column) and the quadrature sum of the linearly combined $Q$ and $U$ map (second column). The next four columns show stacks from different variants of the lensing maps. In the cluster stacks, the central positive feature in the reconstructed convergence maps is the expected CMB-cluster lensing signal \cite{baxter2015,raghunathan2017,raghunathan2019}, while the surrounding oscillatory pattern, most clearly visible for the baseline GMV reconstruction, is a reconstruction bias induced by the cluster-associated temperature signal and map filtering.}
\label{fig:stack}
\end{figure*}

\subsection{Visual inspection of foreground residuals from different estimators}
In Figure~\ref{fig:stack}, we show the four reconstructed lensing maps, GMV, $\gmvprof{}$, $\gmvtsz{}$, and $\gmvxilc{}$, stacked at the locations of galaxy clusters detected below $\xi=10$ and emissive sources detected below 6\,mJy. These stacked images show residual foreground contamination not removed by masking, although the cluster stacks also contain the true cluster lensing signal, visible as a positive convergence peak at the center in the first-row panels of Figure~\ref{fig:stack}.

In the baseline GMV case, we find that the foreground residuals not only leave imprints at the locations of the contaminants but also produce an extended ringing pattern around the central feature in both the cluster and radio-source stacks. No comparable feature can be identified in the infrared-source stack, suggesting that the residual contamination from those sources is small. For $\gmvprof{}$, we test profiles of the form $(C_{\ell}^{\rm tSZ})^{\gamma}$, with $\gamma = \{1, 0.5, 0.25\}$. We find that $\gamma=0.25$ yields the smallest residuals across all three stacks and therefore adopt this value for the $\gmvprof{}$ reconstruction. With this choice, we see no prominent residual structure in any of the three stacks. The $\gmvtsz{}$ and $\gmvxilc{}$ reconstructions show very similar behavior, with substantially reduced residual structure in the cluster and radio-source stacks. Although the LC weighting used for $\gmvtsz{}$ is known to enhance the CIB amplitude, neither reconstruction shows an obvious excess when stacked at the positions of detected infrared sources. While this does not rule out residual CIB contamination more generally, the infrared-source stack provides a useful diagnostic of contamination associated with the brightest detected sources.

\subsection{Debiasing and normalization}
\begin{figure*}
\centering
\includegraphics[width=1.000\textwidth]{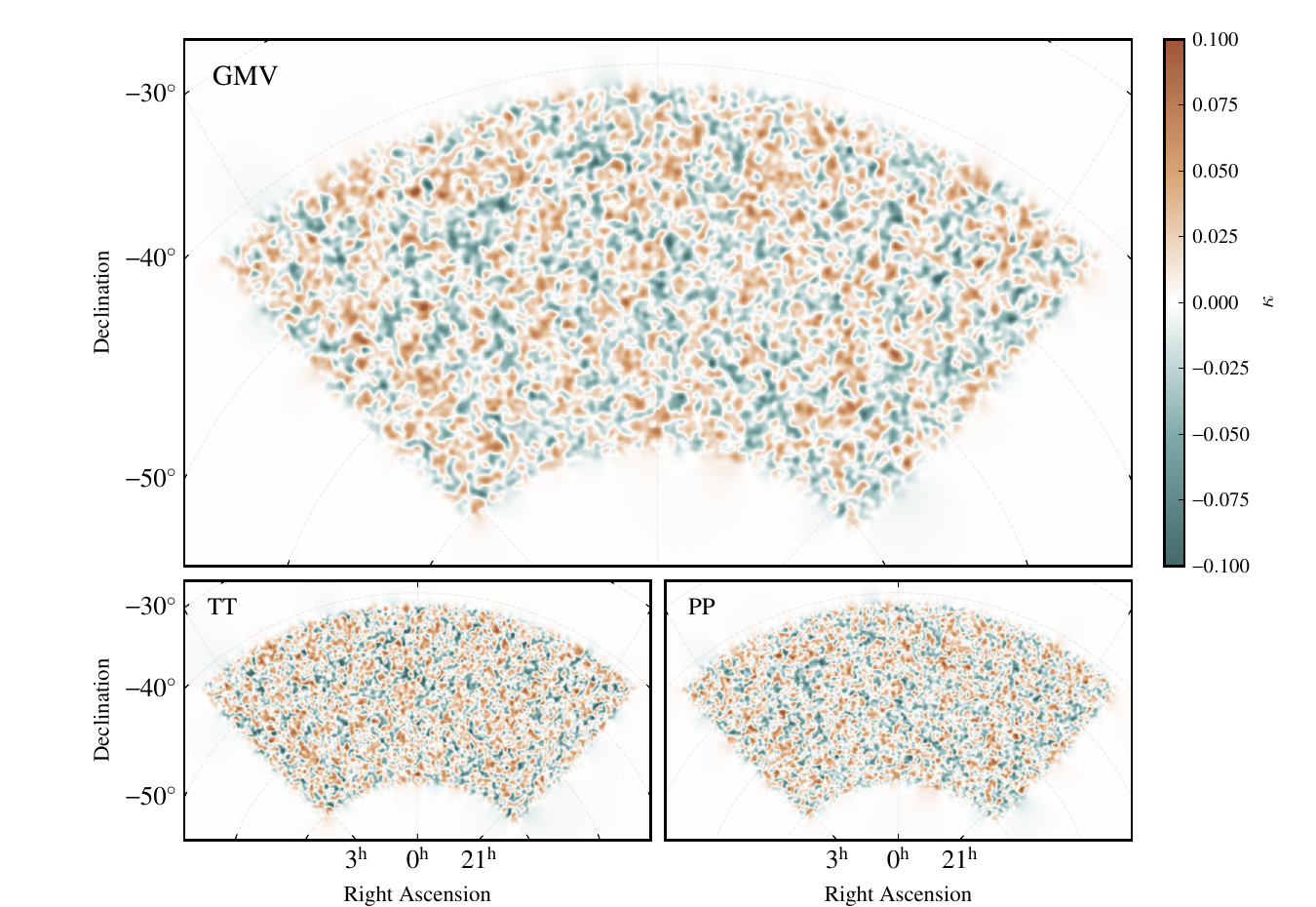} 
\caption{The GMV lensing map reconstructed from the observed SPT-3G D1 maps (top), temperature-only reconstruction (lower left), and polarization-only lensing reconstruction (lower right). These maps have been smoothed by a $30'$ FWHM Gaussian for visualization purposes.}
\label{fig:kappamaps}
\end{figure*}

Once the filtered $\overline{\phi}$ estimates are computed, we subtract the simulation-derived mean-field bias, normalize by the response function, and multiply by factors of $L$ to convert the result into convergence $\kappa$:
\begin{equation}
\hat{\kappa} = \frac{1}{2}L(L+1)\,(\overline{\phi}_{LM}-\overline{\phi}_{LM}^{\rm MF})/\widetilde{\mathcal{R}}^{\alpha\phi}_{L},
\end{equation}
where the mean-field is defined as
\begin{equation}
\overline{\phi}^{\rm MF}_{LM} = \left\langle \overline{\phi}_{LM} \right\rangle_{\rm sim},
\end{equation}
with the average taken over 498 simulation realizations. The resulting convergence maps after applying this normalization are shown in Figure \ref{fig:kappamaps}, separately for GMV, TT (temperature-only), and PP (polarization-only) reconstructions. 

\begin{figure}
\includegraphics[width=1.00\linewidth]{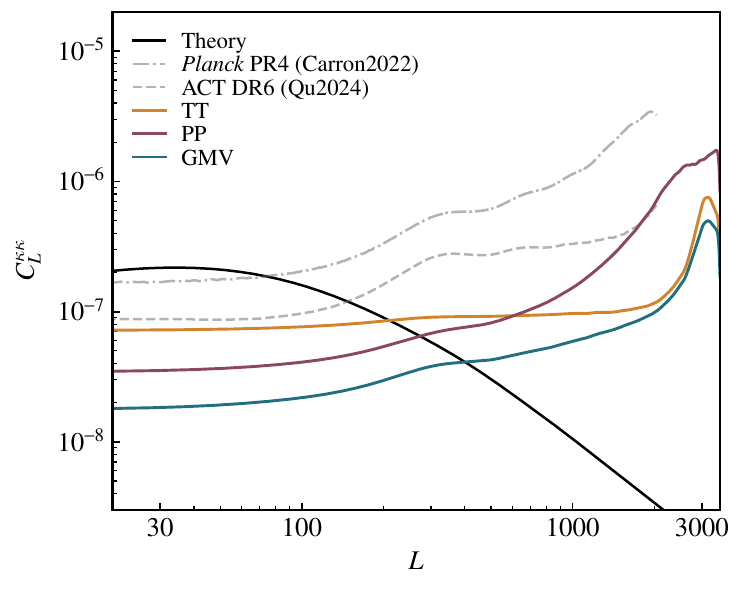} 
\includegraphics[width=1.00\linewidth]{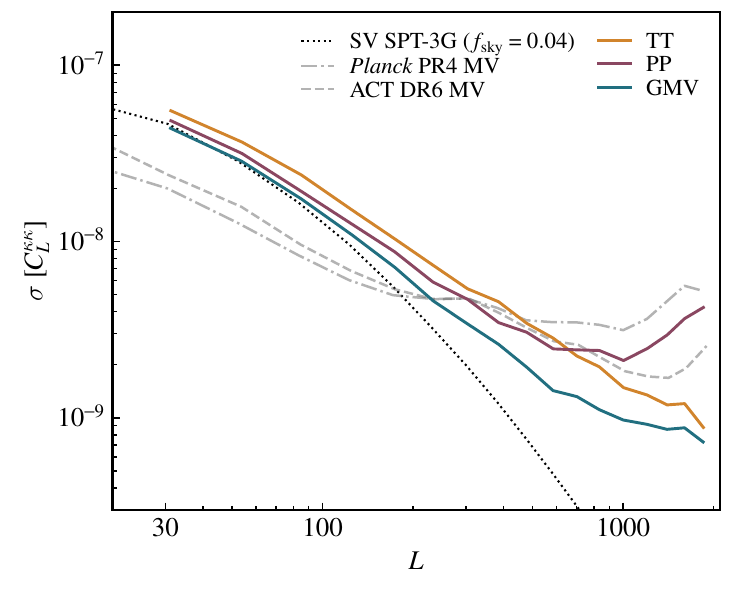} 
\includegraphics[width=1.00\linewidth]{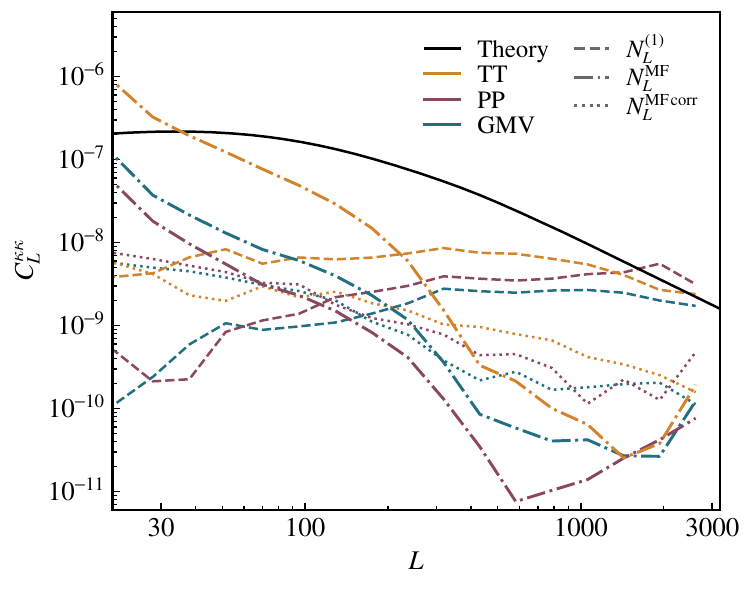} 
\caption{ {\bf Upper:} Comparison of noise levels per-mode from \planck{} PR4 \cite{carron2022}, ACT\,DR6 \cite{qu2024}, and this analysis (colored lines). {\bf Center:} Equivalent but cast in terms of lensing bandpower errors, which take $f_{\rm sky}$ into account. The black dotted line corresponds to the sample variance limit computed from $f_{\rm sky }$ for SPT-3G's $1500 \,\sqdeg$ field. {\bf Bottom:} the amplitudes of the $N_{L}^{(1)}$ bias, the mean-field bias $N_{L}^{\rm MF}$, and the mean-field correlated term $N_{L}^{\rm MFcorr}$.}
\label{fig:lens_noise}
\end{figure}

\subsection{Lensing noise biases}\label{sec:N0N1}

The auto-spectrum of the reconstructed convergence map contains both the
lensing signal and reconstruction-noise biases. Schematically,
\begin{equation}\label{eq:spectrum_debias}
\hat{C}_{L}^{\kappa\kappa}
=C_{L}^{\hat\kappa\hat\kappa}-N_{L}^{(0)}-N_{L}^{(1)}-\cdots ,
\end{equation}
where $C_{L}^{\hat{\kappa}\hat{\kappa}}$ is the raw measured convergence power spectrum and
$N_{L}^{(0)}$ and $N_{L}^{(1)}$ are the conventional disconnected and
secondary-contraction bias terms. The desired signal and the $N_{L}^{(1)}$ bias both originate from the lensing-induced connected CMB four-point function, whereas $N_{L}^{(0)}$ is associated with the disconnected Gaussian contractions.

The $N_{L}^{(0)}$ bias is estimated from the Gaussian simulations by reconstructing lensing from pairs of different realizations and averaging their cross-spectra:
\begin{align}
N_{L}^{(0)}=\bigg\langle &C_{L}([x_{i,a},x_{j,a}],[x_{i,a},x_{j,a}])\nonumber\\
+&C_{L}([x_{i,a},x_{j,a}],[x_{j,a},x_{i,a}])\bigg\rangle,
\end{align}
where $x_{i,a}$ and $x_{j,a}$ are two different realizations from simulation set $a$, with $a$ labeling the \texttt{cmb1phi1} simulation set, and the average is taken over 498 realizations. Similarly, $N_{L}^{(1)}$ is calculated using:
\begin{align}
\hspace{1cm}N_{L}^{(1)}=\bigg\langle &C_{L}([x_{i,a},x_{i,b}],[x_{i,a},x_{i,b}])\nonumber\\
+&C_{L}([x_{i,a},x_{i,b}],[x_{i,b},x_{i,a}])\bigg\rangle
- N_{L}^{(0)},
\end{align}
where $b$ labels the \texttt{cmb2phi1} simulation set, which contains different background CMB realizations but is lensed by the same lensing potential as \texttt{cmb1phi1}, as described in Section~\ref{sec:Gsim}.
For the data spectrum, we instead use the realization-dependent $N_{L}^{(0)}$ estimator~\cite{namikawa2013}, in which lensing maps are reconstructed using combinations of the data and simulation maps:
\begin{align}
N_{L}^{(0),{\rm RD}}&=\bigg\langle C_{L}([d,x_{i,a}],[d,x_{j,a}])+C_{L}([x_{i,a},d],[x_{j,a},d])\nonumber\\
&+\hspace{0.3cm}C_{L}([d,x_{i,a}],[x_{i,a},d])+C_{L}([x_{i,a},d],[d,x_{i,a}])\bigg\rangle\nonumber\\
&-N_{L}^{(0)}.
\end{align}
This construction cancels the leading-order sensitivity of the $N_{L}^{(0)}$ bias to mismatches between the data and simulation covariances, thereby providing a more accurate estimate of the disconnected bias in the raw data lensing power spectrum. Ideally, one would compute $N_{L}^{(0),{\rm RD}}$ for all the simulation realizations too, as this is known to reduce the off-diagonal elements in the covariance matrix~\cite{hanson2011}. However, this requires $N_{\rm sim}\times N_{\rm sim}$ calculations and is computationally expensive, so we instead use an analytic estimate. This is the semi-analytic $N_{L}^{(0)}$, which we denote $N_{L}^{(0),{\rm SA}}$, and is computed directly from the power spectra measured from the maps entering the quadratic estimator~\cite{hanson2011}.
\begin{equation}
N_{L}^{(0),{\rm SA}}=\frac{\widetilde{\mathcal{R}}_{L}^{-2} }{(2L+1)}\sum_{\ell_{1}\ell_{2}} |W_{\ell_{1}\ell_{2}L}^{\alpha,\phi}|^{2} C_{\ell_{1}}^{\alpha,{\rm tot}} C_{\ell_{2}}^{\alpha,{\rm tot}}.
\end{equation}
We subtract one additional bias term correlated with the mean field. We estimate this term from the power spectrum of the input-subtracted convergence maps, which represent the reconstruction residual at the map level, i.e., for $\hat{\kappa}^{\rm res}=\hat{\kappa}-\kappa^{\rm in}$, 
\begin{equation}\label{eq:mfnoise}
N_{L}^{\rm MFcorr} = \big\langle C_{L}(\hat{\kappa}^{\rm res},\hat{\kappa}^{\rm res})\big\rangle - N_{L}^{(0)} -N_{L}^{(1)}. 
\end{equation}
This bias becomes non-negligible at high $L$. As shown in Figure~\ref{fig:lens_noise}, its amplitude ranges from $4$--$7\%$ across the different lensing reconstructions at $L=2000$.\footnote{This term is present only in the lensed simulations; no corresponding signature is observed in the unlensed simulations.} We describe the formulation of this analysis-specific correction in Appendix~\ref{app:mfnoise}.

Finally, the measured debiased lensing spectra are binned by taking a weighted average of $\hat{C}^{\kappa\kappa}_{L}$ over the multipoles within each bin:
\begin{equation}
\hat{C}^{\kappa\kappa}_{ b}=\frac{\sum_{L_{\rm min}}^{{L_{\rm max}}} w_{L}^{b}\hat{C}^{\kappa\kappa}_{L}  }{\sum_{L_{\rm min}}^{{L_{\rm max}}} w_{L}^{b}}.
\end{equation}
We use 17 bins spanning $30 \leq L \leq 3000$: three logarithmically spaced bins between $L=30$ and $L=100$, and 14 logarithmically spaced bins between $L=100$ and $L=3000$, with the bin edges rounded to the nearest integer. The binning function $w_L^b$ is calculated as $1/\sigma^2(\hat{C}_L^{\kappa\kappa})$, where $\sigma^{2}(\hat{C}_L^{\kappa\kappa})$ is computed from the scatter across 498 simulation realizations.

We next compare the noise level of our data with other datasets. The noise levels of the different estimators are shown in  Figure~\ref{fig:lens_noise}. The per-mode reconstruction noise in this work is lower than that of \planck{}\,PR4 and ACT\,DR6. The TT reconstruction is signal-dominated out to $L\sim200$, whereas PP lensing is signal-dominated out to $L\sim300$ and is more sensitive than TT reconstruction out to $L\sim600$. The global minimum-variance combination is signal-dominated out to $L\sim450$, and we find on average a 10\% improvement in the noise level compared to using the suboptimal minimum-variance estimator. 

These noise levels can also be translated into bandpower uncertainties based on the sky coverage of each survey. Given our $1500\,\sqdeg{}$ footprint, our lensing reconstruction is close to sample-variance limited for $L<100$. We therefore expect limited gains from increasing the survey depth on these scales. In contrast, \planck{} and ACT\,DR6 observe a wider area and therefore have a lower sample-variance limit and thus tighter constraints on these modes. On smaller scales, beyond $L=300$, the SPT-3G GMV lensing spectrum has tighter statistical uncertainties. This shows that, while the three lensing measurements have similar overall bandpower signal-to-noise ratios, they probe different scales and are therefore highly complementary.
\section{Pipeline tests, systematic checks, and blinding}\label{sec:validation}

In this section, we perform a series of validation tests to verify the robustness of the reconstruction pipeline and the resulting measurements. We first demonstrate that the analysis pipeline accurately recovers the input lensing signal in simulations. We then carry out a range of null tests and consistency checks designed to identify residual contamination from instrumental effects, foregrounds, and analysis choices. These include reconstructions on unlensed simulations, curl-mode null tests, variations of the CMB multipole range used in the reconstruction, and comparisons between reconstructions with different foreground sensitivities. Finally, we describe the blinding procedure adopted to prevent experimenter bias during the development and validation of the analysis.

\subsection{Simulation mean test}
\begin{figure}
\includegraphics[width=1.00\linewidth]{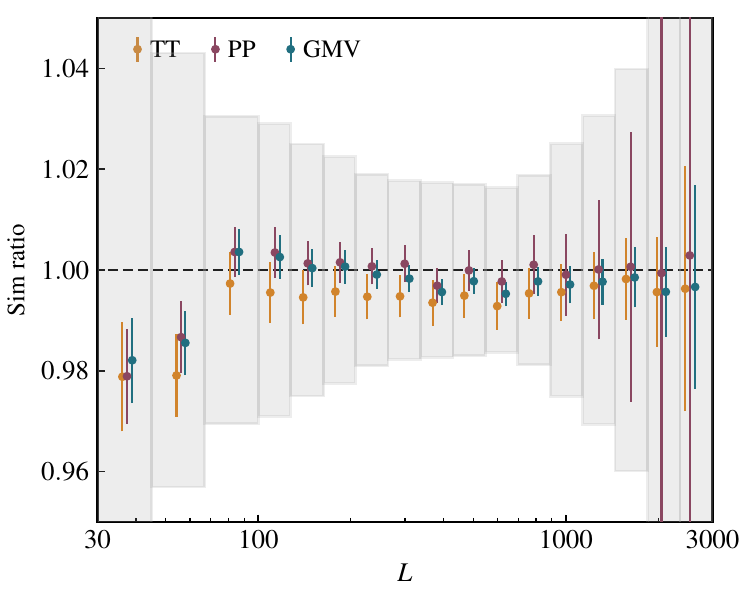} 
\caption{Simulation mean spectra computed from 498 realizations for TT, PP, and GMV. Gray bars denote 0.3$\sigma$ of the GMV reconstruction, which has the tightest error bars of the 3 estimators. }
\label{fig:simmean}
\end{figure}

We first test whether the mean reconstructed lensing spectrum across simulations accurately recovers the known input signal. Any discrepancy between the reconstructed and input signals would indicate a systematic bias introduced during the mock observation, frequency combining, $C^{-1}$ filtering, lensing reconstruction, or power spectrum calculation steps, including biases from masking. This test therefore provides an end-to-end validation that the pipeline is correctly recovering the lensing signal it is designed to measure. As our passing criterion, we require the mean reconstructed spectra from 498 realizations for the TT, PP, and GMV estimators to agree with the input theory to within $0.3\sigma$ in every $L$-bin, where $\sigma$ is the statistical uncertainty of the bandpower measurement for the corresponding estimator. We use this bandpower-level criterion rather than defining the validation threshold in terms of shifts in cosmological parameters, since the latter depends on analysis choices such as the cosmological model and priors adopted in the inference. In contrast, a bandpower-level criterion provides a direct and analysis-independent test of the fidelity of the lensing reconstruction itself. The $0.3\sigma$ threshold should therefore be interpreted as a tolerance for identifying appreciable reconstruction bias in any individual bin, rather than as a prediction for the corresponding shift in a particular parameter constraint. A fully coherent displacement of all bins at the threshold could produce a larger parameter shift, but such a configuration represents an extreme limiting case. As shown in Figure~\ref{fig:simmean}, this criterion is satisfied in all bins for all estimators that we require passing. In practice, the agreement is substantially better than the adopted threshold, with most bins recovering the input theory to much better than $0.1\sigma$ and only a small number reaching deviations of approximately $0.15\sigma$. The reconstructed lensing spectra therefore recover the input theory comfortably within the $0.3\sigma$ per-bin requirement, providing an end-to-end validation that the full mapmaking and lensing reconstruction pipeline introduces no appreciable bias in the recovered signal.

\subsection{Unlensed\ CMB}
We generate 50 realizations of CMB $T/Q/U$ maps drawn from the fiducial {\it lensed} CMB power spectra, but intentionally omit the deflection operation so that the resulting maps contain no lensing-induced mode coupling. Each realization is then passed through the full analysis pipeline in exactly the same manner as the fiducial simulations, including mock observation, LC map construction, $C^{-1}$ filtering, and lensing reconstruction with the quadratic estimator.

Because these maps contain no lensing signal, the reconstructed bandpowers provide a stringent test of the mean-field subtraction and $N_{L}^{(0)}$ debiasing procedures. For this case, the debiased spectrum excludes the $N_{L}^{(1)}$ correction, as $N_{L}^{(1)}$ is proportional to $C_L^{\kappa\kappa}$ but $C_L^{\kappa\kappa}$ is zero. Additionally, we do not subtract $N_L^{\rm MFcorr}$ because $\kappa^{\rm in}$ is zero for this setup. The debiased spectrum is simplified to:
\begin{equation}
\hat{C}_{L}^{uu} = C^{\hat{u}\hat{u}}_{L} - N_{L}^{(0)}.
\end{equation}
The resulting bandpowers are shown in the upper panel of Figure~\ref{fig:unlcurl}. For each estimator, we compute the $\chi^{2}$ of the mean debiased spectrum of the 50 unlensed realizations relative to zero. Since 50 realizations are insufficient for a stable inverse covariance, we instead use the covariance of the 498 signal-subtracted ($\hat{\kappa}-\kappa^{\rm in}$) lensed simulations, rescaled by a single factor to match the variance of the unlensed ensemble. We divide this covariance by 50 to obtain the covariance of the mean unlensed spectrum and apply the Hartlap correction when computing its inverse, using $N_{\rm s}=498$. We obtain PTEs of  0.06, 0.22, and 0.16 for TT, PP, and GMV, respectively, all satisfying our requirement of $\mathrm{PTE}>0.05$: the pipeline generates no spurious lensing power in the absence of a true lensing signal.

\subsection{Systematic checks}

\subsubsection{Curl}
The lensing deflection can be decomposed into a gradient component and a curl component:
\begin{equation}\label{eq:deflection}
\vec{d}(\hat{n})=\vec{\nabla}\phi(\hat{n})+(\star \vec{\nabla})\,\omega(\hat{n}),
\end{equation}
where $\phi$ is the lensing potential, $\omega$ is the curl potential, and $\star$ denotes a $90^{\circ}$ anti-clockwise rotation of the gradient operator.

For scalar density perturbations, and under the Born approximation, the curl component vanishes (\,$\omega=0$\,). This implies that standard large-scale structure lensing produces no curl signal at first order.\footnote{At higher order, post-Born corrections and tensor or vector metric perturbations generate a non-zero $\omega$. For the noise levels relevant to this work, these contributions are expected to be negligible and are therefore ignored. Future datasets, especially when combined with large-scale structure measurements, will enable a detection of this effect \cite{robertson2023,carron2025}. }
Consequently, any apparent detection of a curl signal is an indicator of residual systematics in the data or pipeline, making the curl reconstruction a powerful null test for the lensing analysis.

\begin{figure}
\includegraphics[width=1.00\linewidth]{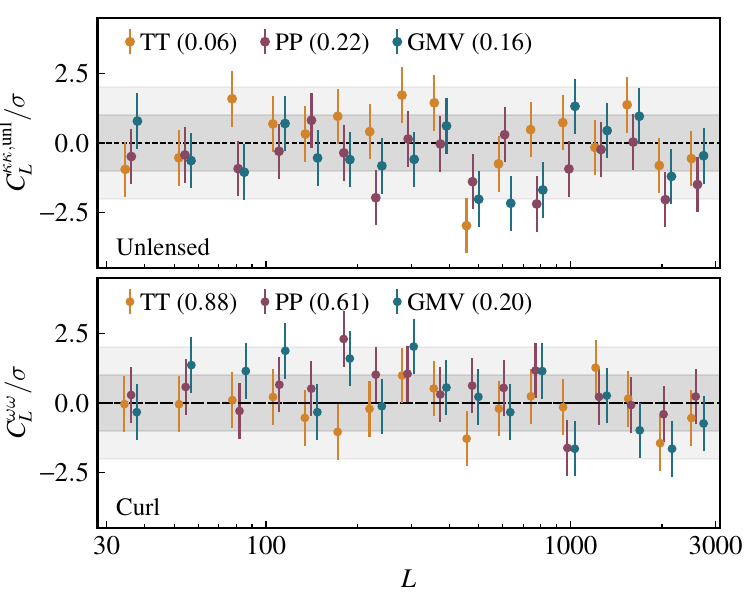} 
\caption{Measurements of the unlensed and curl spectra, divided by their respective statistical uncertainties. For the curl spectra, the corresponding simulation means have been subtracted.}
\label{fig:unlcurl}
\end{figure}

Maps of the reconstructed curl field, $\hat{\omega}$, can be computed using the same pipeline applied to the gradient reconstruction $\hat{\phi}$, with the only modification being the choice of quadratic-estimator weights. Specifically, the standard gradient-mode weight functions are replaced with those appropriate for isolating the curl component in the lensing-induced mode coupling:\footnote{In practice, the curl component is obtained simultaneously with the gradient component. In our implementation, the lensing reconstruction is carried out in position space using the spin-1 deflection field defined in Equation~\eqref{eq:deflection}, and the spin-1 spherical harmonic transform used to obtain $\hat{\phi}$ returns both the gradient and curl components, the latter of which is equivalent to Eq.~(\ref{eq:QE_curl}).}
\begin{multline}\label{eq:QE_curl}
\omega_{L M}^{\alpha}=\frac{(-1)^{M}}{2}\sum_{\ell_{1} m_{1}\ell_{2} m_{2}}\begin{pmatrix}
\ell_{1} & \ell_{2} & L \\
m_{1} & m_{2} & -M \\
\end{pmatrix} \times \\
W_{\ell_{1}\ell_{2}L}^{\alpha,\omega}\overline{X}_{\ell_{1}m_{1}}\overline{Y}_{\ell_{2}m_{2}}.
\end{multline}
We detect a non-zero curl power spectrum $C_{L}^{\omega\omega}$ in both data and simulations, for all the estimators, although the amplitude is much smaller for TT lensing reconstruction.  The amplitude of the data curl spectrum is comparable to that of the simulations.
The close agreement between the data and simulation curl spectra indicates that this feature is captured by our simulation pipeline, and we attribute it to systematic effects introduced by the mapmaking and filtering procedures. Given that the gradient simulation mean-input recovery test passes, this residual systematic does not appreciably bias the reconstructed gradient-mode lensing power spectrum. We therefore subtract the mean curl spectrum measured from the simulations from the data spectrum, interpreting the simulation-derived contribution as a known systematic offset. The uncertainty in this correction, estimated from the scatter among the simulations divided by the square root of the number of simulations, is negligible compared to the statistical uncertainty of the measured curl bandpowers and is therefore not included as an additional contribution to the covariance. The resulting bandpowers are shown in the lower panel of Figure~\ref{fig:unlcurl}. After subtracting the simulation-derived systematic offset, we find that the curl power spectrum is consistent with zero for all TT, PP, and GMV estimators, with $\mathrm{PTE}=0.88$, $0.61$, and $0.20$, respectively.  All three residuals pass the null test, indicating no evidence for additional curl contamination beyond that captured by the simulations.

\begin{figure*}
\includegraphics[width=1.00\linewidth]{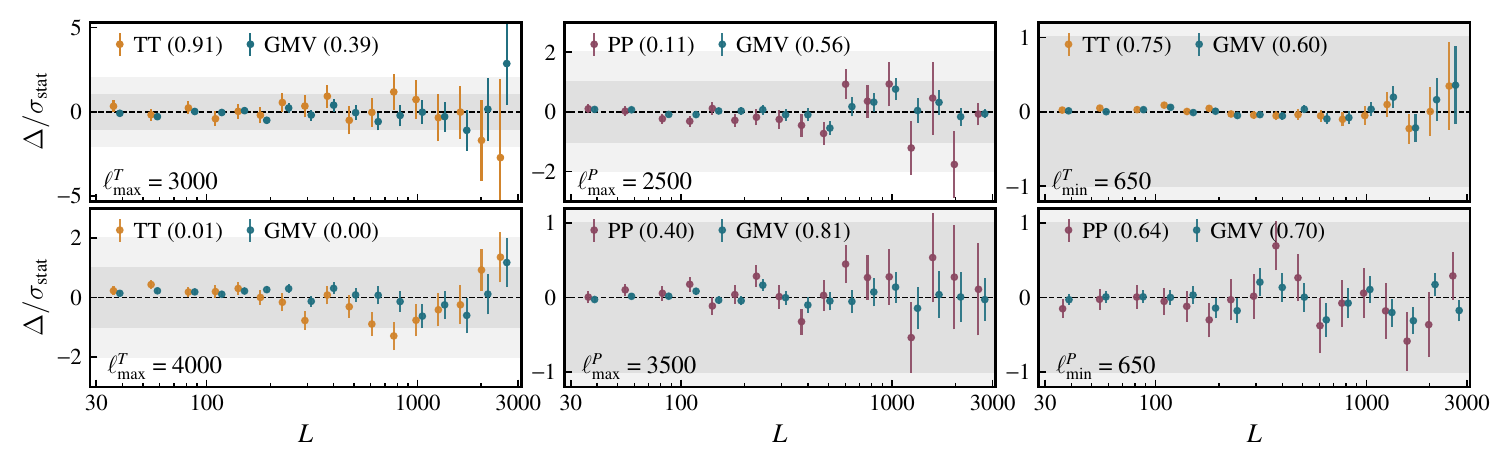} 
\caption{Data-based difference test for different $\ell_{\rm max}^{T}$,  $\ell_{\rm max}^{P}$, $\ell_{\rm min}^{T}$, and $\ell_{\rm min}^{P}$ choices. The plotted differences are normalized by the statistical uncertainties of the baseline analysis, whereas the error bars show the uncertainties on the differences, with the covariance between the two measurements taken into account.}
\label{fig:lrange}
\end{figure*}

\subsubsection{$\ell$ range tests}
We reconstruct CMB lensing from filtered temperature and polarization maps using only a finite range of input CMB multipoles. Varying these input-mode cuts provides a useful diagnostic of possible foreground and map-level systematics. In particular, raising the minimum multipole, $\ell_{\rm min}$, tests sensitivity to large-scale modes that are more affected by filtering, atmospheric noise, Galactic foregrounds, and mask-induced leakage. Lowering the maximum multipole, $\ell_{\rm max}$,  tests sensitivity to small-scale modes where foreground contamination, beam uncertainties, and noise modeling become increasingly important. Therefore, these tests probe whether the reconstructed lensing bandpowers are stable when the quadratic estimator is forced to use different subsets of the temperature and polarization information. We emphasize that the cuts varied here are applied to the input CMB temperature and polarization modes used in the reconstruction, not to the reconstructed lensing multipole $L$.

\noindent {\it Temperature $\ell_{\rm min}$/$\ell_{\rm max}$}: For the baseline analysis, we use temperature multipoles in the range $500 \leq \ell^{T} \leq 3500$ and test the robustness of our results to more restrictive multipole cuts. To test the sensitivity to the lower multipole bound, we increase the cut to $\ell_{\rm min}=650$ and perform a difference test. This yields ${\rm PTE}=0.75$ and $0.60$ for the TT and GMV reconstructions, respectively (the PP reconstruction is unaffected). This strong consistency indicates no detectable shift in the lensing reconstruction associated with large-scale contaminants such as Galactic dust emission and atmospheric noise under this test.

At high multipoles, extragalactic foregrounds become an increasingly important contribution, and extending the $\ell$ range would make the reconstruction more susceptible to foreground-induced biases. We therefore test the robustness of our results to the upper multipole bound by performing an analysis restricted to $\ell_{\rm max}=3000$, which constitutes a more conservative scale cut and reduces sensitivity to foreground contamination.

As shown in the upper left panel of Figure \ref{fig:lrange}, we find the two reconstructions to be consistent, with ${\rm PTE}=0.91$ and $0.39$ for the TT and GMV reconstructions.
This suggests that the extra foreground bias in the baseline $\ell_{\rm max}^{T}$ is not significantly larger than that of the $\ell_{\rm max}^{T} = 3000$ reconstruction. 

We additionally test a more aggressive scale cut of $\ell^{T}_{\rm max}=4000$ to assess whether additional signal-to-noise can be extracted from smaller scales. Increasing $\ell^{T}_{\rm max}$ increases the number of Fourier modes entering the quadratic estimator, thereby increasing the number of available mode pairs that can be used to reconstruct the lensing field. Since the reconstruction signal-to-noise scales roughly with the number of such mode pairs, extending to higher multipoles in principle improves the statistical precision of the lensing measurement. However, these small-scale modes are also where extragalactic foregrounds, such as tSZ, CIB, and radio sources become dominant. As a result, the additional mode pairs at high $\ell$ carry significant non-Gaussian foreground contamination, which biases the lensing reconstruction. The difference test is shown in the lower left panel of Figure \ref{fig:lrange}. As expected, we find that including such high-$\ell$ modes introduces a significant bias, with the resulting consistency test yielding a PTE of $\sim 0$. This test was performed as part of our pre-unblinding validation. Because it did not satisfy our consistency criterion, we retained $\ell^{T}_{\rm max}=3500$ as the baseline choice; if it had passed, we would instead have adopted $\ell^{T}_{\rm max}=4000$.

\noindent {\it Polarization $\ell_{\rm min}$/$\ell_{\rm max}$}:
We perform a similar stability test by varying the $\ell^{P}_{\rm min}/\ell^{P}_{\rm max}$ cuts applied to the polarization modes entering the reconstruction. The resulting difference bandpowers are shown in Figure~\ref{fig:lrange}. Although point-source contamination is expected to be much smaller for polarization than in temperature, this test still probes possible residual contamination from polarized sources, as well as beam-related systematics such as beam-shape uncertainties and $T$-to-$P$ leakage. Varying $\ell_{\rm min}$ primarily tests sensitivity to the sky cut, apodization, possible $E$-to-$B$ leakage, and polarized Galactic foregrounds.

We adopt a baseline choice of $\ell^{P}_{\rm min}$/$\ell^{P}_{\rm max}=500/3000$. We first lower the maximum multipole to $\ell_{\rm max}=2500$ and find results consistent with the baseline, with PTEs of 0.11 and 0.56 for the PP and GMV reconstructions respectively. Increasing the maximum multipole to $\ell^{P}_{\rm max}=3500$ also yields results consistent with the baseline, with PTEs of 0.40 and 0.81 for PP and GMV respectively. This suggests that a higher $\ell^{P}_{\rm max}$ could technically have been used for polarization. However, since we do not find a significant improvement in the signal-to-noise ratio of the final lensing spectrum, we adopt the intermediate choice of $\ell_{\rm max}=3000$ as our fiducial value. We also raise the minimum multipole to a more conservative value of $\ell_{\rm min}=650$ and again find results consistent with the baseline, with PTEs of 0.64 and 0.70 for the PP and GMV reconstructions respectively.

\subsubsection{Difference-spectrum tests for foreground contamination}\label{sec:diff_fg}

The $\ell^{T}_{\rm max}$ tests described above already provide an initial check of foreground contamination: the consistency between the $\ell^{T}_{\rm max}=3000$ and $\ell^{T}_{\rm max}=3500$ reconstructions indicates that any additional foreground bias from modes in the range $3000 \leq \ell \leq 3500$ is small compared to the variance of the difference bandpowers. We now perform a set of more targeted difference-spectrum tests to further assess residual foreground contamination in the baseline reconstruction.

\begin{figure}
\includegraphics[width=1.00\linewidth]{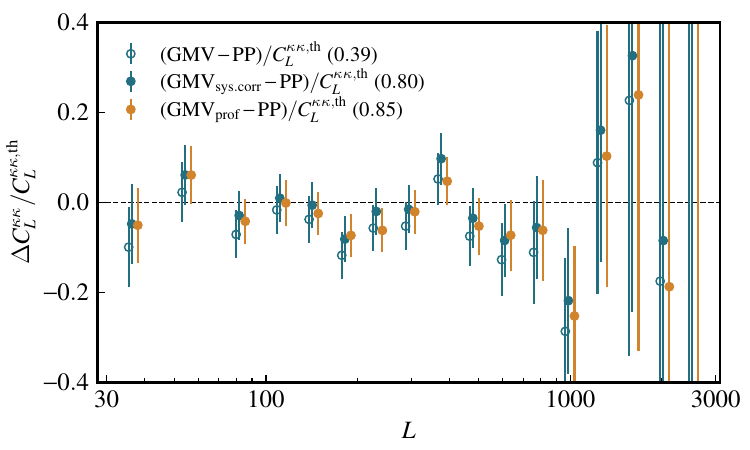} 
\caption{ Data-based fractional difference comparison between the three variants of GMV and the PP bandpowers against the theory lensing amplitude. All the bandpowers have the best-fit values of the instrumental systematic parameters applied. $\gmvsys{}$ additionally applies the best-fit foreground correction to the GMV bandpowers. }
\label{fig:ratio_gmv_pp_fg}
\end{figure}

We compare the standard GMV and profile-hardened GMV reconstructions against the PP reconstruction, which is expected to be substantially less sensitive to extragalactic foreground contamination. The standard GMV bandpowers are known to be shifted by foreground and instrumental systematic effects. We therefore first correct for these expected biases and then test whether the corrected GMV spectrum is consistent with the PP reconstruction. We estimate the corrections using the emulator described in Section~\ref{sec:emulator_train}, evaluated at the best-fit foreground parameters from the high-$\ell$ $TT$ measurement and the best-fit temperature and polarization calibration parameters from the primary CMB analysis. We show the resulting fractional difference spectra in Figure~\ref{fig:ratio_gmv_pp_fg}.

We obtain PTE values of $0.80$ and $0.85$ for GMV and profile-hardened GMV, respectively, demonstrating that both reconstructions are consistent with the polarization-only result given the size of the difference-spectrum error bars. We note that the raw GMV--PP points without any systematic corrections are systematically below zero; as such, a signed $\chi$ statistic would be sensitive to a small coherent shift in the GMV reconstruction that is not strongly captured by the $\chi^2$ test. This is not surprising given the bias observed at the source/cluster location stacks in Figure~\ref{fig:stack}. Overall, the inferred foreground bias is small ($\lesssim 10\%$ of the signal) for $L\leq1000$, where the difference-spectrum measurement has statistical power.

Given that the magnitude of the inferred foreground bias is small on the scales where the measurement has statistical power, we adopt the standard GMV lensing reconstruction as our baseline for cosmological inference. Residual foreground and instrumental systematic uncertainties are accounted for during parameter inference through the model-based correction described in Section~\ref{sec:inference}.

\begin{figure}
\includegraphics[width=1.00\linewidth]{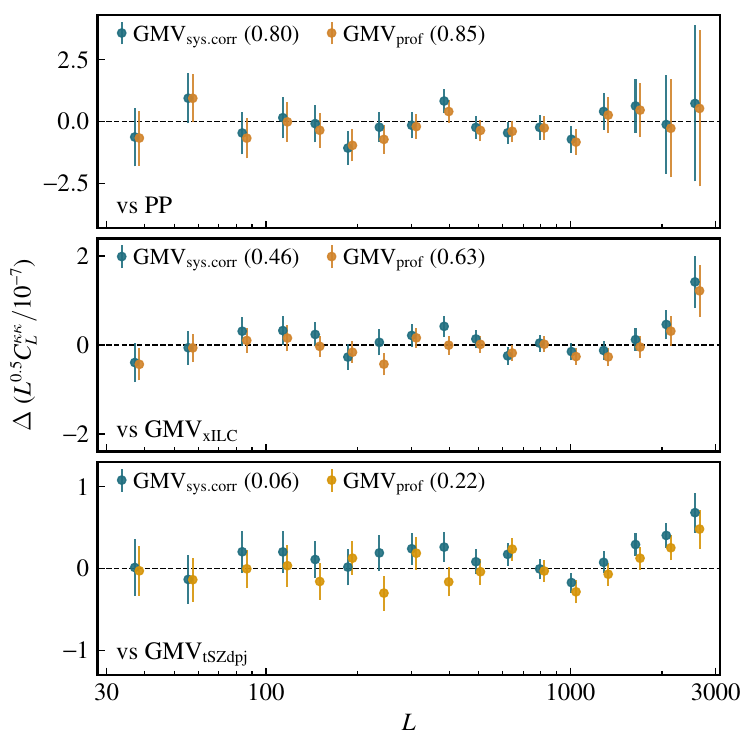} 
\caption{Differences of the $\gmvsys{}$ and $\gmvprof{}$ bandpowers relative to the PP, $\gmvxilc{}$, and $\gmvtsz{}$ reconstructions. The numbers shown in brackets correspond to the PTE values relative to the null.}
\label{fig:diff_fg}
\end{figure}

In Figure~\ref{fig:diff_fg}, we show the difference spectra between the systematics-corrected GMV bandpowers and the PP, $\gmvxilc{}$, and $\gmvtsz{}$ bandpowers, together with the corresponding difference-bandpower PTEs. We find ${\rm PTE}=0.80$, $0.46$, and $0.06$, respectively. These PTE values show that, after applying the best-fit foreground and instrumental-systematic corrections, the GMV reconstruction is consistent with the foreground-mitigated ($\gmvxilc$/$\gmvtsz$) reconstructions, with no evidence for significant residual foreground bias. We do, however, observe a trend at high $L$ when comparing our baseline GMV reconstruction with the $\gmvxilc{}$ and $\gmvtsz{}$ variants. The same trend is not statistically significant in comparisons with the PP reconstruction because of its larger uncertainties on these scales. Therefore, we test the impact of removing the last two bins on the inferred cosmological results in Section~\ref{sec:ResultsLensingonly}.

\subsection{Blinding}\label{sec:blinding}
To prevent experimenter bias, we followed a two-step blinding scheme. In the first step, we refrained from plotting the data bandpowers alongside theory predictions until we had passed the simulation mean test, curl-mode test, unlensed test, and $\ell_{\rm min}/\ell_{\rm max}$ tests described in Section~\ref{sec:validation}.\footnote{The teams involved in~\cite{ge2025} worked in parallel up to the point when those results were published, at which time the polarization-only bandpowers were unblinded, since MUSE used only polarization. The temperature-only, MV, and GMV lensing reconstructions remained blinded until the tests were passed.} 
The cosmological parameters were unblinded after we passed both the alternate-cosmology test and the unbiasedness test using \textsc{Agora} simulation realizations, described in Sections \ref{sec:agora_test} and \ref{sec:altcosmo_test}. Before fully unblinding, we also checked for tension between the CMB lensing reconstruction and the primary CMB and BAO datasets, since some results require consistency between datasets before they are combined.\\

Following unblinding, we made a few minor modifications to the analysis, which we summarize below.
\begin{itemize}[leftmargin=0.7em]

\item[-] In the initial unblinding, the so-called physical beam model was assumed for the polarization component of the beam.  The initial model had the main-lobe beam normalized at $\ell=800$.  The updated model is based on the physical normalization at small angular scales, meaning the normalization is no longer fixed at $\ell=800$.  This change also affects the specific $P_{\rm cal}$ values used.
\item[-] In the original pipeline, the measured transfer functions were first deconvolved from the maps before the three frequency channels were combined. The transfer function was then reapplied prior to lensing reconstruction to suppress the low-$m$ modes, which become noisy in the transfer-function-deconvolved maps. In the updated procedure, by contrast, only the beam is deconvolved, while the transfer function is left in place and the frequency channels are combined directly in transfer-function-convolved space. This simplification is possible because the SPT-3G transfer function is effectively the same across all frequency channels. The effect of the transfer function is instead incorporated at the $C^{-1}$ filtering stage through an analytic model that imposes a declination-dependent $m$-cut. Although this treatment is conceptually more direct and formally more optimal, and was therefore expected to improve the signal-to-noise ratio, the change made a negligible difference to the result.

\item[-] In the $C^{-1}$ filtering step, it was found that incorrect factors of the pixel-window function were applied to the data model in the $C^{-1}$ filter. Although this error should only make the lensing reconstruction suboptimal, we recalculated all lensing reconstructions using the correct pixel-window-function factors.
\end{itemize}
After these modifications were implemented, we reran all pre-unblinding validation tests and confirmed that the updated analysis continued to satisfy the original validation criteria.

\section{Inference framework and validation}\label{sec:inference}
In this section, we describe and validate the cosmological parameter inference framework used in this analysis.

\subsection{Cosmological models and inference settings}
Our baseline cosmology is a six-parameter $\Lambda$CDM model with purely adiabatic scalar perturbations. The free parameters are the physical baryon density, $\Omega_{\rm b} h^2$; the physical cold-dark-matter density, $\Omega_{\rm c} h^2$; the approximate angular size of the sound horizon at recombination, $\theta_{\rm MC}$; the optical depth to reionization, $\tau$; the amplitude of the primordial scalar curvature power spectrum at $k=0.05\,{\rm Mpc}^{-1}$, $A_{\rm s}$; and its spectral index, $n_{\rm s}$. We assume one massive neutrino species with a total mass of $\sum m_\nu = 60\,{\rm meV}$.\footnote{When the sum of neutrino masses is allowed to vary, we set {\tt nu\_mass\_eigenstates}, the number of non-degenerate massive-neutrino eigenstates, to 3. We adopt the standard neutrino radiation density, split into a massless contribution of $N_{\rm eff}^{\rm massless}=2.0293$  and a massive-neutrino contribution.}

In addition, we report constraints on derived parameters such as the Hubble constant $H_{0}$ and $\sigma_8$, defined as the square root of the variance of the density field smoothed with a spherical top-hat filter of radius $8\,h^{-1}$\,Mpc and calculated in linear perturbation theory~\citep{peebles1980}. Finally, we examine a series of extensions to $\Lambda$CDM, including modified lensing amplitudes, the sum of the neutrino masses $\summnu$, the spatial curvature $\omk$, and the evolving dark energy model $w_0w_a$CDM.

The lensed CMB and CMB lensing-potential power spectra are computed using \texttt{CAMB}.\footnote{\url{https://camb.info} (\texttt{v1.4.1}).} For the nonlinear matter power spectrum, we use the \texttt{HMcode2020} prescription \citep{mead2021}. We test the impact of alternative nonlinear prescriptions and baryonic feedback treatments in Appendix~\ref{app:tagn}. The \texttt{CAMB} accuracy settings, BBN predictor, nonlinear prescription, and parameter choices adopted for each data combination are summarized in Table~\ref{tab:theory_settings}. For analyses that include primary CMB data, either alone or as part of a combined likelihood, we follow the settings of \cite{louis2025}, except that we use \texttt{RECFAST} rather than \texttt{CosmoRec} as the recombination model.

For chains involving CMB lensing, primary CMB, or BAO, we sample the posterior using the Metropolis-Hastings MCMC sampler implemented in \texttt{Cobaya} \citep{torrado2021}, with fast dragging enabled to explore the parameter space efficiently. We declare convergence when the Gelman-Rubin statistic for the posterior means reaches $R-1 < 0.01$, and when the corresponding statistic for the bounds of the $95\%$ credible intervals reaches $R-1 < 0.05$, thereby requiring stability both in the posterior bulk and in the tails.

For chains involving optical galaxy surveys, we sample the posterior using \texttt{CosmoSIS}. We use the publicly available likelihood modules distributed with the \texttt{cosmosis-standard-library}, adopting cosmological priors matched to those of the DES+KiDS joint analysis~\cite{deskids2023}. For experiment-specific nuisance parameters, we adopt the systematic priors specified by each survey.

\begin{table*}[t]
\caption{Summary of theory settings used for different data combinations.}
\centering
\footnotesize
\begin{tabular}{p{0.20\textwidth} p{0.25\textwidth} p{0.25\textwidth} p{0.22\textwidth}}
\toprule
Setting 
& CMB lensing only 
& CMB lensing + primary CMB 
& Large-scale-structure probes \\
\midrule

Theory code
& \texttt{CAMB} through \texttt{Cobaya}
& \texttt{CAMB} through \texttt{Cobaya}
& \texttt{CAMB} through \texttt{CosmoSIS} \\

Sampler
& \texttt{Metropolis-Hastings} 
& \texttt{Metropolis-Hastings} 
& \texttt{Nautilus} \\

\texttt{lens\_potential\_accuracy}
& \texttt{4}
& \texttt{8}
& \texttt{4} \\

\texttt{lens\_margin}
& \texttt{1250}
& \texttt{2050}
& \texttt{1250} \\

\texttt{AccuracyBoost}
& \texttt{1.0}
& \texttt{1.0}
& \texttt{1.1} \\

\texttt{lSampleBoost}
& \texttt{1.0}
& \texttt{1.0}
& \texttt{1.0} \\

\texttt{lAccuracyBoost}
& \texttt{1.0}
& \texttt{1.2}
& \texttt{1.0} \\

\texttt{kmax}
& \texttt{5}
& \texttt{10}
& \texttt{100} \\

\texttt{k\_per\_logint}
& \texttt{0}
& \texttt{130}
& \texttt{0} \\

\texttt{min\_l\_logl\_sampling}
& \texttt{5000}
& \texttt{6000}
& \texttt{5000} \\

\texttt{nonlinear}
& \texttt{True}
& \texttt{True}
& \texttt{True} \\

\texttt{halofit\_version}
& \texttt{mead2020}
& \texttt{mead2020}
& \texttt{mead2020\_feedback} \\

$H_{0}$ range
& $[40,100]$
& $[40,100]$
& $[40,100]$ \\

\bottomrule
\end{tabular}
\label{tab:theory_settings}
\end{table*}

\subsection{Lensing likelihood}\label{sec:lensing_like}

In this analysis, the parameter vector $\boldsymbol{\Theta}$ consists of the cosmological parameters $\boldsymbol{\theta}^{c}$, the foreground parameters $\boldsymbol{\theta}^{f}$, and the systematic parameters $\boldsymbol{\theta}^{s}$. We first describe the CMB lensing likelihood, which depends only on the cosmological parameters $\boldsymbol{\theta}^{c}$. We then describe how the effects of foregrounds and instrumental systematics are incorporated through the additional parameters $\boldsymbol{\theta}^{f}$ and $\boldsymbol{\theta}^{s}$ using an emulator.

Similar to previous SPT lensing analyses, we adopt a Gaussian likelihood for the measured lensing bandpowers $\hat{C}_{b}^{\kappa\kappa}$ and write the CMB lensing log-likelihood as
\begin{equation}
\label{eq:cmblike}
\begin{aligned}
& -2 \ln \mathcal{L}_\kappa(\boldsymbol{\Theta})= \\
& \hspace{0.3cm}\sum_{bb'}\left[\hat{C}_{b}^{\kappa \kappa}-C_{b}^{\kappa \kappa,\mathrm{model}}(\boldsymbol{\Theta})\right] \mathbb{C}_{bb'}^{-1}\left[\hat{C}_{b'}^{\kappa \kappa}-C_{b'}^{\kappa \kappa, \mathrm{model}}(\boldsymbol{\Theta})\right],
\end{aligned}
\end{equation}
where $\hat{C}_{b}^{\kappa\kappa}$ denotes the measured lensing bandpowers, $C_{b}^{\kappa\kappa,\mathrm{model}}(\boldsymbol{\Theta})$ is the corresponding binned model lensing power spectrum evaluated for the parameter values $\boldsymbol{\Theta}$, and $\mathbb{C}_{bb'}$ is the bandpower covariance matrix. We form $\mathbb{C}_{bb'}$ from the covariance of 498 reconstructed lensing spectra from simulations. We apply the Hartlap factor $\alpha_{\rm H} = (N_{\rm s} - N_{\rm b} - 2) / (N_{\rm s} - 1)$ to the inverse covariance matrix, where $N_{\rm s}$ is the number of simulation realizations and $N_{\rm b}$ is the number of data bins. We do not apply any further conditioning to the covariance matrix, based on the tests presented in Appendix~\ref{app:bpcm_cond} and the fact that the number of simulations is sufficient to estimate the $17\times17$ covariance matrix.\footnote{
We find $< 0.15 \sigma$ differences in $\som$ recovery with two other versions of smoothed covariance matrices (see Appendix~\ref{app:bpcm_cond}).} 

We begin by describing the cosmological dependence of $C_{b}^{\kappa\kappa,\mathrm{model}}$. In addition to the direct cosmological dependence of the theoretical lensing spectrum, the model includes corrections associated with the cosmology assumed when computing the reconstruction response and $N_{L}^{(1)}$ bias. As shown in Equations~\eqref{eqn:simresp} and \eqref{eq:spectrum_debias}, both of these quantities are computed using simulations generated at a fixed fiducial cosmology. Because this fiducial cosmology may not coincide with the true underlying cosmology, we include corrections that capture how the reconstruction response and $N_{L}^{(1)}$ bias vary with cosmology relative to their values at the baseline model.
Accordingly, we vary three ingredients with cosmology: the underlying lensing signal $C_{L}^{\kappa\kappa}$, the $N_{L}^{(1)}$ bias, and the reconstruction response, which sets the normalization of both terms. Since the $N_{L}^{(0)}$ bias is replaced by the realization-dependent estimate $N_{L}^{(0),{\rm RD}}$, which is anchored to the observed data realization, we do not vary this term with cosmology. The cosmological dependence of the lensing model spectrum is thus given by: 
\begin{equation}\label{eq:clkk_model}
C_L^{\kappa \kappa, {\rm model}} (\boldsymbol{\theta^c})=\frac{\mathcal{R}_L^2(\boldsymbol{\theta^c})}{\mathcal{R}_L^2(\boldsymbol{\theta^c}_{\rm fid})} C_{L}^{\kappa \kappa} ({\boldsymbol{\theta^c}})+ N_{L}^{(1)}(\boldsymbol{\theta^c}) - N_L^{(1)}(\boldsymbol{\theta^c}_{\rm fid}),
\end{equation}
where $\boldsymbol{\theta}_{\rm fid}^{c}$ denotes the fiducial cosmology of the Gaussian simulations used to compute the reconstruction response function and the $N_{L}^{(1)}$ bias.

By applying the ratio of the analytic response functions $\mathcal{R}_L$ evaluated at $\boldsymbol{\theta^c}$ and at the fiducial cosmology $\boldsymbol{\theta^c}_{\rm fid}$  and the difference in $N^{(1)}_L$ at the two cosmologies, 
the \texttt{CAMB} output $C_L^{\kappa \kappa} ({\boldsymbol{\theta^c}})$ is shifted to look like the measured $\hat{C}_{b}^{\kappa \kappa}$ if $\boldsymbol{\theta^c} \equiv \boldsymbol{\theta^{\rm data}}$.

Evaluating Equation~\eqref{eq:clkk_model} on-the-fly as we evaluate the likelihood is costly, especially when simulation-based corrections are needed on top of the analytical calculations of $\mathcal{R}_{L}$ and $N^{(1)}_{L}$. Therefore, we follow previous lensing works~\cite{simard2018,bianchini2020,planck2015lens,madhavacheril2024}, and approximate their cosmological dependence using a first-order Taylor expansion around the fiducial cosmology $\boldsymbol{\theta^c}_{\rm fid}$. Terms involving products or higher powers of the spectral deviations are second order or higher and are neglected. Although the correction associated with $\mathcal{R}_L$ is multiplicative, its fractional variation around the fiducial cosmology is small, such that its leading-order effect in the Taylor expansion can be expressed as an additive correction to the lensing spectrum. Rather than expanding directly in the cosmological parameters, we express these first-order corrections in terms of deviations of the relevant power spectra from their fiducial values. Specifically, we compute how the response  changes with variations in the primary CMB spectra $C_{\ell}^{TT}$, $C_{\ell}^{TE}$, and $C_{\ell}^{EE}$, and how the $N_{L}^{(1)}$ correction changes with variations in the lensing spectrum $C_{L}^{\kappa\kappa}$, with all derivatives evaluated at the fiducial spectra. We collect these derivatives into the matrices $M^{\kappa\kappa}_{b\ell}$ and $M^{x}_{b\ell}$, with $x\in\{TT,TE,EE\}$, which describe how changes in the underlying spectra propagate to the lensing bandpowers.

The cosmological dependence of the model spectrum is then approximated as~\cite{planck2018lens}:
\begin{equation}
\label{eq:lincorr}
\begin{aligned}
C_b^{\kappa \kappa, {\rm model}}({\boldsymbol{\theta^c}}) &= C_b^{\kappa \kappa}({\boldsymbol{\theta^c}}) \\
&+ M^{\kappa\kappa}_{bL} \left(C_{L}^{\kappa\kappa}(\boldsymbol{\theta^c})-C_{L}^{\kappa\kappa}(\boldsymbol{\theta^c}_{\rm fid})\right)\\
&+ \sum_{x} M^{x}_{b\ell} \left(C_{\ell}^{x}(\boldsymbol{\theta^c})-C_{\ell}^{x}(\boldsymbol{\theta^c}_{\rm fid})\right).
\end{aligned}
\end{equation}
For joint analyses combining the primary CMB and CMB lensing likelihoods, the primary CMB spectra $C_{\ell}^{x}(\boldsymbol{\theta^c})$ are constrained directly by the primary CMB block of the likelihood. The final correction term in Equation~\eqref{eq:lincorr} therefore propagates the allowed variations in these spectra into the CMB lensing model. In such cases, we use Equation~\eqref{eq:lincorr} as written. For analyses of CMB lensing alone, however, the primary CMB spectra are not directly constrained, allowing the response correction to vary over an unphysically broad range. We therefore adopt the modified treatment described in the following section.

\subsubsection{Lensing-only likelihood}\label{sec:likelihood_lensonly}
For the lensing-only likelihood, we evaluate the response correction using the observed primary CMB spectra rather than the spectra predicted at each sampled cosmology. In this sense the ``lensing-only'' likelihood is not fully independent of the primary
CMB: although the primary CMB likelihood is not included as a cosmological data set in the inference, the measured primary CMB spectra are still used to determine the correction to the reconstruction response $\mathcal{R}_{L}$. Evaluating the correction at the observed spectra accounts for the difference between the fiducial primary CMB spectra used to compute the reconstruction response and those measured on the observed sky.

For this purpose, we use the measured SPT-3G {\it Lite} bandpowers derived for the {\it Lite} likelihood from \cite{camphuis2026}. The {\it Lite} likelihood is a CMB-only version of the full SPT-3G primary CMB likelihood that takes cosmological and calibration parameters as inputs to the model. It uses CMB-only bandpowers in which the foreground and beam-systematic parameters have already been marginalized, with the corresponding uncertainty incorporated into the covariance matrix. The temperature and polarization calibration parameters remain explicit and are sampled in the likelihood.
We replace the model primary CMB spectra entering the response correction with the corresponding {\it Lite} bandpowers, so that

\begin{equation}\label{eqn:modelspec_lensonly}
\begin{aligned}
C_{b}^{\kappa \kappa,\mathrm{model}}(\boldsymbol{\Theta})
&= C_b^{\kappa \kappa}({\boldsymbol{\theta^c}})\\
&+ M^{\kappa\kappa}_{bL}
\left(
C_{L}^{\kappa\kappa}(\boldsymbol{\theta^c})
-C_{L}^{\kappa\kappa}(\boldsymbol{\theta^c}_{\rm fid})
\right)\\
&+ \sum_{x} M^{x}_{b\ell}
\left(
\underbracket[0.8pt]{\hat C_{\ell}^{{Lite},x}}_{\text{lensing-only}}
-\,C_{\ell}^{x}(\boldsymbol{\theta^c}_{\rm fid})
\right).
\end{aligned}
\end{equation}
In this construction, the last term becomes a fixed correction term that does not depend on the sampled cosmological parameters and fixes the response to the one evaluated at the measured CMB bandpowers.

To account for the uncertainties of the {\it Lite} bandpowers in the lensing-only likelihood, we add
an extra term to the covariance matrix~\cite{madhavacheril2024, planck2018lens}, 
\begin{equation}\label{eqn:cmb_marg}
	\mathbb{C}_{bb'} \rightarrow \mathbb{C}_{bb'}
	+ \underbracket[0.8pt]{
	\Delta_{\ell_b} M^{x}_{b\ell_b} \,
	\Sigma^{\rm CMB; {\it xy}}_{\ell_b \ell'_b}
	M^{y}_{\ell'_b b'} \Delta_{\ell'_b}
	}_{\text{lensing-only}} ,
\end{equation}
where $\Sigma^{\rm CMB}$ denotes the binned primary CMB spectrum covariance, $x, y \in \{TT, TE, EE\}$, $\Delta_{\ell_b}$ represents the size of the bin, and repeated indices are summed. 
For $\Sigma^{\rm CMB}$, we concatenate the {\it Lite}/CMB-only covariance matrices of {\it Planck} PR3~\cite{planck2018_likelihood}, ACT DR6~\cite{louis2025},\footnote{https://github.com/ACTCollaboration/DR6-ACT-lite} and SPT-3G D1~\cite{camphuis2026}.\footnote{https://github.com/SouthPoleTelescope/spt\_candl\_data; based on~\cite{candl}.}
We concatenate the covariance matrices at the following multipole ranges such that the diagonal elements of the resultant matrix are the smallest of the three individual matrices: 
\begin{itemize}
\item $TT$: {\it Planck} for $\ell < 1500$ and ACT otherwise; 
\item $TE$: {\it Planck} for $\ell < 1000$, ACT for $\ell = [1000, 2000]$, and SPT for $\ell > 2000$;
\item $EE$: {\it Planck} for $\ell < 800$, ACT for $\ell = [800, 2300]$, and SPT for $\ell > 2300$. 
\end{itemize}
We neglect the small covariance between experiments induced by the common lensing signal. The magnitude and impact of this covariance are discussed in Appendix~\ref{app:cmb_marg}.

\subsubsection{Lensing + primary CMB likelihood}
When the lensing likelihood is sampled jointly with the primary CMB likelihoods, the primary CMB data directly constrain the $TT$, $TE$, and $EE$ spectra entering the response function through $M^{x}_{b\ell}$. We can therefore evaluate the response correction in Equation~\eqref{eq:lincorr} directly at each sampled point in parameter space. The uncertainties in the primary CMB spectra are propagated through the shared cosmological and nuisance parameters, so no additional  contribution is added to the lensing bandpower covariance.

We neglect correlations between the primary CMB and CMB lensing bandpowers, as including these correlations does not affect cosmological parameter uncertainties at the current noise levels~\cite{trendafilova2023}. We therefore evaluate the joint likelihood by summing the lensing and primary CMB log-likelihoods. For the primary CMB data, we use the {\it Planck} PR3, ACT\,DR6, and SPT-3G D1 $TT/TE/EE$ CMB-only/{\it Lite} likelihoods. For SPT-3G D1 specifically, the effects of beam systematics are already marginalized in the CMB-only bandpowers and included in the CMB-only covariance matrix. 
Thus, we do not vary the beam parameters in the joint lensing + primary CMB analysis and instead fix them to their fiducial values. We vary calibration parameters in the joint likelihood.

The priors adopted for the cosmological, foreground, and instrumental systematic parameters in the lensing-only and lensing + primary CMB cases are listed in Table~\ref{tab:priors}.

\refstepcounter{footnote}
\edef\systematicsfootnote{\arabic{footnote}}
\begin{table}[t]
\caption{Priors imposed on the cosmological, instrumental\protect\hyperlink{fn:systematics}{\textsuperscript{\systematicsfootnote}}, and foreground parameters investigated in this work, when considering either lensing-only datasets or also including primary CMB measurements. Parameters that are fixed are reported by a single number. $\mathcal{U}(a,b)$ denotes a uniform distribution between $[a,b]$, while $\mathcal{N}(\mu,\sigma)$ indicates a Gaussian distribution with mean $\mu$ and standard deviation $\sigma$.}
\centering
\begin{tabular}{c|c|c}
\toprule
Parameter               & Lensing (+BAO)         & Lensing + CMB      \\
\midrule
\multicolumn{3}{c}{$\Lambda$CDM} \\
\midrule
$\Omega_{\rm b} h^2$    & $\mathcal{N}(0.02238,0.0005)$    & $\mathcal{U}(0,0.1)$      \\
$\Omega_{\rm c} h^2$    & $\mathcal{U}(0.005,0.99)$        & $\mathcal{U}(0.005,0.99)$ \\
$100\,\theta_{\rm MC}$  & $\mathcal{U}(0.5,10)$            & $\mathcal{U}(0.5,10)$     \\
$\tau$                  & --     & $\mathcal{U}(0.01,0.1)$   \\
$n_{\rm s}$             & $\mathcal{N}(0.96,0.02)$         & $\mathcal{U}(0.2,2.0)$    \\
$\ln (10^{10}A_s)$      & $\mathcal{U}(1.61,4.0)$          & $\mathcal{U}(1.61,4.0)$   \\
\midrule
\multicolumn{3}{c}{Extensions} \\
\midrule
$\sum m_{\nu}$ [eV]            & 0.06                         & 0.06 or $\mathcal{U}(0,5)$   \\
$\omk$                         & 0 or $\mathcal{U}(-0.3,0.3)$ & 0 or $\mathcal{U}(-0.3,0.3)$ \\
$w_0$ & $-1$ or $\mathcal{U}(-2,1)$ & $-1$ or $\mathcal{U}(-2,1)$ \\
$w_a$ & $0$ or $\mathcal{U}(-5,2)$ & $0$ or $\mathcal{U}(-5,2)$ \\
$A_{\rm 2pt},\,A_{\rm recon}$  & 1 or $\mathcal{U}(0,10)$     & 1 or $\mathcal{U}(0,10)$     \\
\midrule
\multicolumn{3}{c}{Instrumental} \\
\midrule
$\eta_{1,2,3,4}$               & 0                              & 0                            \\
$\beta_{\rm pol}^{95,150,220}$ & 0.536,\,0.685,\,0.658          & 0.536,\,0.685,\,0.658        \\
$T_{\rm cal}$                  & $\mathcal{N}(0.9992,0.0027)$ & $\mathcal{N}(1.0000,0.0036)$ \\
$P_{\rm cal}$                  & $\mathcal{N}(1.0072,0.0040)$ & $\mathcal{U}(0.8,1.2)$       \\
\midrule
\multicolumn{3}{c}{Foregrounds} \\
\midrule
$A_{\rm tSZ}$       & $\mathcal{N}(0.978,0.022)$ & $\mathcal{N}(0.978,0.022)$ \\
$A_{\rm CIB}^{150}$ & $\mathcal{N}(0.973,0.003)$ & $\mathcal{N}(0.973,0.003)$ \\
$A_{\rm CIB}^{220}$ & $\mathcal{N}(1.000,0.002)$ & $\mathcal{N}(1.000,0.002)$ \\
$A_{\rm rad}^{95}$  & $\mathcal{N}(0.963,0.008)$ & $\mathcal{N}(0.963,0.008)$ \\
$A_{\rm rad}^{150}$ & $\mathcal{N}(0.950,0.018)$ & $\mathcal{N}(0.950,0.018)$ \\
\bottomrule
\end{tabular}

\label{tab:priors}
\end{table}

\footnotetext[\value{footnote}]{\hypertarget{fn:systematics}{}The instrumental-systematics settings differ slightly from those reported in C26. The $\beta_{\rm pol}$ values used in this work are derived from the full SPT-3G primary CMB likelihood using an updated beam definition. The $T_{\rm cal}$ and $P_{\rm cal}$ constraints are derived from the $\cmbspa$ posteriors using the SRoll2-based treatment described in Appendix~\ref{app:tau_prior}, whereas C26 adopted a Gaussian prior based on \cite{planck2020LFIHFItau}.}

\subsubsection{Emulator for systematics modeling}\label{sec:emulator_train}

Thus far, we have described the cosmological dependence of the lensing model. We now turn to the effects of instrumental systematics and astrophysical foregrounds, which must also be modeled in the likelihood.

We have two approaches for modeling the effects of instrumental systematics on the lensing spectrum. In the first, we rerun the full lensing reconstruction pipeline for a set of varied instrumental systematic parameters and use the resulting spectra to train an emulator that interpolates the corresponding changes in the lensing spectrum between the sampled parameter values. In the second, their effects are included analytically through the primary CMB spectra used to compute the lensing response. We describe the emulator approach here and the analytic approach in Appendix~\ref{app:analytic_sys_model}.

The dependence on foreground parameters is modeled through an emulator trained on the \textsc{Agora} simulations. Although foreground biases are reduced on large angular scales because polarization dominates the statistical weight of the baseline GMV reconstruction, residual contamination from temperature foregrounds remains and must be modeled. Instrumental systematics can also modify the amplitude and shape of the reconstructed lensing spectrum, leading to degeneracies with residual foreground effects. We therefore vary the foreground and instrumental systematic parameters jointly in the cosmological likelihood. The emulator propagates these variations through the full reconstruction pipeline, allowing correlations between their effects to be captured.

The complete expression for the emulator-based model spectrum is:
\begin{equation}\label{eqn:modelspec_emu}
\begin{aligned}
C_{b}^{\kappa \kappa,\mathrm{model}}(\boldsymbol{\Theta}) & = C_b^{\kappa \kappa}({\boldsymbol{\theta^c}}) \overbracket[0.8pt]{\frac{C_b^{\kappa \kappa}({\boldsymbol{\theta^c}_{\rm fid}}, \boldsymbol{\theta^f, \theta^s})}{C_b^{\kappa \kappa}({\boldsymbol{\theta^c}_{\rm fid}}, {\boldsymbol{\theta^s}_{\rm fid}, {\boldsymbol{\theta^f}=0 }} )}}^{\rm emulated} \\
&+ M^{\kappa\kappa}_{bL}
\left(
C_{L}^{\kappa\kappa}(\boldsymbol{\theta^c})
-C_{L}^{\kappa\kappa}(\boldsymbol{\theta^c}_{\rm fid})
\right)\\
 &+ \sum_{x\in \{TT,TE,EE\}} M^{x}_{b\ell} \left(C_{\ell}^{x}(\boldsymbol{\theta^c})-C_{\ell}^{x}(\boldsymbol{\theta^c}_{\rm fid})\right),
\end{aligned}
\end{equation}
where the effects of foreground and systematic calibration parameters are included in the emulated ratio. This emulator-based approach is our baseline lensing model and is used for the main results, including both the lensing-only likelihoods and the lensing likelihoods combined with primary CMB data.

To build the emulator, we begin by generating input $T/Q/U$ maps at 95/150/220\,GHz for 200 training points sampled using a Latin-hypercube design over the parameter ranges listed in Table~\ref{tab:emulprior}. Each of these points is defined by a set of temperature and polarization calibration parameters ($T_{\rm cal}$, $P_{\rm cal}$), amplitudes of the first four temperature beam uncertainty eigenmodes ($\eta_{1},\eta_{2},\eta_{3},\eta_{4}$), polarized beam parameters ($\beta_{\rm pol}^{95}, \beta_{\rm pol}^{150}, \beta_{\rm pol}^{220}$), and foreground amplitude parameters\footnote{Other parameters such as the $A_{\rm kSZ}$, $A_{\rm rad}^{220}$, and $A_{\rm CIB}^{95}$ are fixed to 1.0 to reduce the dimensionality of the problem.} $A_{\rm tSZ}$, $A_{\rm CIB}^{150}$, $A_{\rm CIB}^{220}$, $A_{\rm rad}^{95}$, $A_{\rm rad}^{150}$. The temperature and polarization maps at frequency band $\nu$ can be written as
\begin{align}
T^{\nu} &= \hspace{1.65em}T^{\nu}_{\rm cal}\left[T_{\rm CMB} + \sum_{\alpha}A_{\alpha}^{\nu} M_{\alpha}^{\nu}\right]\otimes B_{\ell,\nu}^{T},\\
Q^{\nu} &= T^{\nu}_{\rm cal}P^{\nu}_{\rm cal}\hspace{0.08cm}\left[Q_{\rm CMB} + A_{\rm rad}^{\nu} M_{\rm rad}^{\nu}\right]\otimes B_{\ell,\nu}^{P},\\
U^{\nu} &= T^{\nu}_{\rm cal}P^{\nu}_{\rm cal}\hspace{0.08cm}\left[U_{\rm CMB}\hspace{0.05cm} + A_{\rm rad}^{\nu} M_{\rm rad}^{\nu}\right]\otimes B_{\ell,\nu}^{P},
\end{align}
where $T^{\nu}_{\rm cal}$ and $P^{\nu}_{\rm cal}$ are calibration parameters, $M$ denotes the foreground-component maps, $\alpha$ indexes tSZ, kSZ, CIB, and radio, and $B_{\ell,\nu}^{T}$, $B_{\ell,\nu}^{P}$ are the temperature and polarization beams, respectively. 
For the calibration parameters, we vary the absolute temperature and polarization calibration at 150\,GHz, ${T^{150}_{\rm cal}}$ and ${P^{150}_{\rm cal}}$, and keep the inter-frequency temperature calibration fixed to the fiducial values.
The polarized beam parameters enter $B_{\ell,\nu}^{P}$ as in Equation~\eqref{eq:betapol}. The temperature beam parameters transform $B_{\ell,\nu}^{T} \rightarrow B_{\ell,\nu}^{T}+\sum_{i=1}^{4}\eta_i\,dB_{\ell,\nu,i}^{T},$ (see also Equation~\eqref{eqn:beams}).

At each of the 200 training points, the simulated single-frequency maps are treated in the same way as the data maps and combined using our fiducial LC frequency weights to produce the minimum-variance CMB map. We perform the lensing reconstruction without adding noise and debias the resulting spectra by subtracting $N_{L}^{(0),{\rm RD}}$ and $N_{L}^{(1)}$ computed for this simulation set. Because the reconstruction is performed without noise, its response differs from that of the baseline reconstruction, leading to incorrect relative weighting between the estimators when constructing the minimum-variance lensing map. We correct for this mismatch using ratios of the response functions to recover the appropriate relative weighting of the individual estimators:
\begin{equation}
\hat{\kappa}^{\rm MV}_L
=
\frac{1}{2}L(L+1)
\frac{
\sum_{\alpha}
\bar{\phi}^{\alpha}_{L}
\frac{\widetilde{\mathcal R}^{\rm \alpha,fid}_{L}}
{\widetilde{\mathcal R}^{\rm \alpha,noiseless}_{L}}
}{
\sum_{\alpha}
\widetilde{\mathcal R}^{\rm \alpha,fid}_{L}
},
\end{equation}
where $\mathcal{\widetilde{R}}_{\rm noiseless}$ represents the simulation-based response function constructed from noiseless simulations, whereas $\mathcal{\widetilde{R}}_{\rm fid}$ represents the fiducial response function. Since the difference between the MV and GMV estimators is small, we use the MV-estimator-derived emulator for the GMV likelihood.

\begin{table}[t]
\centering
\caption{Parameter distributions used to construct the emulator training set.}
\label{tab:emulprior}
\begin{tabular}{c|c}
\toprule
Parameter & Training range \\
\midrule
$T_{\rm cal}$ & $\mathcal{U}(0.99,1.01)$ \\
$P_{\rm cal}$ & $\mathcal{U}(0.97,1.03)$ \\
$\eta$ & $\mathcal{U}(-3,3)$ \\
$\beta_{\rm pol}^{\nu}$ & $\mathcal{U}(0,1)$ \\
$A_{\rm tSZ}$ & $\mathcal{U}(0,1.3)$ \\
$A_{\rm CIB}^{150,220}$ & $\mathcal{U}(0,1.3)$ \\
$A_{\rm radio}^{95,150}$ & $\mathcal{U}(0,1.3)$ \\
\bottomrule
\end{tabular}
\end{table}

\begin{figure*}
\includegraphics[width=1.00\linewidth]{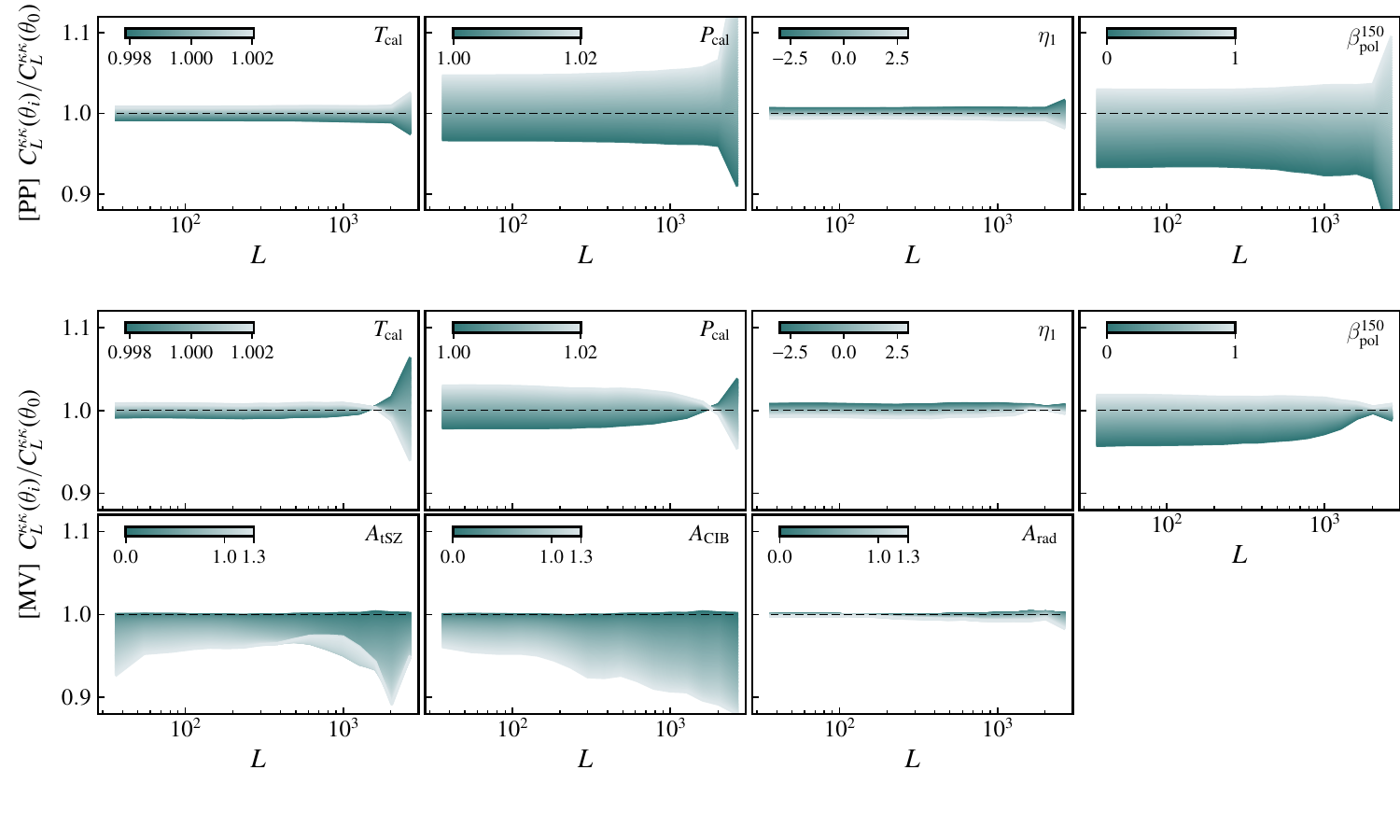} 
\caption{Changes to the model spectrum by varying one systematic parameter at a time. Top panels are for the polarization-only case, which is largely unaffected by foregrounds, while the lower panels are for the minimum-variance reconstruction, which is affected by foreground contamination. The variations shown here for the calibration parameters are around the posterior mean  ($T_{\rm cal}$, $P_{\rm cal})=(0.9992,\,1.0072)$ obtained from the analysis of \citet{camphuis2026}. }
\label{fig:emul}
\end{figure*}

Once the debiased minimum-variance bandpowers have been computed for all 200 training points, we use \texttt{GPJax} to construct an emulator that predicts the ratio\footnote{ Since the lensing reconstruction is based on the same input CMB realization, this allows us to reduce the sample variance.} $C_b^{\kappa \kappa}({\boldsymbol{\theta^c}_{\rm fid}}, \boldsymbol{\theta^f, \theta^s}) / C_b^{\kappa \kappa}({\boldsymbol{\theta^c}_{\rm fid}}, {\boldsymbol{\theta^s}_{\rm fid}, {\boldsymbol{\theta^f}=0 }} )$.
Once trained, we generate 50 test samples drawn from the same parameter ranges but at locations distinct from the training points to validate the emulator. We quantify its accuracy by comparing the predicted and true perturbed lensing spectra across the test samples. The residuals are small compared with the bandpower uncertainties at all $L$: the largest deviations occur in the highest-$L$ bins, where they reach $\sim\!0.1\sigma$, and are well below this elsewhere, making the emulator error negligible in our analysis. In Figure~\ref{fig:emul}, we show the emulated correction factor as we vary the calibration and foreground parameters for both PP and GMV reconstructions. 

Although this emulator is built assuming a fiducial $\Lambda$CDM cosmology, we expect it to be sufficient for the present analysis: the effects of instrumental systematics are independent of cosmology, while the foreground amplitudes are calibrated to the power measured in the observed data rather than varied with the cosmological parameters sampled in the likelihood. A fully cosmology-dependent forward model of the foregrounds could in principle provide additional cosmological constraining power, but we do not include such information here and leave this for future work.

We validate the modeling of the systematic parameters in the emulator framework by comparing it with the analytical approach. The parameters recovered using both approaches are described in Appendix~\ref{app:pipe_vad}. We adopt the emulator-based approach in our baseline analysis pipeline based on the discussion in Appendix~\ref{app:emu_sys}. 

\subsection{Validation}

\subsubsection{Agora test}\label{sec:agora_test}
We validate the full inference pipeline using ten patches extracted from one full-sky \textsc{Agora} simulation,\footnote{The validation of the inference pipeline using Gaussian simulations (with foreground parameters set to zero) is documented in Appendix~\ref{app:pipe_vad}.} each processed through the full lensing reconstruction pipeline identically to the data. For each realization $i$, we obtain a posterior on $\theta=\som$ with mean $\hat{\theta}_{i}$ and standard deviation $\sigma_{i}$, and define
$z_{i} = (\hat{\theta}_{i}-\theta_{\rm true})/\sigma_{i}$,
where $\theta_{\rm true}$ is the input cosmology. We assess consistency with the input cosmology using the $\chi^2$ of the $z_i$ values relative to zero. 

\begin{figure}
\includegraphics[width=1.00\linewidth]{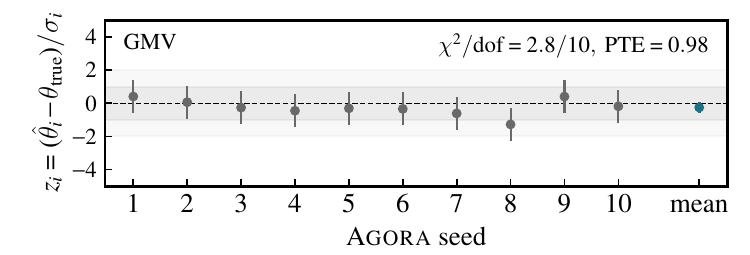} 
\caption{Deviations of $\sigma_8\Omega_{\rm m}^{0.25}$ from truth recovered from 10 \textsc{Agora} patches for the standard GMV reconstruction, shown as $z_i=(\hat{\theta}_i-\theta_{\rm true})/\sigma_i$. The shaded bands indicate the expected $1\sigma$ and $2\sigma$ scatter under the null hypothesis, assuming independent patches. The teal point shows the mean $z$ across the 10 patches, with the uncertainty scaled by $1/\sqrt{10}$ under the same assumption. Since the \textsc{Agora} patches are not strictly independent, these intervals and the corresponding PTEs are intended as approximate diagnostics.}
\label{fig:test_agora}
\end{figure}

The results are shown in Figure~\ref{fig:test_agora}. We find a mean deviation of $\bar{z}=-0.25\pm0.32$, consistent with zero. The ten realizations yield $\chi^2/{\rm dof}=2.8/10$, corresponding to a $\mathrm{PTE}=0.98$. We note that both the quoted uncertainty on the mean and the PTE assume independent realizations. Because the \textsc{Agora} simulation lightcones are  constructed by tiling simulation boxes, structures repeat across the sky and therefore the ten patches are not fully independent. The quoted significances should therefore be interpreted approximately. Nevertheless, the test provides an end-to-end validation of the inference pipeline and is sensitive to coherent biases in the recovered cosmology.

\begin{figure}
\includegraphics[width=1.00\linewidth]{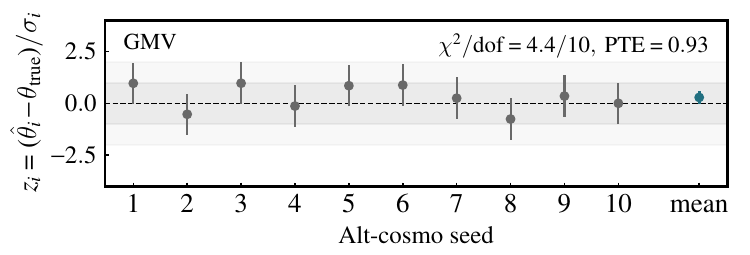} 
\caption{
Distribution of deviations in $\sigma_8\Omega_{\rm m}^{0.25}$ from the true value, recovered from 10 independent simulations using the standard GMV estimator. The simulations are generated at an input cosmology lying $5\sigma$ from the fiducial cosmology in the $\Omega_{\rm m}$--$\som$ plane of the posterior from \cite{qu2026}.
}
\label{fig:test_altcosmo}
\end{figure}

\begin{figure}
\includegraphics[width=1.00\linewidth]{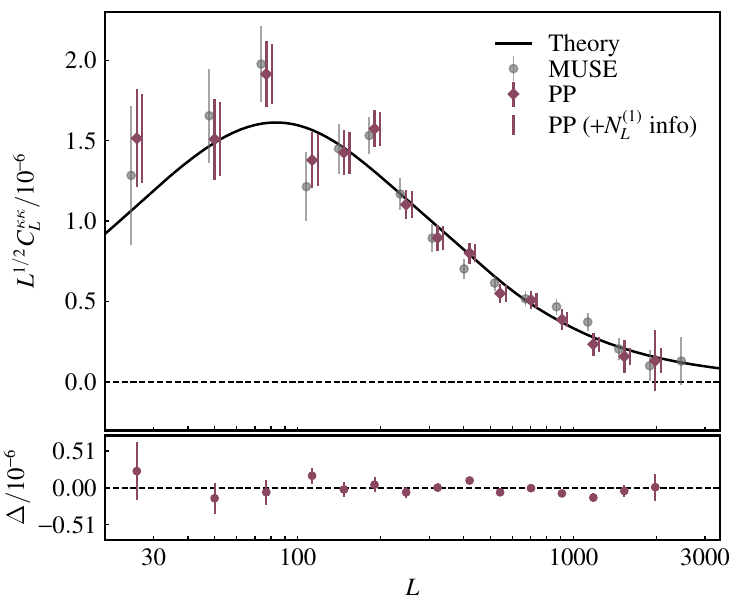} 
\caption{Comparison of the lensing auto-spectrum from \textsc{MUSE} (gray) and our polarization-only lensing spectrum (PP; purple). Note that the bandpower window functions are different for MUSE and our reconstruction. In this plot we adopt the MUSE window functions, which are not necessarily optimal for our reconstruction. The last bin is also partially outside of our $L$-range and is therefore omitted. We note that the MUSE bandpowers already incorporate the cosmological dependence of $N_L^{(1)}$, whereas the native PP error bars do not. To reflect the extra sensitivity to cosmology from the $N_L^{(1)}$ information, we include the error bars on the per-bin amplitude estimated from the Fisher matrix using the full likelihood. {\bf Lower:} the difference between the two bandpowers adopting the same bandpower window function on our measurements.}
\label{fig:vsmuse_bandpowers}
\end{figure}

\begin{figure*}
\includegraphics[width=0.7\linewidth]{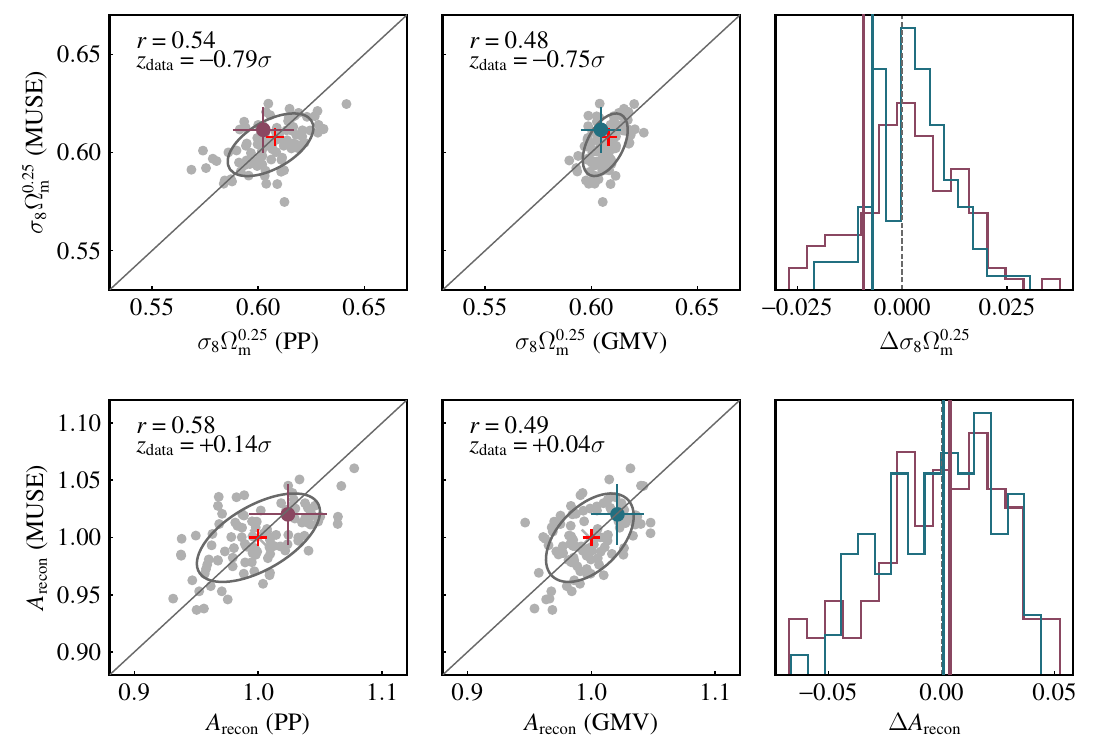} 
\caption{Best-fit values for the parameters $\sigma_{8}\Omega_{\rm m}^{0.25}$ and $A_{\rm recon}$, derived from 100 simulation realizations (gray points) for PP and GMV and compared with those for MUSE. The simulation input is shown in red crosses, and the ellipses indicate the $1\sigma$ confidence regions determined from the distributions. In the left and middle panels, the PP and GMV data constraints, obtained by fixing the cosmological parameters to the $\cmbspa$ best-fit values, are shown in purple and teal, respectively. The right panels show the distributions of the differences between the PP or GMV best-fit values and the corresponding MUSE values across the simulations, with the vertical purple and teal lines indicating the differences measured in the data.}
\label{fig:muse_sim_comparison}
\end{figure*}

\subsubsection{Alternate cosmology test}\label{sec:altcosmo_test}
We test whether the full inference pipeline accurately recovers the input cosmology of a mock data map when it differs from the fiducial cosmology used in the rest of this analysis. We choose the mock-data cosmology to lie in the $\mathord{\sim}5\sigma$ region of the \citet{qu2026} posterior in the $\Omega_{\rm m}$--$\som$ parameter space, while keeping the simulations used to compute the reconstruction response and $\None$ bias at the fiducial cosmology. One aspect tested by this setup is therefore the accuracy of the linear cosmology corrections described in Equation~\eqref{eq:lincorr}, which account for the mismatch between the cosmology of the mock data and that assumed in the simulations.

At this alternate cosmology, we generate 10 pairs of unlensed CMB and lensing potential realizations, perform the deflection operation, add foregrounds, mock-observe, and add sign-flip noise. These maps are treated as the data maps, while the simulation maps assuming the fiducial cosmology are still used to compute the mean-field and response function. The only component that changes is the $\rdNzero$ calculation, where the alternate cosmology maps are used in one leg of the quadratic estimator.

Since the purpose of this test is to validate the sufficiency of the $\rdNzero$ correction paired with the linear corrections for the response and $\None{}$, we fix the foreground and systematic parameters to their fiducial values. In the likelihood, similar to the lensing-only case where we replace $C_{\ell}^{x}(\boldsymbol{\theta^c})$ with the measured CMB bandpowers, we use the alternative cosmology CMB spectrum $C_{\ell}^{x}(\boldsymbol{\theta^c}_{\rm alt})$:

\begin{equation}
\begin{aligned}
C_{b}^{\kappa \kappa,\mathrm{model}}(\boldsymbol{\Theta}) & =  C_b^{\kappa \kappa}({\boldsymbol{\theta^c}}) \\
 &+ M^{\kappa\kappa}_{bL} \left(C_{L}^{\kappa\kappa}(\boldsymbol{\theta^c})-C_{L}^{\kappa\kappa}(\boldsymbol{\theta^c}_{\rm fid})\right)\\
 &+ \sum_{x\in \{TT,TE,EE \}} M^{x}_{b\ell} \left({C}_{\ell}^{x}(\boldsymbol{\theta^c}_{\rm alt})-C_{\ell}^{x}(\boldsymbol{\theta^c}_{\rm fid})\right),
\end{aligned}
\end{equation}
where $\boldsymbol{\theta^c}_{\rm alt}$ corresponds to a cosmology with $\sigma_{8}=0.773$ and $\Omega_{\rm m}=0.305$, with all other parameters matching the fiducial cosmology.

The results for this alternative-cosmology test are shown in Figure~\ref{fig:test_altcosmo}. Averaging over the ten realizations, we find a mean deviation of $\bar{z}=0.28\pm0.32$ relative to the input cosmology, consistent with zero. The ten realizations yield $\chi^2/{\rm dof}=4.4/10$, corresponding to a $\mathrm{PTE}=0.93$. These results demonstrate that the pipeline recovers the input cosmology without significant bias even when it differs from the fiducial model by approximately $5\sigma$.

\subsubsection{Comparison with  MUSE}\label{sec:vsMUSE}
In addition to testing the inference pipeline on simulations, we check the consistency of our polarization results from the data with those from MUSE at both the bandpower and parameter levels.\footnote{The bandpower-level comparison was performed as part of our pre-unblinding validation. For the parameter-level comparison, the simulation-based distribution was established before unblinding, while the location of the observed data point relative to that distribution was evaluated after unblinding.}
In Figure~\ref{fig:vsmuse_bandpowers}, we compare the lensing bandpowers measured by the PP reconstruction in this work with those measured by MUSE, which uses  polarization data only. Both measurements are in good agreement with the $\cmbspa$ best-fit prediction. For this comparison, we apply the MUSE bandpower window functions to the PP spectrum, rather than those computed specifically for it, to enable a like-for-like comparison in a common binning scheme. The lower panel shows the difference between the two estimates, $\Delta = C_L^{\rm PP} - C_L^{\rm MUSE}$, with the mean difference and its covariance estimated from 100 matched simulations by subtracting the PP and MUSE bandpowers realization by realization. The residuals are consistent with zero across the full multipole range, $L \sim 30$--$3000$, yielding $\chi^2 = 19.8$ for 15 degrees of freedom (${\rm PTE} = 0.18$). This agreement supports that our quadratic estimator pipeline does not introduce significant biases relative to the MUSE framework.

To quantify the consistency of our lensing results with those from the MUSE pipeline at the parameter level, we compare the recovered best-fit values of $\som$ and $A_{\rm recon}$ (the relative amplitude of the reconstructed spectra against input lensing spectra) with the distribution of differences obtained from matched-seed simulations, as shown in Figure~\ref{fig:muse_sim_comparison}. For each of the 100 matched-seed realizations, we compute
\begin{equation}
\Delta_\theta^{(i)} = \theta_{\rm MUSE}^{(i)} - \theta_{\rm PP/GMV}^{(i)},
\end{equation}
where $\theta$ denotes the MAP estimate of either $\som$ or $A_{\rm recon}$ obtained by minimizing the negative log-posterior, and $i$ is the realization index. From these simulations, we estimate the scatter of $\Delta_\theta$, $\sigma_{\Delta}$, and evaluate the normalized statistic
\begin{equation}
z = \frac{\theta_{\rm MUSE}-\theta_{\rm PP/GMV}}{\sigma_{\Delta}}.
\end{equation}
We then convert these values to PTE values. For $A_{\rm recon}$, we find ${\rm PTE}$ values of $0.89$ and $0.97$ for the PP and GMV reconstructions relative to the MUSE reconstruction, respectively, indicating full consistency among all three estimators. For $\som$, the PP and GMV reconstructions are consistent with the MUSE reconstruction, with ${\rm PTEs}=0.43$ and $0.45$, respectively. Overall, these results show that the cosmological constraints derived from the PP reconstruction based on the quadratic estimator are statistically consistent with those obtained from MUSE, with no evidence for a significant discrepancy beyond the level expected from realization scatter.

\section{Results}\label{sec:results}

In this section we present the main cosmological constraints that we obtain. Throughout this section, we adopt the notation summarized in Table~\ref{table:notation} for the datasets and data combinations used in the analysis.

\begin{table*}
\caption{Table summarizing the shorthand notation used throughout this work to describe the various datasets and dataset combinations. For the ACT DR6 CMB-lensing, we use the baseline likelihood rather than the extended $L$-range likelihood, which uses lensing bandpowers to higher $L$.  For $\cmbspa$ we use the $lite$ version of the likelihood in this analysis using the same scale cuts as used in \cite{louis2025}. }
\begin{tabular}{lcc}
\toprule
Notation   & Dataset  & Ref \\
\midrule
$\kgmv$ & Global minimum variance CMB lensing & This work  \\
$\kgmvprof$ & Profile-hardened global minimum-variance CMB lensing & This work \\
$\kpp$  & Polarization-only CMB lensing & This work  \\
$\kgpa$ & GMV +\planck{}\,PR4 + ACT\,DR6 CMB lensing & This work  \\
$\kmuse$ & SPT-3G D1 MUSE CMB lensing   & \cite{ge2025}  \ \\
$\kact$ & ACT\,DR6 CMB lensing  & \cite{qu2024}  \ \\
$\kplk$ & \planck{}\,PR4 lensing & \cite{carron2022}  \ \\
$\kpa$ & \planck{}\,PR4 + ACT\,DR6 CMB lensing & \cite{qu2024}  \ \\
$\kmpa$ & MUSE+\planck{}\,PR4 + ACT\,DR6 CMB lensing  & \cite{qu2026}  \ \\
\midrule
$\cmbspt$ & SPT-3G D1 $TT/TE/EE$  & \cite{camphuis2026}  \ \\
$\cmbspa$ & SPT-3G D1 + ($\planck$ PR3 + ACT $TT/TE/EE$)  & \cite{louis2025,camphuis2026}   \ \\
\midrule
$\baodesi$ & DESI\,DR2 BAO  & \cite{abdulkarim2025a,abdulkarim2025b}   \\
$\baopredesi$ & 6dF+SDSS+eBOSS  BAO  & \cite{beutler2011,ross2015,alam2017,alam2021}    \\
\bottomrule
\end{tabular}
\label{table:notation}
\end{table*}

\subsection{Lensing amplitude $A_{\rm recon}$ (fixed cosmology)}
\label{ssec:resultsAlensFixcosmo}

We first report the lensing amplitude $A_{\rm recon}^{\theta_{\rm fix}}$, defined relative to the $C_L^{\kappa\kappa}$ spectrum predicted by the best-fit $\Lambda$CDM parameters from the $\cmbspa$ dataset (i.e., without lensing)~\cite{camphuis2026}, where the superscript $\theta_{\rm fix}$ denotes that the cosmological parameters are held fixed. For this fit, we also fix all nuisance parameters to their central values listed in Table~\ref{tab:priors}.

We use the lensing bandpower covariance matrix \emph{without} including the CMB-marginalization term~(Equation~\eqref{eqn:cmb_marg}) for this quantity to have the closest correspondence with the SNR values quoted in previous CMB lensing work~\cite{carron2022, qu2024, qu2026}.\footnote{Previous CMB lensing measurements typically report SNR as $\sqrt{d^{\dagger} C^{-1} d}$, where $d$ is the measured lensing bandpowers and $C$ is the bandpower covariance matrix prior to including the CMB-marginalization term. While we can omit inclusion of the CMB-marginalization term in the lensing bandpower covariance of $\kappa_{\rm ACT}$ and $\kappa_{\rm Planck}$ straightforwardly, the $\kmuse$ covariance natively has primary CMB and systematics marginalized. Thus $A_{\rm recon}^{\theta_{\rm fix}}$ of $\kmuse$ has a slightly larger uncertainty than otherwise.}
For the baseline GMV lensing reconstruction, we obtain:
\begin{equation}
A_{\rm recon}^{\theta_{\rm fix}}=1.015 \pm 0.021\hspace{0.2cm} (\kappa_{\rm GMV},\ {\rm sys.\, fixed}),\nonumber
\end{equation}
which is 1$\sigma$ consistent with unity. This corresponds to an amplitude-fit signal-to-noise ratio, ${\rm SNR}\equiv A_{\rm recon}/\sigma(A_{\rm recon})$, of 48. By measuring SNR this way, we account for the lensing bandpowers' sensitivity to $N_{L}^{(1)}$ as $A_{\rm recon}$ changes. Since $N_L^{(1)}$ depends on $C_L^{\kappa\kappa}$, its cosmology dependence provides additional sensitivity to the overall lensing amplitude, complementing the debiased bandpower measurements. Because our baseline measurement has high sensitivity to small angular scales where $N_{L}^{(1)}$ makes a non-negligible contribution, including this dependence increases the amplitude-fit SNR.

To assess the robustness of this amplitude measurement to foreground modeling, we compare results obtained with the foreground corrections included and omitted in the likelihood, thereby bracketing the resulting change in amplitude. For the cases where the foreground amplitudes are fixed to 0 (no foreground correction) or 1 (fiducial amplitude), we find $A_{\rm recon}^{\theta_{\rm fix}}=0.972\pm0.022$ and $1.015\pm0.022$, respectively, both within $1.3\sigma$ of the $\cmbspa$ cosmology. Foreground contamination primarily appears as a negative amplitude bias, driven mainly by the correlation between tSZ and $\kappa$. The measured amplitude is therefore expected to be slightly suppressed when no correction is applied, as observed here. We emphasize that the $A_{\rm fg}=0$ case is not intended as a realistic foreground model, since foreground contamination is known to be present, but is included only to illustrate the impact of neglecting the foreground correction entirely.

We now turn to the polarization-only lensing reconstruction, for which we measure:
\begin{equation}
A_{\rm recon}^{\theta_{\rm fix}}=1.017 \pm 0.030 \hspace{0.2cm} (\kappa_{\rm PP}).
\nonumber
\end{equation}
This corresponds to a signal-to-noise ratio of 34 and is also consistent with unity to within $1\sigma$. Since the polarization-only reconstruction is  nearly free of foreground biases, it provides an important cross-check of the baseline GMV result. In particular, the close agreement between the amplitudes measured from $\kgmv$ and $\kpp$ provides an additional consistency check on the foreground treatment, consistent with the difference-spectrum tests presented in Section~\ref{sec:diff_fg}. This result is also in good agreement with the polarization-only lensing amplitude obtained from the MUSE pipeline,
\begin{equation}
A_{\rm recon}^{\theta_{\rm fix}}=1.011 \pm 0.026 \hspace{0.2cm} (\kappa_{\rm MUSE}),
\nonumber
\end{equation}
although the uncertainty from the quadratic-estimator reconstruction is larger by $\sim15\%$, as expected, given that MUSE provides a more optimal lensing reconstruction.

Lastly, we also evaluate the lensing amplitude for the profile-hardened GMV, setting $A_{\rm fg}=0$. In this case we obtain
\begin{equation}
A_{\rm recon}^{\theta_{\rm fix}}=0.995 \pm 0.022\hspace{0.2cm} (\kgmvprof), \nonumber
\end{equation}
which lies between the amplitudes obtained with fiducial foreground correction and with no foreground correction in the likelihood. This behavior is not surprising. Profile hardening is designed to remove the leading-order contamination from Poisson-distributed foregrounds matching the assumed profile. Because the adopted profile matches only a limited range of source shapes, we expect some leakage from sources whose profiles differ from the assumed one. The profile-hardening  procedure is therefore not expected to remove all foreground contributions. The resulting amplitude suggests that profile hardening mitigates some fraction of the foreground-induced bias while leaving residual contamination.

\begin{figure*}
\includegraphics[width=1.00\linewidth]{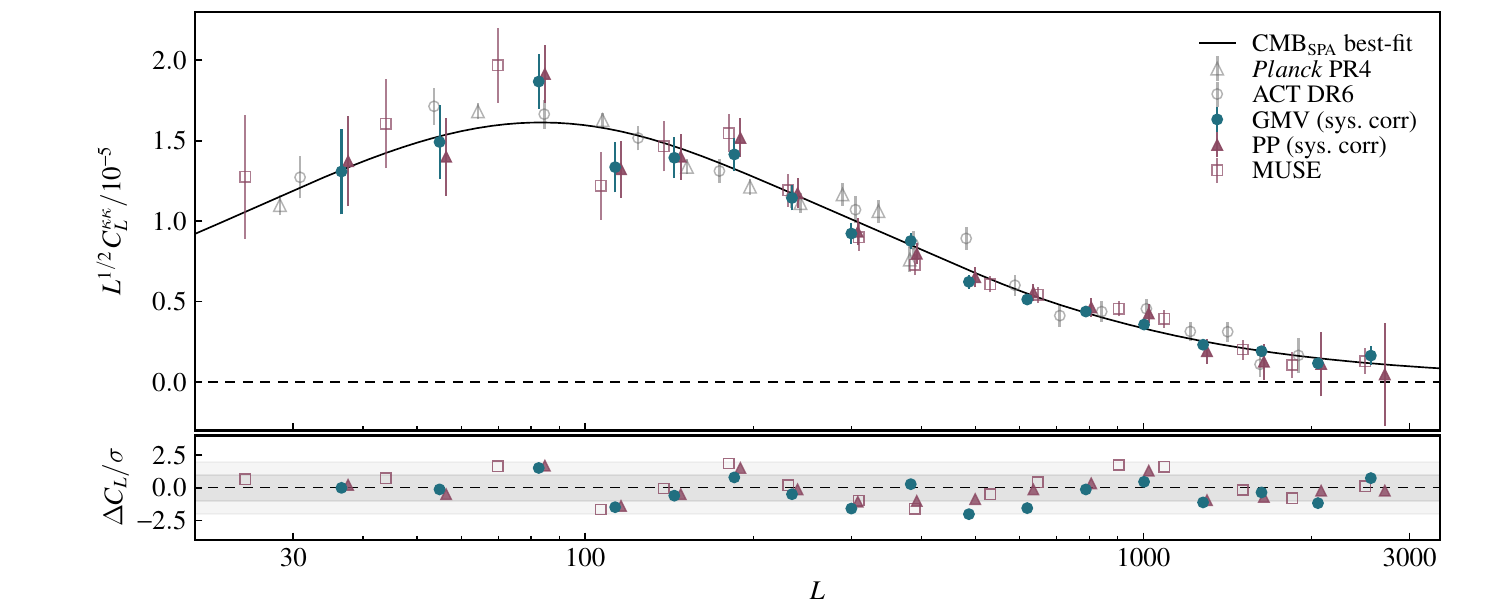} 
\caption{Measured lensing bandpowers for systematics-corrected GMV (teal), systematics-corrected PP (purple triangles), and MUSE (open purple squares) compared with measurements from \planck{} PR4 \citep{carron2022} (gray triangles) and ACT DR6 \citep{qu2024} (gray circles). The solid line is derived from the $\cmbspa$ best-fit cosmology (i.e., without lensing). The bottom panel shows the difference between each measurement and the $\cmbspa$ model, normalized by the corresponding standard deviation.}
\label{fig:clkk_main}
\end{figure*}

In Table~\ref{table:Akk}, we summarize our results, including the amplitude measured for other single-experiment lensing measurements in the literature. All the reconstructions are consistent with unity to within $1\sigma$. The bandpowers from this work are shown in Figure \ref{fig:clkk_main}, and are compared with the theoretical prediction from the $\cmbspa$ best-fit cosmology. As expected from the $A_{\rm recon}^{\theta_{\rm fix}}$ measurements, our baseline model of standard GMV with systematic correction is shown to be consistent with the $\cmbspa$ model spectrum.

\begin{table}
\caption{Summary of the best-fit fixed-cosmology lensing amplitude relative to the $\cmbspa$ best-fit cosmology.
}
\begin{tabular}{lcc}
\toprule
 Dataset   &  $A_{\rm recon}^{\theta_{\rm fix}}$ \\
\midrule
SPT-3G D1 $\kgmv$   & $1.015\pm0.021$ \\
SPT-3G D1 $\kpp$    & $1.017 \pm 0.030$ \\
SPT-3G D1 $\kgmvprof$ & $0.995 \pm 0.022$ \\
SPT-3G D1 $\kmuse$  & $1.011 \pm 0.026$ \\
\planck{} PR4  $\kappa_{\rm Planck}$ & $0.992 \pm 0.020$  \\
ACT DR6 $\kappa_{\rm ACT}$  & $1.004 \pm 0.023$  \\
\bottomrule
\end{tabular}
\label{table:Akk}
\end{table}

\subsection{Constraints from CMB lensing alone}\label{sec:ResultsLensingonly}

We now examine the cosmological constraints inferred from the lensing measurement alone. As CMB lensing probes the integrated matter distribution along the line of sight, its power spectrum is sensitive to a weighted projection of the matter power spectrum across a broad range of redshifts.

Therefore, the lensing power spectrum is sensitive to the amplitude of matter fluctuations $\sigma_{8}$ and the matter density $\Omega_{\rm m}$, with additional mild sensitivity to $H_{0}$ through the angular scale associated with matter-radiation equality. In particular, it is sensitive to the combination $\sigma_{8}\Omega_{\rm m}^{\alpha}$, where $\alpha$ is the degeneracy slope, for which the optimal direction is $\alpha\!\sim\!0.25$ for CMB lensing.

\begin{figure}
\includegraphics[width=1.00\linewidth]{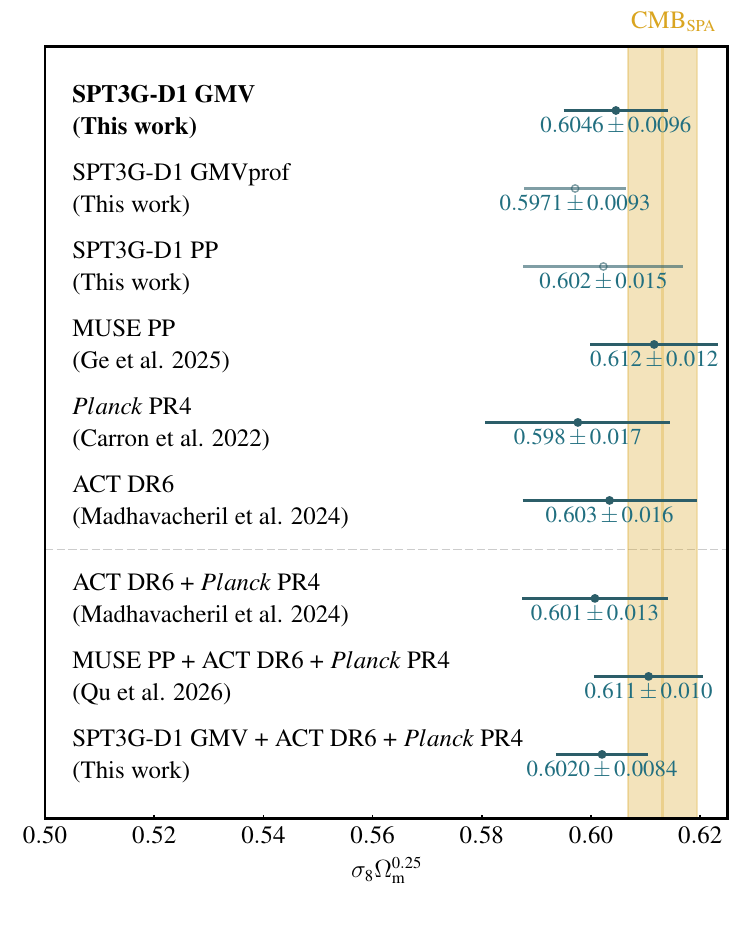} 
\caption{Summary of constraints on $\som$ using CMB-lensing-only measurements from different experiments and experiment combinations. The gold band represents the $\cmbspa$ measurement using primary CMB information without lensing. Our baseline measurement $\kgmv$ represents the most precise measurement of $\som$ using a single CMB lensing dataset, while the combined $\kgpa$-derived constraint on $\som$ reaches 1.4\% precision. All CMB lensing-derived $\som$ constraints are consistent with the $\cmbspa$ result. }
\label{fig:som_allexpt}
\end{figure}

For our baseline reconstruction, using the GMV estimator with calibration and
foreground parameters marginalized over informative priors
(see Appendix~\ref{sec:sim_calibration} for how these priors were chosen), we obtain
\begin{align}\nonumber
\som&=0.6046\pm0.0096 \hspace{0.3cm} (\kappa_{\rm GMV}).
\end{align}
This corresponds to a $1.6\%$ measurement and is the tightest current constraint
on $\som$ from a single CMB-lensing dataset. In Figure~\ref{fig:som_allexpt}, we compare the $\som$ constraints from this work with previous measurements.

To test the dependence of this result on the adopted foreground priors, we repeat the analysis with foreground prior widths enlarged by a factor of ten relative to their fiducial values. We find $\sigma_8\Omega_{\rm m}^{0.25}=0.6030\pm0.0098$, corresponding to only a modest degradation in constraining power. Removing the informative foreground priors and replacing them with uniform priors over the emulator training range gives $\sigma_8\Omega_{\rm m}^{0.25}=0.600\pm0.010$, with a further increase in the marginalized uncertainty and a $\sim0.2\sigma$ shift in the central value. This shows that the width of the foreground priors affects the marginalized constraint modestly. This is likely because foregrounds modify the shape of the lensing spectrum in a way that limits their degeneracy with $\sigma_8\Omega_{\rm m}^{0.25}$.

\begin{figure}
\includegraphics[width=1.00\linewidth]{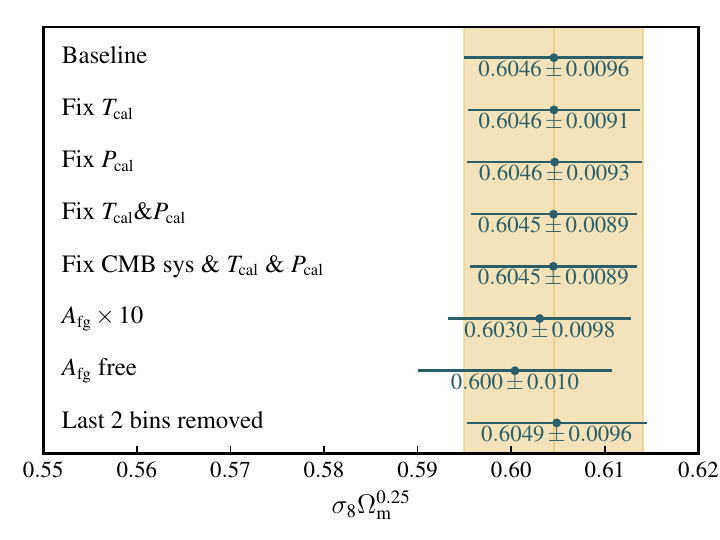} 
\caption{Summary of constraints on $\som$ from the GMV estimator under variations in the treatment of the systematic and foreground parameters, and in the multipole range. The baseline result is shown at the top; subsequent rows fix the calibration and CMB systematic parameters, broaden or remove the foreground priors, and discard the two highest-$L$ bandpowers.}
\label{fig:som_sys}
\end{figure}

In Figure~\ref{fig:som_sys}, we show the changes in the central values and uncertainties of $\som$ when the priors on systematic parameters are modified. When fixing the temperature calibration, the polarization calibration, both calibration parameters simultaneously, or both calibration parameters together with the CMB systematic parameters, we find negligible changes in both the central values and uncertainties. We also test the case of removing the two highest-$L$ bins, motivated by the hints of discrepancies between our baseline GMV reconstruction and the $\gmvxilc$ and $\gmvtsz{}$ variants as discussed in Section \ref{sec:diff_fg}, and find a negligible difference in the inferred constraints on $\som$.

For PP reconstruction, we obtain:
\begin{equation}\nonumber
\som=
0.602\pm0.015 \hspace{0.3cm} (\kappa_{\rm PP}),
\end{equation}
which is consistent with our baseline result. In contrast, for the profile-hardened reconstruction (fixing $A_{\rm fg}$ to 0), we obtain:
\begin{equation}
\som =
0.5971 \pm 0.0093 \hspace{0.3cm} (\kappa_{\rm GMVprof}), \nonumber
\end{equation}
which is unsurprisingly lower than the baseline result. Applying the same treatment as in the baseline analysis (marginalizing over foreground contamination) to the profile-hardened reconstruction is expected to shift the central value slightly upward and broaden the corresponding uncertainty.

Connecting the $\som$ results to the $A_{\rm recon}^{\theta_{\rm fix}}$ results in the previous section, we note that while $\som$ is a parameter related to the overall lensing amplitude, the distribution of angular multipoles that contribute to the $\som$ constraint differs from that contributing to $A_{\rm recon}^{\theta_{\rm fix}}$. For $A_{\rm recon}^{\theta_{\rm fix}}$, the relative contribution of each bandpower is determined primarily by its inverse variance, whereas for $\som$, it also depends on the scale-dependent response of the lensing spectrum to the cosmological parameters. In particular, on nonlinear scales, the matter power spectrum depends more strongly on $\sigma_{8}$ than the linear-theory scaling $P_{\rm m}\propto\sigma_{8}^{2}$, giving higher-$L$ bandpowers additional cosmological constraining power. Our measurement is appreciably sensitive to these higher-$L$, mildly nonlinear modes, whose sensitivity to changes in $\som$ differs from a simple scale-independent rescaling of the lensing power spectrum. Additionally, cosmological and nuisance parameters are marginalized over in the $\som$ constraints, while they are fixed in the determination of $A_{\rm recon}^{\theta_{\rm fix}}$. As such, a higher value of $A_{\rm recon}^{\theta_{\rm fix}}$ does not necessarily translate to a higher value of $\som$.

\subsubsection{Comparison with other CMB lensing measurements }\label{sec:otherlensing}

We compare our results with previous lensing measurements from SPT, including the analyses of \cite{simard2018,bianchini2020,pan2023}. These works used progressively improved datasets: the $\mathord{\sim}2500\,\sqdeg$ SPT-SZ survey using temperature-only reconstruction \cite{simard2018}, the $\mathord{\sim}500\,\sqdeg$ SPTpol survey using both temperature and polarization \cite{bianchini2020}, and an analysis of the first half-season of SPT-3G temperature data (2018) over $\mathord{\sim}1500\,\sqdeg$ 
\cite{pan2023}. The $\som$ values obtained from these datasets are:
\begin{equation}
\som=
\begin{cases}
0.598\pm0.024 \hspace{1cm} ({\rm SPT\textnormal{-}SZ}) \\[0.1cm]
0.593\pm0.025 \hspace{1cm} ({\rm SPTpol}) \\[0.1cm]
0.595\pm0.026 \hspace{1cm} ({\rm SPT\textnormal{-}3G}\textnormal{-}2018),\nonumber
\end{cases}
\end{equation}
corresponding to a $4$--$4.5\%$ uncertainty on $\som$. Compared with these constraints, our new measurement is fully consistent with the previous SPT results while improving the precision to $1.6\%$.

For the two {\it Planck} lensing analyses~\cite{planck2018lens,carron2022}, the constraints obtained are:
\begin{equation}
\som=
\begin{cases}
0.589\pm0.020 \hspace{0.3cm} ({\it Planck}\textnormal{-}{\rm PR3}), \\[0.1cm]
0.598\pm0.017 \hspace{0.3cm} ({\it Planck}\textnormal{-}{\rm PR4})\, \nonumber
\end{cases}
\end{equation}
which are in good agreement with our baseline results. Similarly, the lensing analysis from ACT\,DR6 \cite{madhavacheril2024} obtains:
\begin{equation}\nonumber
\som=0.603\pm0.016 \hspace{0.5cm} ({\rm ACT\,DR6}).
\end{equation}
Given that the {\it Planck} and ACT measurements cover a much larger area and obtain most of their signal-to-noise from larger angular scales and temperature information, while the SPT-3G result draws much of its constraining power from smaller scales and polarization data, the agreement suggests that the inferred amplitude of matter fluctuations at intermediate redshifts is relatively insensitive to the particular angular scales or CMB fields used for the lensing reconstruction.

In \cite{madhavacheril2024}, the ACT\,DR6 and \planck{} PR4 lensing bandpowers were jointly fit at the likelihood level, with scale cuts chosen such that the two experiments contribute over different multipole ranges. The covariance matrix accounts for correlations induced by the overlapping sky coverage, yielding a combined constraint of:
\begin{equation}\nonumber
\som = 0.601 \pm 0.013 \hspace{0.5cm} (\kpa) .
\end{equation}
This analysis was subsequently extended in \cite{qu2026} to include the SPT-3G D1 MUSE lensing reconstruction, giving:
\begin{equation}\nonumber
\som = 0.611 \pm 0.010 \hspace{0.5cm} (\kmpa) ,
\end{equation}
which was the tightest reported CMB lensing constraint on $\som$ at the time. 

Finally, we combine our baseline SPT-3G GMV lensing measurement with ACT\,DR6 and \planck{} PR4 following the approach in \cite{qu2026} to obtain an updated joint constraint on $\som$. In \cite{qu2026}, the correlations between MUSE and ACT DR6, and between MUSE and \planck{} PR4, were estimated analytically and validated with simulations. These correlations were found to be small: below $10\%$ with \planck{} PR4 and below $15\%$ with ACT DR6. This can be explained by the limited sky overlap, the different multipole ranges that dominate the measurements, and the fact that MUSE is a polarization-only reconstruction while the ACT DR6 and \planck{} PR4 reconstructions are primarily temperature-driven. Because our baseline GMV reconstruction includes temperature information, we expect its correlations with ACT DR6 and \planck{} PR4 to be somewhat larger than those found for MUSE. Even so, their impact on the combined constraint should remain small, since polarization provides most of the SPT-3G constraining power on the large angular scales where ACT DR6 and \planck{} PR4 carry the most statistical weight. Furthermore, \cite{qu2026} showed that there is minimal difference in $\som$ constraints between the case where no correlation is included between MUSE and the other two experiments and the pessimistic case where unrealistic amounts of correlations are included.  For these reasons, we do not include the cross-covariance between GMV and ACT DR6/\planck{} PR4 in the combination. The combined $\som$ constraint from GMV, \planck{} PR4, and ACT DR6 is:
\begin{equation}\nonumber
\som = 0.6020\pm0.0084 \hspace{0.3cm} (\kgpa),
\end{equation}
a $1.4\%$ measurement from CMB lensing alone. This precision is on par with \planck's primary CMB constraint on $\som$.

\subsection{CMB lensing + BAO}
CMB lensing alone constrains the parameters to a narrow, elongated region in the three-dimensional space spanned by $\Omega_{\rm m}$, $\sigma_{8}$, and $H_{0}$. In comparison, BAO observations constrain $D_M(z)/r_{\rm d}$ and $D_H(z)/r_{\rm d}$ across multiple redshifts. The overall distance scale is set primarily by $(h r_{\rm d})^{-1}$, while the redshift dependence is governed by $E(z)\equiv H(z)/H_0$, which depends on $\Omega_{\rm m}$ and, in extended cosmological models, on parameters such as curvature and $w$. Adding BAO to the CMB lensing measurement pins down $\Omega_{\rm m}$ and substantially tightens the allowed parameter space, yielding a late-time cosmological constraint that is largely independent of primary CMB measurements, and provides a powerful consistency test of our cosmological model between early- and late-time observations.

We use the combination of our baseline lensing measurement, $\kappa_{\rm GMV}$, with the Dark Energy Spectroscopic Instrument data release 2 (DESI\,DR2; \cite{abdulkarim2025a,abdulkarim2025b}) BAO measurements as our baseline. From this combination we obtain: 
\begin{equation}\nonumber
\left.
\begin{array}{l@{\quad}l}
\som&=0.6103\pm0.0084  \\
\Omega_{\rm m}&=0.2937 \pm 0.0076\\
\sigma_{8}&=0.829 \pm 0.012\\
H_{0} &= 68.60 \pm 0.56 {\rm \,[km/s/Mpc]}\\
\end{array}
\right\}
\left(
\begin{array}{@{}c@{}}
\kappa_{\rm GMV}\\[-2pt]
+\\[-2pt]
\baodesi
\end{array}
\right).
\end{equation}
Within $\Lambda$CDM, the $\baodesi$ measurements are known to favor slightly lower values of $\Omega_{\rm m}$ and higher values of $H_{0}r_{\rm d}$ than those inferred from primary CMB temperature and polarization data and earlier BAO analyses.

\begin{figure*}[!t]
\centering
\includegraphics[width=0.65\linewidth]{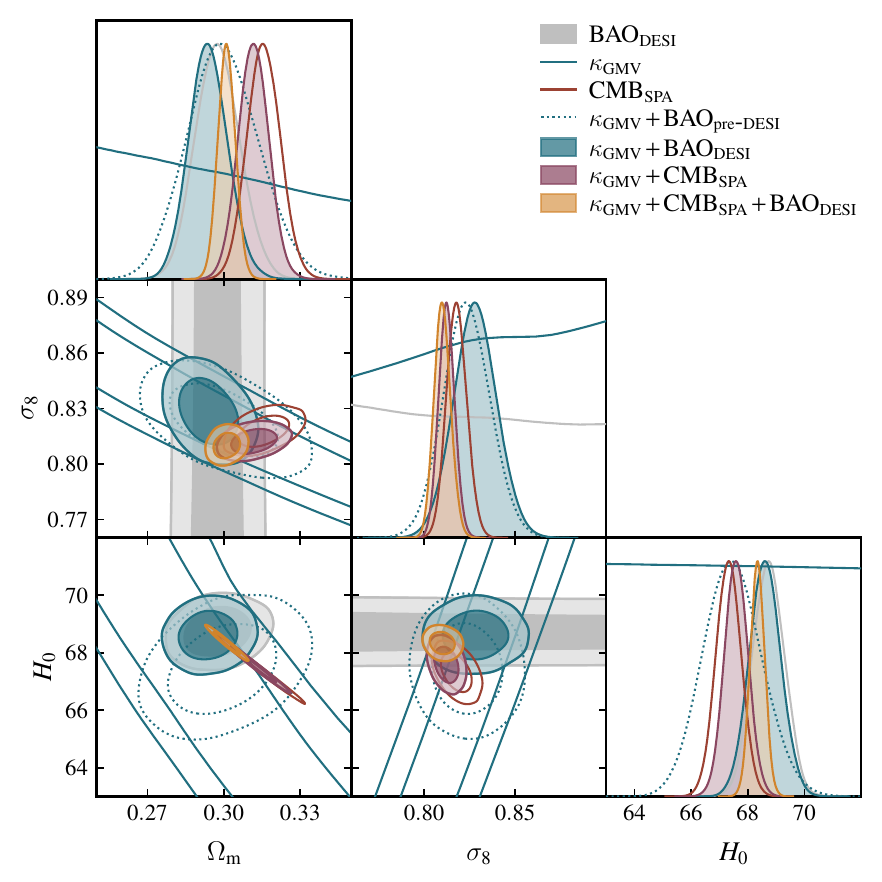}
\caption{
Comparison of cosmological constraints in the $\Omega_{\rm m}$-$\sigma_{8}$-$H_{0}$ parameter space from $\baodesi$ alone, $\kgmv$ alone, $\cmbspa$ alone, $\kgmv$ combined with BAO (both DESI and pre-DESI), $\kgmv$ combined with $\cmbspa$, and the full combination of $\kgmv$, $\cmbspa$, and $\baodesi$. The addition of BAO/CMB pins down the position along the lensing degeneracy direction, significantly tightening constraints on $\Omega_{\rm m}$ and $H_{0}$.}
\label{fig:lenscmb_omegam_sigma8_H0}
\end{figure*}

The preference for lower $\Omega_{\rm m}$ in the DESI data is evident when we compare our results with those obtained using pre-DESI BAO measurements. These consist of 6dFGS \cite{beutler2011}, the SDSS DR7 Main Galaxy Sample (MGS; \citealt{ross2015}), BOSS DR12 luminous red galaxies \cite{alam2017}, and eBOSS DR16 LRGs \cite{alam2021}.
This combination yields:
\begin{equation}\nonumber
\left.
\begin{array}{l@{\quad}l}
\som&=0.6093\pm0.0085  \\
\Omega_{\rm m}&=0.299 \pm 0.013\\
\sigma_{8}&=0.824 \pm 0.013\\
H_{0} &= 67.43 \pm 1.01 \,{\rm \,[km/s/Mpc]}
\end{array}
\right\}
\left(
\begin{array}{@{}c@{}}
\kappa_{\rm GMV}\\[-2pt]
+\\[-2pt]
\baopredesi
\end{array}
\right).
\end{equation}
Relative to DESI, the pre-DESI datasets intersect the CMB lensing $\Omega_{\rm m}$–$\sigma_{8}$–$H_0$ posterior at higher $\Omega_{\rm m}$, with correspondingly lower preferred values of $\sigma_{8}$ and $H_0$. 
The shifts in the posteriors can be seen in Figure~\ref{fig:lenscmb_omegam_sigma8_H0} with $\kgmv$+pre-DESI in dotted teal and $\kgmv$+DESI in filled teal contours.
To determine whether these shifts affect the overall agreement with primary CMB constraints, we quantify the statistical distance between the combined lensing+BAO results and primary $\cmbspa$ measurements using \texttt{tensiometer}~\cite{raveri2020,raveri2021}; see Appendix~\ref{app:tensiometer} for details of the estimator. In the $\Omega_{\rm m}$--$\sigma_{8}$--$H_{0}$ parameter space, we find distances of 1.8 and 0.4$\sigma$ for $\kgmv+\baodesi$ versus $\cmbspa$ and for $\kgmv+\baopredesi$ versus $\cmbspa$, respectively. Performing the same comparison in the $\Omega_{\rm m}$--$h r_{\rm d}$ plane yields distances of 2.3 and 0.6$\sigma$, respectively. Thus, the pre-DESI combination remains in closer agreement with the primary CMB, while the DESI combination shows a larger difference, most apparent in the $\Omega_{\rm m}$--$h r_{\rm d}$ plane.

\subsection{CMB lensing + CMB primary}\label{sec:cmb_lens}
Primary CMB measurements probe the physics in the early Universe around the epoch of recombination and provide precise constraints on the cosmological parameters that set the initial conditions and background evolution. CMB lensing, on the other hand, probes the intermediate-redshift Universe through the integrated matter distribution along the line of sight and is therefore a more direct tracer of late-time structure growth. Because the two observables probe different epochs and are sensitive to cosmological parameters in different ways, their combination provides complementary constraints on cosmological parameters.

For the primary CMB constraints, we take the combination of SPT-3G\,D1~\citep{camphuis2026}, \planck{} PR3~\citep{planck2018_likelihood}, and ACT\,DR6~\citep{louis2025} temperature and polarization measurements, which we denote as $\cmbspa$. For runs involving $\cmbspa$, we additionally include the SRoll2 low-$\ell$ $EE$ likelihood~\citep{delouis2019} to constrain $\tau$. We combine our baseline $\kgmv$ measurement with $\cmbspa$ and obtain:
\begin{equation}\nonumber
\left.
\begin{array}{l@{\quad}l}
\som&=0.6070\pm0.0050  \\
\Omega_{\rm m}&=0.3117 \pm 0.0059\\
\sigma_{8}&=0.8124 \pm 0.0044\\
H_{0} &= 67.60 \pm 0.42 {\rm \,[km/s/Mpc]}\\
\end{array}
\right\}
\left(
\begin{array}{@{}c@{}}
\kappa_{\rm GMV}\\[-2pt]
+\\[-2pt]
\cmbspa
\end{array}
\right).
\end{equation}
The results are shown in Figure~\ref{fig:lenscmb_omegam_sigma8_H0}, and the best-fit values are summarized in Table~\ref{tab:all_parameters}. We find that adding $\kgmv$ to $\cmbspa$ improves the primary CMB constraints by $9.2\%$, $21.4\%$, and $6.7\%$ in $\Omega_{\rm m}$, $\sigma_{8}$, and $H_{0}$, respectively. 

We also combine GMV+ACT DR6+\planck{} PR4 lensing with $\cmbspa$, obtaining
\begin{equation}\nonumber
\left.
\begin{array}{l@{\quad}l}
\som&=0.6074\pm0.0044  \\
\Omega_{\rm m}&=0.3117 \pm 0.0057\\
\sigma_{8}&=0.8130 \pm 0.0039\\
H_{0} &= 67.59 \pm 0.40 {\rm \,[km/s/Mpc]}\\
\end{array}
\right\}
\left(
\begin{array}{@{}c@{}}
\kgpa\\[-2pt]
+\\[-2pt]
\cmbspa
\end{array}
\right),
\end{equation}
which yields constraints that are $12\%$, $30\%$, and $11\%$ tighter in $\Omega_{\rm m}$, $\sigma_{8}$, and $H_{0}$, respectively, than those from $\cmbspa$ alone. These results highlight the increasingly important role of CMB lensing in sharpening cosmological parameter constraints.

\begin{figure}[t]
\centering
\includegraphics[width=\columnwidth]{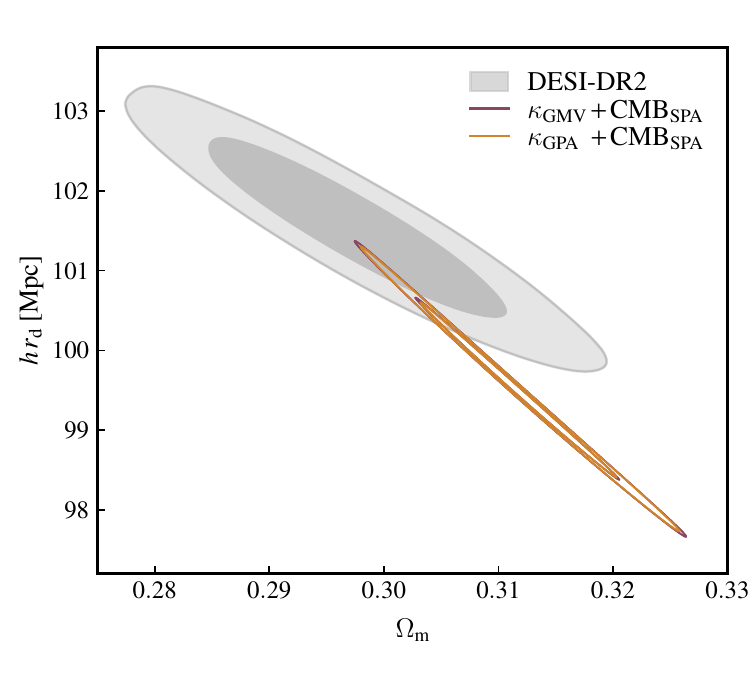}
\caption{Constraints in the $\Omega_{\rm m}$-$hr_{\rm d}$ plane for $\baodesi$, $\kgmv+\cmbspa$, and $\kgpa+\cmbspa$. The distances between the CMB+lensing constraints and $\baodesi$ are 1.8 and 1.9$\sigma$ for $\kgmv$ and $\kgpa$ combined with $\cmbspa$, respectively. }
\label{fig:omegam_hrdrag}
\end{figure}

Before combining $\kappa_{\rm GMV/GPA}+\cmbspa$ with BAO datasets, we quantify the consistency of these constraints with BAO measurements. We evaluate the separation between the constraints in the $\Omega_{\rm m}$--$h r_{\rm d}$ plane, which are the parameters directly constrained by BAO within $\Lambda$CDM. Relative to $\baodesi$, we find distances of $1.8\sigma$ and $1.9\sigma$ for $\kgmv+\cmbspa$ and $\kgpa+\cmbspa$, respectively. Using $\baopredesi$, the corresponding distances are $1.1\sigma$ for both combinations. These results indicate no statistically significant discrepancy between the CMB lensing + primary CMB and BAO constraints. The corresponding posterior contours are shown in Figure~\ref{fig:omegam_hrdrag}. 

These distances are also overall smaller than those measured in C26, who used the $\kmpa+\cmbspa$ combination\footnote{$\kmpa$ denotes the lensing measurements from MUSE, \planck, and ACT; see Table~\ref{table:notation}.} and found a $2.8\sigma$ difference relative to $\baodesi$. There are two factors contributing to this difference. 
The first is our use of SRoll2 rather than the $\tau$ prior $\mathcal{N}(0.051,0.006)$ adopted in C26~\cite{planck2020LFIHFItau}. SRoll2 favors a higher $\tau$, which shifts $\Omega_{\rm m}$ lower and improves agreement with $\baodesi$.
To make a more direct comparison with C26, we repeat the analysis using the same Gaussian $\tau$ prior. As shown in Appendix~\ref{app:tau_prior}, this increases the $\kgpa+\cmbspa$ distance from $\baodesi$ to $2.3\sigma$ from $1.9\sigma$. The remaining difference arises from the slightly lower $\som$ preferred by $\kgpa$ compared with $\kmpa$: lower $\som$ from lensing leads to lower $\Omega_{\rm m}$ when combined with primary CMB, because $\sigma_8$ and $\Omega_{\rm m}$ are positively correlated within the parameter space constrained by $\cmbspa$, as shown in Figure~\ref{fig:lenscmb_omegam_sigma8_H0}. This shifts the combined primary CMB + lensing constraint toward the DESI-derived $\Omega_{\rm m}$, reducing the distance to approximately $2\sigma$.

The sensitivity of the CMB primary+lensing versus BAO comparison to the treatment of $\tau$ can also propagate into shifts of other parameters in extended models of $\Lambda$CDM (see e.g., Sections~\ref{sec:mnu}, \ref{sec:omk}, \ref{sec:w0wa}, and C26). This motivates us to directly constrain $\tau$ using the combination of CMB lensing and primary CMB, without including a low-$\ell$ polarization likelihood or an external $\tau$ prior. This is possible because the primary CMB constrains $A_{\rm s}e^{-2\tau}$, while CMB lensing provides an independent constraint on $A_{\rm s}$ through the amplitude of matter fluctuations, thereby breaking the $A_{\rm s}$--$\tau$ degeneracy. For the $\kgpa$+$\cmbspa$ dataset, we measure $\tau$ to be:
\begin{equation}\nonumber
\tau = 0.072 \pm 0.014  \hspace{0.5cm}  (\kgpa + \cmbspa, \text{no\ SRoll2}),
\end{equation}
consistent with $\tau = 0.076 \pm 0.013$ obtained from $\kmpa$+$\cmbspa$ in C26, and also with the SRoll2-derived $\tau = 0.0566^{+0.0053}_{-0.0062}$~\cite{pagano2020} at 1.0$\sigma$. The distance between $\kgpa$+$\cmbspa$ and $\baodesi$ in the $\Omega_{\rm m}$--$h r_{\rm d}$ plane is 0.9$\sigma$ in this case.

\subsection{Combined constraints: primary CMB  + lensing + BAO}\label{sec:te_lens_bao}

Since the CMB lensing, primary CMB, and BAO measurements are mutually
consistent at the $\lesssim 2\sigma$ level, we combine the three probes and
report the resulting joint constraints. For our baseline combination of
$\kgmv$, $\cmbspa$, and $\baodesi$, we find
\begin{equation}\nonumber
\left.
\begin{array}{l}
  \Omega_{\rm m} = 0.3011\pm{0.0034} \\[0.1cm]
  \sigma_{8} \hspace{0.15cm} = 0.8102\pm{0.0045} \\[0.1cm]
  H_{0} \hspace{0.05cm} = 68.34\pm{0.26}
  \,{\rm \,[km/s/Mpc]}
\end{array}
\right\}
\left(
\begin{array}{@{}c@{}}
  \kappa_{\rm GMV}\\[-2pt]
  +\\[-2pt]
  \cmbspa\\[-2pt]
  +\\[-2pt]
  \baodesi
\end{array}
\right).
\end{equation}
When we instead use the combined lensing result, we obtain:
\begin{equation}\nonumber
\left.
\begin{array}{l@{\quad}l}
\Omega_{\rm m} &= 0.3014 \pm 0.0034 \\[0.1cm]
\sigma_{8} &= 0.8117 \pm 0.0042 \\[0.1cm]
H_{0} &= 68.32 \pm 0.26 {\rm \,[km/s/Mpc]}
\end{array}
\right\}
\left(
\begin{array}{@{}c@{}}
\kgpa\\[-2pt]
+\\[-2pt]
\cmbspa\\[-2pt]
+\\[-2pt]
\baodesi
\end{array}
\right).
\end{equation}
This combined constraint provides our most precise determination of the late-time cosmological parameters. Its stability under alternative choices of the CMB lensing input shows that the inferred parameters are not driven by any single lensing dataset. Since the primary CMB and BAO data are held fixed throughout, this comparison specifically tests the consistency of the different CMB lensing measurements within the $\Lambda$CDM framework.

\begin{table}[H]
\caption{
Cosmological constraints from CMB lensing reconstructions combined with $\cmbspa$ and $\baodesi$ constraints. Quoted uncertainties correspond to $68\%$ confidence intervals.
}
\centering
\begin{tabular}{l c c c}
\toprule

& $\Omega_{\rm m}$ 
& $\sigma_{8}$ 
& $H_{0}$ [km/s/Mpc] \\
\midrule
$\cmbspa$ & & & \\
\hspace{0.1cm}+$\kgmv$   &  $0.3117 \pm 0.0059$ & $0.8124 \pm 0.0044$ & $67.60 \pm 0.42$ \\
\hspace{0.1cm}+$\kgpa$   &  $0.3117 \pm 0.0057$ & $0.8130 \pm 0.0039$ & $67.59 \pm 0.40$ \\
\hspace{0.1cm}+$\kmuse$  &  $0.3169 \pm 0.0057$ & $0.8191 \pm 0.0047$ & $67.24 \pm 0.39$ \\
\hspace{0.1cm}+$\kmpa$    &  $0.3152 \pm 0.0055$ & $0.8168 \pm 0.0039$ & $67.35 \pm 0.38$ \\
\hspace{0.1cm}+$\kact$    &  $0.3145 \pm 0.0061$ & $0.8166 \pm 0.0047$ & $67.40 \pm 0.43$ \\
\hspace{0.1cm}+$\kplk$ &  $0.3128 \pm 0.0059$ & $0.8150 \pm 0.0047$ & $67.52 \pm 0.41$ \\
\midrule
\multicolumn{4}{l}{$\cmbspa$+$\baodesi$} \\
\hspace{0.1cm}+$\kgmv$   &  $0.3011 \pm 0.0034$ & $0.8102 \pm 0.0045$ & $68.34 \pm 0.26$ \\
\hspace{0.1cm}+$\kgpa$   &  $0.3014 \pm 0.0034$ & $0.8117 \pm 0.0042$ & $68.32 \pm 0.26$ \\
\hspace{0.1cm}+$\kmuse$  &  $0.3034 \pm 0.0034$ & $0.8173 \pm 0.0051$ & $68.19 \pm 0.25$ \\
\hspace{0.1cm}+$\kmpa$    &  $0.3032 \pm 0.0034$ & $0.8164 \pm 0.0043$ & $68.20 \pm 0.25$ \\
\hspace{0.1cm}+$\kact$    &  $0.3020 \pm 0.0035$ & $0.8137 \pm 0.0048$ & $68.28 \pm 0.26$ \\
\hspace{0.1cm}+$\kplk$ &  $0.3017 \pm 0.0035$ & $0.8127 \pm 0.0049$ & $68.30 \pm 0.26$ \\

\bottomrule
\end{tabular}
\label{table:results_lens_cmb_bao}
\end{table}

\subsubsection{Scalar spectrum tilt $n_{\rm s}$}
Adding $\baodesi$ to combinations of \planck{} PR3 and ACT\,DR6 data shifts the inferred $n_{\rm s}$ to higher values relative to previous \planck{} results~\cite{louis2025,planck2018_likelihood}, an effect that has attracted considerable attention. This shift brings some previously disfavored monomial inflation models into better agreement with observations, while models such as Starobinsky and Higgs inflation become less favored, depending on the assumed reheating history~\cite{louis2025,balkenhol2025,inflation_book,inflation_encyl,inflation_review2025}.

For our baseline lensing combination, we find
\begin{equation}
n_{\rm s}=
\begin{cases}
0.9701 \pm 0.0034\hspace{0.3cm} (\cmbspa+\kgpa);\\
0.9742 \pm 0.0030\hspace{0.3cm} (\cmbspa+\kgpa+\baodesi).
\end{cases}
\end{equation}
The shift toward higher $n_{\rm s}$ when $\baodesi$ is included can be understood from the anticorrelation between $n_{\rm s}$ and $\Omega_{\rm m}$~\cite{Ferreira2026}: the lower $\Omega_{\rm m}$ favored by $\baodesi$ shifts the combined constraint toward higher $n_{\rm s}$. Since the difference between the CMB and BAO constraints in the $\Omega_{\rm m}$--$h r_{\rm d}$ plane is smaller in this work than in C26, the corresponding upward shift in $n_{\rm s}$ is also reduced, although the preference for higher $n_{\rm s}$ remains.

\subsection{Constraints on $A_{\rm lens}$, $A_{\rm 2pt}$, and $A_{\rm recon}$}\label{ssec:a2ptarecon}

In Section~\ref{ssec:resultsAlensFixcosmo}, we measured the lensing amplitude relative to the $\Lambda$CDM prediction at the $\cmbspa$ best-fit cosmology. In this section, we combine our lensing measurement with primary CMB and BAO data and investigate whether the lensing amplitude measured from the lensing reconstruction is consistent with that inferred from the peak-smoothing effect in the primary CMB spectra.

Following~\cite{ge2025,camphuis2026}, we introduce two independent parameters to characterize the amplitude of lensing. We use $A_{\rm 2pt}$ to rescale the lensing-induced smoothing of the acoustic peaks in the primary CMB spectra, and $A_{\rm recon}$ to rescale the theoretical CMB lensing power spectrum. This parameterization generalizes the commonly used $A_{\rm lens}$ parameter, in which a single amplitude rescales the lensing potential power spectrum and therefore simultaneously controls both the peak-smoothing effect in the primary CMB spectra and the reconstructed CMB lensing power spectrum. In our analysis, we vary the standard $\Lambda$CDM parameters jointly with $A_{\rm 2pt}$ and $A_{\rm recon}$.\footnote{The $A_{\rm recon}$ in this section is different from the $A_{\rm recon}^{\theta_{\rm fix}}$ in Section~\ref{ssec:resultsAlensFixcosmo} in that $A_{\rm recon}$ has $\Lambda$CDM and nuisance parameters marginalized over and is sampled jointly with the lensing datasets and the primary CMB (and BAO) datasets. Because of different parameter degeneracies, the differences in $A_{\rm recon}^{\theta_{\rm fix}}$  between datasets in the earlier section do not have to track the differences in $A_{\rm recon}$ or $A_{\rm lens}$ in this section. For example, the uncertainty on $A_{\rm recon}$ is the largest for $\kgmv$, while its constraint on  $A_{\rm recon}^{\rm \theta_{\rm fix}}$ is the second tightest amongst the different measurements. This is because  calibration parameters are  degenerate with $A_{\rm recon}$.}

\begin{figure}
\includegraphics[width=0.99\linewidth]{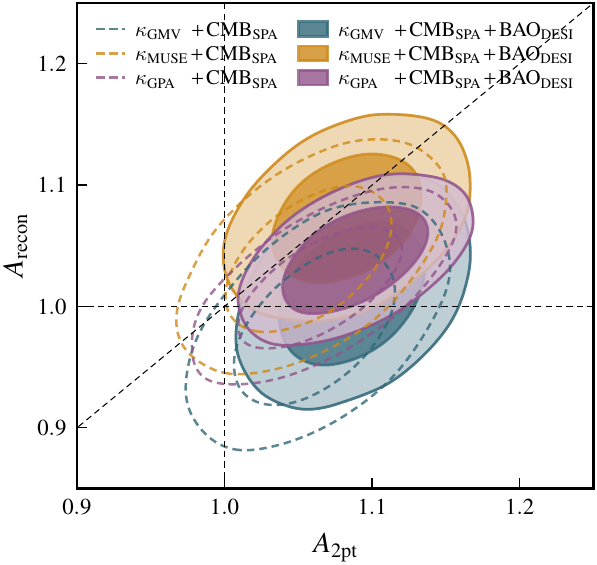} 
\caption{Constraints on $A_{\rm 2pt}$ and $A_{\rm recon}$ from jointly analyzing  $\kgmv/\kgpa/\kmuse$, $\cmbspa$, and $\baodesi$. Dashed black lines represent $A_{\rm 2pt}=1$, $A_{\rm recon}=1$, and $A_{\rm 2pt}=A_{\rm recon}$.  }
\label{fig:a2ptarec}
\end{figure}

We show in Figure~\ref{fig:a2ptarec} the $A_{\rm 2pt}$--$A_{\rm recon}$ posteriors for $\kgmv$, $\kmuse$, $\kgpa$, and $\kmpa$, combined with either $\cmbspa$ alone or $\cmbspa+\baodesi$. The recovered $A_{\rm 2pt}$ values are largely insensitive to the choice of lensing dataset. They lie approximately $1.6$--$1.8\sigma$ above the $\Lambda$CDM expectation of $A_{\rm 2pt}=1$ when combined with $\cmbspa$ alone, with the preference increasing to approximately $2.5$--$3\sigma$ after adding $\baodesi$. The constraints are driven primarily by $\cmbspa$ and $\baodesi$ datasets. Turning to $A_{\rm recon}$, when the lensing measurements are combined with $\cmbspa$ alone, all of the inferred values are consistent with unity within $\sim1\sigma$. The $\kmuse$ result lies slightly higher than the others, as also noted in~\cite{ge2025}, but the different lensing measurements remain mutually consistent. Further adding $\baodesi$ shifts the $A_{\rm recon}$ posteriors upward for all of the lensing measurements. The inclusion of $\baodesi$ shifts the cosmological parameters toward a lower predicted CMB lensing amplitude, primarily through $\sigma_8$ and $\Omega_{\rm m}$, which is compensated by a larger value of $A_{\rm recon}$. While the other lensing combinations show a clearer preference for $A_{\rm recon}>1$, the $\kgmv$ and $\kgpa$ results show only mild upward shifts, at $0.3\sigma$ and $1.3\sigma$ above unity, respectively:
\begin{equation}
A_{\rm recon}=
\begin{cases}
1.011 \pm 0.039\hspace{0.3cm} (\kgmv + \cmbspa + \baodesi)\nonumber,\\
1.039 \pm 0.029\hspace{0.3cm} (\kgpa \hspace{0.1cm} + \cmbspa + \baodesi)\nonumber.\\
\end{cases}
\end{equation} 
Taken together, while the primary CMB spectra consistently prefer $A_{\rm 2pt}>1$, the corresponding behavior of $A_{\rm recon}$ depends on the lensing reconstruction. A coherent excess in both parameters would be more suggestive of a common physical origin, since explaining both the peak smoothing and lensing reconstruction through systematics would require correlated effects in two distinct observables. Conversely, an excess confined primarily to $A_{\rm 2pt}$ (as we see with $\kgmv$) would more naturally point to effects specific to the primary CMB spectra. Given the spread among the lensing reconstructions, however, we do not draw a strong conclusion from the current measurements.

In addition to varying $A_{\rm 2pt}$ and $A_{\rm recon}$ independently, we also consider a single-parameter model in which $A_{\rm lens}$ jointly rescales the lensing contribution to the primary CMB power spectra and the CMB lensing power spectrum. We find
\begin{equation}
A_{\rm lens}=
\begin{cases}
1.060 \pm 0.029\hspace{0.3cm} (\kgmv + \cmbspa + \baodesi),\nonumber\\
1.058 \pm 0.027\hspace{0.3cm} (\kgpa \hspace{0.1cm}+\cmbspa +\baodesi).\nonumber\\
\end{cases}
\end{equation}
Both results lie approximately $2\sigma$ above the $\Lambda$CDM expectation of $A_{\rm lens}=1$. The preference for $A_{\rm lens}>1$ is driven primarily by the primary-CMB constraint on $A_{\rm 2pt}$ and, as discussed above, is enhanced by the lower value of $\Omega_{\rm m}$ preferred by $\baodesi$. Table~\ref{table:a2ptarec} summarizes the results from these analyses and also lists analogous constraints obtained by replacing $\kgmv$ or $\kgpa$ with other CMB lensing datasets for reference. Since a preference for excess lensing can artificially tighten the upper bound on the sum of neutrino masses, we examine its impact on the neutrino-mass constraints in the following section.

\begin{table}
\caption{Constraints on $A_{\rm 2pt}$, $A_{\rm recon}$, and $A_{\rm lens}$ using the combination lensing + CMB primary or lensing + CMB primary + BAO.  }
\label{table:a2ptarec}
\begin{tabular}{lccc}
\toprule
 Dataset   & $A_{\rm 2pt}$ &  $A_{\rm recon}$ &   $A_{\rm lens}$ \\
\midrule
$\cmbspa$ & & \\
  \hspace{0.1cm}+$\kgmv$   & $1.062\pm0.036$ & $0.984\pm0.042$ & $1.037 \pm0.034$   \\
 \hspace{0.1cm}+$\kmuse$  &  $1.058\pm0.037$ & $1.040\pm0.040$ & $1.054\pm 0.032$ \\
 \hspace{0.1cm}+$\kgpa$  &  $1.065\pm0.037$ & $1.016\pm0.033$ & $1.036 \pm0.031$  \\
  \hspace{0.1cm}+$\kmpa$  &  $1.066\pm0.036$ & $1.029\pm0.031$  & $1.040\pm 0.030$\\
 \hspace{0.1cm}+$\kplk$  &  $1.064\pm0.036$ & $1.010\pm0.032$ & $1.030 \pm 0.030$  \\
  \hspace{0.1cm}+$\kact$  &  $1.064\pm0.036$ & $1.025\pm0.037$  & $1.046 \pm 0.032$\\
\midrule

\multicolumn{4}{l}{$\cmbspa$+$\baodesi$} \\
\hspace{0.1cm}+$\kgmv$ & $1.087\pm0.033$ &  $1.011\pm0.039$ & $1.060\pm0.029$\\
 \hspace{0.1cm}+$\kmuse$ & $1.083\pm0.034$ &  $1.073\pm0.035$ & $1.082 \pm 0.027$ \\
 \hspace{0.1cm}+$\kgpa$ & $1.088\pm0.032$ &  $1.039\pm0.029$ & $1.058\pm0.027$\\
  \hspace{0.1cm}+$\kmpa$ & $1.094\pm0.032$ &  $1.058\pm0.026$ & $1.068 \pm 0.024$\\
 \hspace{0.1cm}+$\kplk$ & $1.089\pm0.033$ &  $1.034\pm0.028$ & $1.054 \pm 0.025$\\
  \hspace{0.1cm}+$\kact$ & $1.090\pm0.033$ &  $1.050\pm0.034$ & $1.072 \pm 0.028$\\
 \bottomrule
\end{tabular}
\end{table}

\subsection{Massive neutrinos}\label{sec:mnu}
The combination of CMB lensing, primary CMB, and BAO data constrains $\summnu$ through their complementary sensitivity to the initial conditions, expansion history, and growth of structure. Primary CMB measurements constrain the physical matter densities and primordial fluctuation amplitude, while BAO measurements constrain the late-time distance scale and break geometric degeneracies. CMB lensing then measures the integrated growth of structure relative to these constraints. Since massive neutrinos suppress structure growth below their free-streaming scale, the combined comparison provides sensitivity to $\summnu$~\cite{Lesgourgues_Mangano_Miele_Pastor_2013}.

To understand how these datasets constrain $\summnu$, we first consider the parameter combination constrained jointly by primary CMB and BAO data, and then the additional information provided by CMB lensing. The CMB acoustic peaks precisely determine the angular size of the sound horizon at recombination, $\theta_{\rm s}$,\footnote{In our parameterization, we sample $\theta_{\rm MC}$, an approximation to the angular acoustic scale $\theta_{\rm s}$~\citep{planck2013params}.} while BAO measurements determine low-redshift distances relative to the closely related sound horizon at the drag epoch, $r_{\rm d}$. Requiring a single expansion history to reproduce both measurements tightly constrains the combination $\omega_{\rm m} r_{\rm d}^2$, where $\omega_{x} \equiv \Omega_{x} h^{2}$ denotes a physical density and $\omega_{\rm m} = \omega_{\rm c} + \omega_{\rm b} + \omega_{\nu}$ and $\omega_{\nu} \simeq \sum m_{\nu}/(93.14\,{\rm eV})$~\cite{loverde2024,lynch2025}.
The combination of primary CMB and BAO data therefore already provides sensitivity to $\summnu$ through its contribution to the physical matter density. CMB lensing adds complementary information through its sensitivity to a combination of $\sigma_8$ and $\Omega_{\rm m}$. When combined with primary CMB and BAO data, this additional constraint on the late-time matter distribution helps reduce parameter degeneracies and improves the resulting constraint on $\summnu$.\footnote{\citet{lynch2025} characterize the complementary CMB-lensing information as an approximate constraint on the combination $\omega_{\rm cb}-b\omega_\nu$, with $b\simeq0.6$ and $\omega_{\rm cb} \equiv \omega_{\rm c} + \omega_{\rm b}$.}

We begin with the 95\% C.L. upper limit obtained using ${\cmbspa}$ and $\baodesi$ without including CMB lensing:
\begin{equation}\nonumber
\summnu < 0.067\,{\rm eV}\qquad (\cmbspa+\baodesi).
\end{equation}
As discussed in \cite{camphuis2026, ge2025, qu2026, green2025, loverde2024}, this surprisingly\footnote{Given the constraining power of the two datasets individually, a weaker upper limit on $\summnu$ would be expected if their preferred $\Lambda$CDM parameters were in closer agreement. In particular, if the prior were extended to allow $\summnu<0$, a substantial fraction of the posterior density would lie below zero, reflecting the difference between the matter densities preferred by the CMB and BAO data. Imposing the physical prior $\summnu\geq0$ truncates this part of the posterior and yields an unusually low upper limit.} stringent upper limit arises from the different values of $\Omega_{\rm m}$ preferred by the two datasets. In the $\Omega_{\rm m}$--$\summnu$ plane, the degeneracy directions of $\cmbspa$ and $\baodesi$ intersect such that the lower value of $\Omega_{\rm m}$ preferred by $\baodesi$ shifts the combined posterior toward lower $\summnu$~\cite{lynch2025}. Since the posterior is truncated at the physical boundary $\summnu=0$, this shift compresses the allowed posterior volume and yields a tighter one-sided upper limit.

Next we look into adding CMB lensing. We obtain:
\begin{equation}\nonumber
\summnu < 0.075\, {\rm eV}\ (\kgmv + \cmbspa + \baodesi), 
\end{equation}
when our CMB lensing measurement $\kgmv$ is added and 
\begin{equation}\nonumber
\summnu < 0.072\, {\rm eV}\ ( \kgpa + \cmbspa + \baodesi),
\end{equation}
if both $\planck{}$ PR4 and ACT\,DR6 lensing measurements are additionally included.
We therefore observe that our lensing measurement relaxes the neutrino mass upper limit. In contrast, in the case of using $\planck{}$ PR4 and ACT\,DR6 lensing alone without $\kgmv$ (i.e., $\kpa$), the constraint remains at $\summnu<0.067\,{\rm eV}$, and in the case of replacing $\kgpa$  with the combination of MUSE, $\planck{}$ PR4, and ACT\,DR6 lensing ($\kmpa$), we obtain $\summnu<0.058\,{\rm eV}$, lower than the no-CMB-lensing case.

To understand the origin of this shift, we next consider combinations of primary CMB and CMB lensing data without BAO. Adding our CMB lensing measurement to $\cmbspa$, we obtain
\begin{equation}\nonumber
\summnu < 0.268\,{\rm eV}\qquad (\kgmv+\cmbspa).
\end{equation}
Using $\kgpa$ gives a slightly lower upper limit of $0.265\,{\rm eV}$, while replacing $\kgmv$ with $\kpa$ yields $\summnu<0.230\,{\rm eV}$. For comparison, $\cmbspa$ alone yields $\summnu<0.206\,{\rm eV}$. Thus, even without $\baodesi$, adding CMB lensing shifts the posterior toward larger values of $\summnu$, with a larger shift when $\kgmv$ is included.

\begin{figure}
\includegraphics[width=1.00\linewidth]{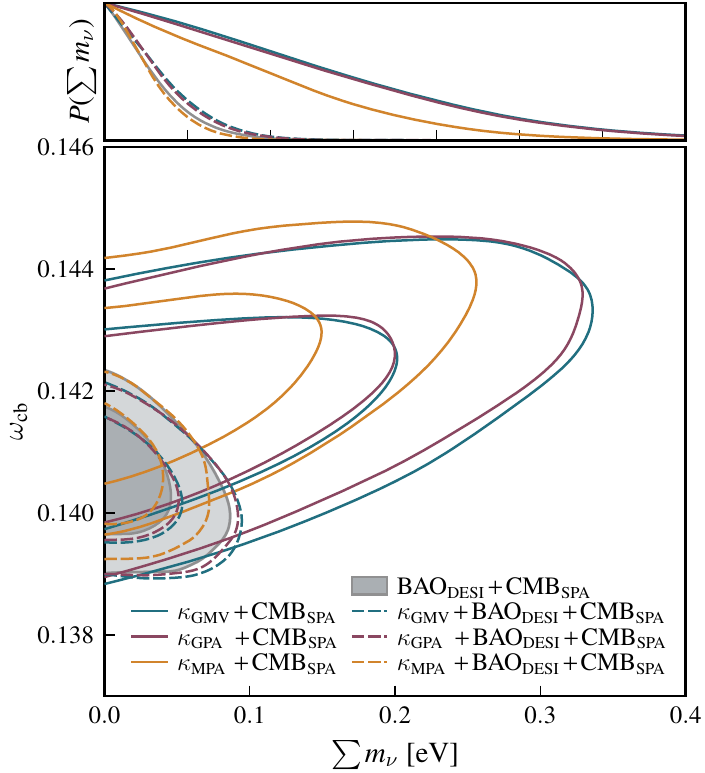} 
\caption{{\bf Upper}: 1D $\summnu$ upper limit with and without $\baodesi$ (dashed and solid lines, respectively). {\bf Lower}: Degeneracy in the $\omega_{\rm cb}$--$\summnu$ plane
for CMB lensing + $\cmbspa$ (solid contours) and CMB lensing + $\cmbspa$  + $\baodesi$ (dashed contours). 
The $\kgmv$+$\cmbspa$ and $\kgpa$+$\cmbspa$ solid contours have better overlap with the gray filled contour of $\cmbspa$+$\baodesi$. As a result, the $\summnu$ upper limit is not as artificially tightened as the $\kmpa$ (yellow contours) case, which has larger discrepancy with  $\cmbspa$+$\baodesi$ along $\omega_{\rm cb}$.
}
\label{fig:omegacb_mnu_2d}
\end{figure}

This behavior is illustrated in Figure~\ref{fig:omegacb_mnu_2d}. The gray shaded contour shows the $\cmbspa+\baodesi$ posterior, while the solid open contours show the posteriors obtained by combining $\cmbspa$ with the different CMB lensing measurements. The directions of the posterior contours from the two sets differ, allowing for degeneracy breaking when the datasets are combined, but also leading to tight upper limits when the individual datasets prefer slightly different regions of the parameter space.

$\cmbspa$+$\kgmv$ and $\cmbspa$+$\kgpa$ prefer a lower $\omega_{\rm cb} \equiv
\omega_{\rm c} + \omega_{\rm b}$
compared to $\cmbspa$ alone. Therefore, when both are combined with $\baodesi$, which sets $\Omega_{\rm m}$, the upper limits from the with-CMB lensing cases (teal and purple solid lines) shift up compared to the no-CMB lensing cases (gray filled contours). This is also a reflection of the $\Omega_{\rm m}$ from $\cmbspa$+$\kgmv$ and $\cmbspa$+$\kgpa$ being more consistent with $\baodesi$ within $\Lambda$CDM (as discussed in Section~\ref{sec:cmb_lens}).

A smaller factor that also shifts the $\summnu$ upper limits is the lower lensing amplitude preferred by our CMB lensing measurements relative to the prediction from $\cmbspa+\baodesi$. To isolate this amplitude difference from the direct effect of neutrino mass on the lensing spectrum, we introduce an amplitude parameter $A_{\rm recon}^{m_{\nu}=0}$ that rescales the lensing spectrum while keeping its shape fixed to that predicted by the best-fit $\Lambda$CDM model with $\summnu=0~{\rm eV}$ from $\cmbspa+\baodesi$.\footnote{Specifically, we fit the $\kappa+\cmbspa+\baodesi$ data combinations in the $\Lambda$CDM+$A_{\rm recon}$ model, fixing $\summnu=0~{\rm eV}$ and $A_{\rm 2pt}=1$ and imposing the standard priors. The remaining $\Lambda$CDM parameters remain close to those preferred by $\cmbspa+\baodesi$, apart from the small additional dependence on the shape of the lensing spectrum.} Thus, $A_{\rm recon}^{m_{\nu}=0}$ directly measures the amplitude of the observed lensing spectrum relative to this reference prediction.

For both $\kgmv$ and $\kgpa$, we find $A_{\rm recon}^{m_{\nu}=0}$ to be $2-4\%$ below unity, indicating that these measurements prefer less lensing than predicted by $\cmbspa+\baodesi$. Since increasing $\summnu$ suppresses the lensing power, this lower preferred amplitude corresponds to having larger neutrino masses. It therefore contributes to the higher upper limits obtained from $\kgmv/\kgpa+\cmbspa+\baodesi$ relative to $\cmbspa+\baodesi$ alone.

We further note that the 95\% C.L. upper limit on $\summnu$ is about 25\% higher for $\kgpa$+$\cmbspa$+$\baodesi$ than when $\kmpa$ is used for CMB lensing\footnote{In C26, the upper limit for the $\kmpa+\cmbspa+\baodesi$ data combination is $\summnu < 0.055\,{\rm eV}$. The difference comes from the choice of using a Gaussian $\tau$ prior in C26 vs. using SRoll2 to fit $\tau$ in this work.}:
\begin{equation}\nonumber
\summnu < 0.058\,{\rm eV}\quad (\kmpa+\cmbspa+\baodesi).
\end{equation}
This follows from the same argument discussed above. Relative to our baseline case of using $\kgmv$, the combination using $\kmpa$ selects a region of parameter space that is less consistent (higher $\omega_{\rm cb}$) with that preferred by $\cmbspa$+$\baodesi$, as reflected by the smaller overlap between the $\kmpa+\cmbspa$ and $\cmbspa+\baodesi$ contours in Figure~\ref{fig:omegacb_mnu_2d}. Since the constraint on $\summnu$ is quoted as a one-sided upper limit, this reduced overlap shifts the posterior toward lower values of $\summnu$, leading to a tighter bound. The preference of $\kmpa$ for $A_{\rm recon}^{m_{\nu}=0}>1$ when run in combination with $\cmbspa$ and $\baodesi$ in the same model as described in the previous paragraph acts in the same direction: at fixed primary-CMB constraints, a preference for larger $A_{\rm recon}^{m_{\nu}=0}$ acts to lower the preferred value of $\summnu$.

The difference in the upper limits between $\kmpa$ and $\kgmv/\kgpa$, when each is combined with $\cmbspa+\baodesi$, reflects both the level of tension between the CMB data (primary + lensing) and $\baodesi$, and the statistically unsurprising differences in the lensing amplitudes preferred by the various lensing measurements, with the former providing the dominant contribution.\footnote{As noted in Section~\ref{sec:te_lens_bao}, the distance between $\kgpa+\cmbspa$ and $\baodesi$, quantified in the $\Omega_{\rm m}$-$h r_{\rm d}$ plane within $\Lambda$CDM, is $1.9\sigma$, whereas the corresponding distance for $\kmpa+\cmbspa$ relative to  $\baodesi$ was reported to be $\sim 2.8\sigma$ in C26.}

Lastly, we consider the effect of baryonic feedback on the interpretation of the neutrino-mass constraint. Like massive neutrinos, baryonic feedback suppresses the matter power spectrum on nonlinear scales and could therefore be partially degenerate with the effects of $\summnu$ on the lensing signal. To assess this effect, we replace our fiducial nonlinear prescription, {\tt mead2020}, with {\tt mead2020\_feedback} and allow the parameter {\tt HMCode\_logT\_AGN} to vary between 7.0 and 8.5~\cite{mccarthy2021}. For $\kgpa+\cmbspa+\baodesi$, the upper limit changes only from $0.072$ to $0.071\,{\rm eV}$, indicating that baryonic feedback has a negligible effect on the inferred neutrino-mass constraint for the current data combination.

We conclude this section by noting that, although our upper limit on $\sum m_\nu$ is higher than that reported by C26, this does not necessarily reflect weaker constraining power. The upper limit is sensitive to the physical boundary at $\sum m_\nu=0$, and the C26 posterior is pushed more strongly against this boundary, resulting in a tighter one-sided upper limit. As discussed above, this tighter bound is driven in part by the differing values of the physical and fractional matter densities preferred in $\Lambda$CDM by $\cmbspa$+CMB lensing and by $\cmbspa+\baodesi$. Consistent with this interpretation, replacing $\baodesi$ with $\baopredesi$ or with Type Ia supernova data yields higher upper limits on $\summnu$~\cite{qu2026}. More generally, various analysis choices and model extensions that could potentially reduce the mismatch between $\cmbspa+$CMB lensing and $\baodesi$ have been explored, including modifications to $\tau$~\cite{jhaveri2025, sailer_disputauble}, allowing nonzero spatial curvature~\cite{chen2025}, and extensions involving evolving dark energy~\cite{abdulkarim2025a}.\footnote{For the $\kgpa+\cmbspa+\baodesi$ data combination, allowing $\summnu$ to vary in $w_0w_a$CDM yields a 95\% C.L. upper limit of $\summnu<0.214\,{\rm eV}$.} An independent probe with constraining power on $\Omega_{\rm m}$ comparable to that of $\baodesi$ would therefore be particularly valuable for clarifying the origin of the current level of difference between CMB data and $\baodesi$.

\begin{table*}
\caption{Constraints on sampled and derived $\Lambda$CDM parameters for $\kgmv$ and $\kgpa$ combined with $\cmbspa$, with and without $\baodesi$. Quoted values are posterior means with marginalized $68\%$ confidence intervals.}
\begin{tabular}{lccccc}
\toprule
 & $\kappa_{\rm GMV}$ & $\kappa_{\rm GMV}$ & $\kappa_{\rm GPA}$ & $\kappa_{\rm GPA}$ & \multirow{2}{*}{${\rm CMB}_{\rm SPA}$} \\
Parameter & $+\,{\rm CMB}_{\rm SPA}$ & $+\,{\rm CMB}_{\rm SPA}+{\rm BAO}_{\rm DESI}$ & $+\,{\rm CMB}_{\rm SPA}$ & $+\,{\rm CMB}_{\rm SPA}+{\rm BAO}_{\rm DESI}$ &  \\
\midrule
\multicolumn{6}{l}{\emph{Sampled}} \\
$10^{4}\theta_{\rm MC}$ & $104.075 \pm 0.024$ & $104.092 \pm 0.022$ & $104.075 \pm 0.024$ & $104.092 \pm 0.022$ & $104.071 \pm 0.024$ \\
$100\,\Omega_{\rm b} h^{2}$ & $2.2445 \pm 0.0095$ & $2.2504 \pm 0.0092$ & $2.2447 \pm 0.0095$ & $2.2506 \pm 0.0092$ & $2.2422 \pm 0.0096$ \\
$100\,\Omega_{c} h^{2}$ & $11.93 \pm 0.10$ & $11.748 \pm 0.062$ & $11.930 \pm 0.098$ & $11.755 \pm 0.061$ & $12.00 \pm 0.11$ \\
$n_{\rm s}$ & $0.9702 \pm 0.0035$ & $0.9744 \pm 0.0029$ & $0.9701 \pm 0.0034$ & $0.9742 \pm 0.0030$ & $0.9694 \pm 0.0035$ \\
$\ln(10^{10} A_{\rm s})$ & $3.050 \pm 0.011$ & $3.057 \pm 0.011$ & $3.052 \pm 0.010$ & $3.061 \pm 0.011$ & $3.059 \pm 0.012$ \\
$\tau_{\rm reio}$ & $0.0587 \pm 0.0058$ & $0.0632 \pm 0.0060$ & $0.0592 \pm 0.0058$ & $0.0643 \pm 0.0060$ & $0.0601 \pm 0.0060$ \\
\midrule
\multicolumn{6}{l}{\emph{Derived}} \\
$H_{0}\,[\mathrm{km/s/Mpc}]$ & $67.60 \pm 0.42$ & $68.34 \pm 0.26$ & $67.59 \pm 0.40$ & $68.32 \pm 0.26$ & $67.32 \pm 0.45$ \\
Age [Gyr] & $13.793 \pm 0.015$ & $13.774 \pm 0.012$ & $13.793 \pm 0.015$ & $13.775 \pm 0.012$ & $13.799 \pm 0.015$ \\
$10^{9} A_{\rm s} e^{-2\tau_{\rm reio}}$ & $1.8781 \pm 0.0079$ & $1.8743 \pm 0.0078$ & $1.8789 \pm 0.0071$ & $1.8765 \pm 0.0072$ & $1.889 \pm 0.011$ \\
$\Omega_{\Lambda}$ & $0.6883 \pm 0.0059$ & $0.6988 \pm 0.0034$ & $0.6882 \pm 0.0057$ & $0.6985 \pm 0.0034$ & $0.6842 \pm 0.0065$ \\
$\Omega_{\rm m}$ & $0.3117 \pm 0.0059$ & $0.3011 \pm 0.0034$ & $0.3117 \pm 0.0057$ & $0.3014 \pm 0.0034$ & $0.3157 \pm 0.0065$ \\
$r_{d}\,[\mathrm{Mpc}]$ & $147.21 \pm 0.26$ & $147.62 \pm 0.18$ & $147.20 \pm 0.25$ & $147.60 \pm 0.18$ & $147.05 \pm 0.27$ \\
$\sigma_{8}$ & $0.8124 \pm 0.0044$ & $0.8102 \pm 0.0045$ & $0.8130 \pm 0.0039$ & $0.8117 \pm 0.0042$ & $0.8179 \pm 0.0056$ \\
\bottomrule
\end{tabular}
\label{tab:all_parameters}
\end{table*}

\subsection{Spatial curvature $\omk$}\label{sec:omk}
Although current data are broadly consistent with a spatially flat Universe, allowing $\omk$ to vary provides an important test of both the standard cosmological model and the inflationary expectation of near-flatness. A nonzero $\omk$ modifies the distance-redshift relation and the angular-diameter distance to last scattering, and therefore can be partially degenerate with other cosmological parameters in analyses based primarily on background geometry.

In this context, CMB lensing provides complementary information to primary CMB and distance-based probes. In addition to its geometric sensitivity through the lensing kernel, CMB lensing directly probes the projected matter distribution and the growth of structure over a broad range of redshifts. This additional sensitivity helps break the well-known geometric degeneracy that limits curvature constraints from the primary CMB alone~\cite{stomper_efstathious1999}. Put differently, models with different values of $\omk$ can often produce similar primary CMB anisotropy spectra by compensating shifts in other parameters, but they generically predict different late-time structure growth and projection effects, to which CMB lensing is sensitive.

We report our constraint on $\omk$ when combining $\kgpa$ and $\cmbspa$: 
\begin{align}
100\omk&=-0.48 \pm 0.48 \hspace{0.5cm} ( \kgpa+\cmbspa),\nonumber
\end{align}
which is consistent with spatial flatness at 1.0$\sigma$. This result pulls $\omk$ closer to spatial flatness than the most recent result from C26, which reported $100 \omk = -0.85 \pm 0.48$. This can be understood from the posterior degeneracies in this data combination. The lower $A_{\rm recon}$ from the CMB lensing measurement in this work compared to that used in C26 shifts the posterior toward lower $\Omega_{\rm m}$. Since $\Omega_{\rm m}$ and $\omk$ are negatively correlated, this in turn shifts $\omk$ toward less negative values.

When we combine $\kgpa$+$\cmbspa$ with $\baodesi$, we obtain 
\begin{align}
100\omk&=0.21 \pm 0.11 \hspace{0.5cm} ( \kgpa+\cmbspa+\baodesi ),\nonumber
\end{align}
at 1.9$\sigma$ from spatial flatness. Given the difference between $\kgpa$+$\cmbspa$ and $\baodesi$ within $\Lambda$CDM (with $\baodesi$ preferring lower $\Omega_{\rm m}$) and the anti-correlation between $\Omega_{\rm m}$ and $\omk$, the positive shift is expected. We note that this is a milder shift than was seen in C26.

\subsection{Evolving dark energy $w_{0}/w_{a}$}\label{sec:w0wa}
The latest results from DESI have raised interest in models with a time-varying dark-energy equation of state. In the following, we allow the equation of state of dark energy to evolve according to
\begin{align}
w(z) = w_0 + w_a \frac{z}{1+z}.
\end{align}
Relative to $\Lambda$CDM, the DESI~DR2 distance measurements, when combined with CMB and Type Ia supernova (SNIa) observations, favor regions of the resulting $(w_0,w_a)$ parameter space that correspond to an evolving dark-energy equation of state, although the statistical significance remains dependent on the choice of datasets being combined.

\begin{figure}
\includegraphics[width=0.98\linewidth]{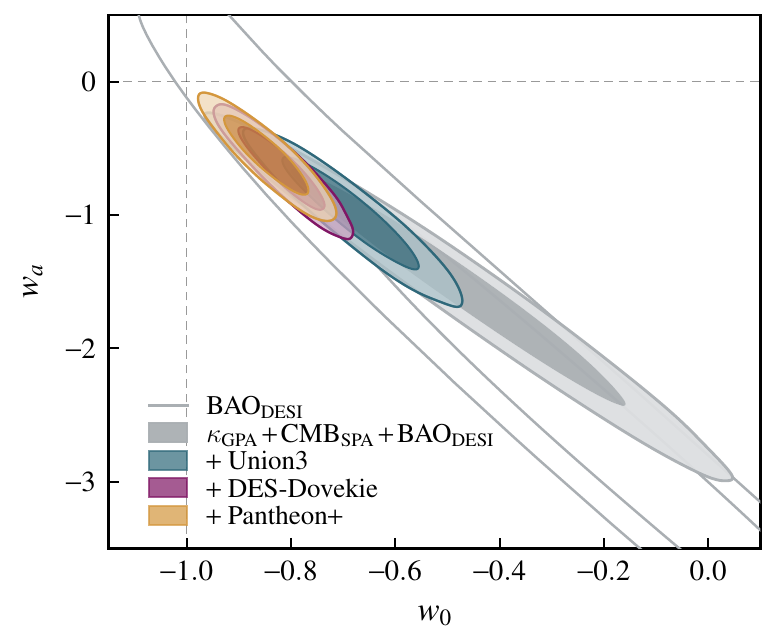} 
\caption{Constraints in the $(w_0,w_a)$ plane from $\baodesi$ alone, $\kgpa+\cmbspa+\baodesi$, and the latter combined with Union3, DES-Dovekie, or Pantheon+ SNIa data.}
\label{fig:w0wa}
\end{figure}

CMB lensing provides only weak constraints on $(w_{0},w_{a})$, since its kernel peaks at higher redshift than the epoch where dark energy has its strongest influence. However, when combined with primary CMB measurements, CMB lensing can narrow the overall cosmological parameter space and thereby indirectly improve constraints on $(w_{0},w_{a})$. When combined with $\baodesi$ we obtain:
\begin{equation}\nonumber
\left.
\begin{array}{l@{\quad}l}
w_0 &= -0.47 \pm 0.22\hspace{0.1cm} \nonumber\\
w_a &= -1.55 \pm 0.61 \hspace{0.1cm}\nonumber
\end{array}
\right\}
\left(
\begin{array}{@{}c@{}}
\kgpa + 
\cmbspa+
\baodesi
\end{array}
\right).
\end{equation}
This is 2.1$\sigma$ away from the $\Lambda$CDM expectation of $(w_0,w_a) = (-1,0)$. As shown in Figure~\ref{fig:w0wa}, the addition of primary CMB and CMB lensing to $\baodesi$ also shifts the posterior closer to the $\Lambda$CDM expectation.

Similar to previous work, we find that including DES-Dovekie~\cite{popovic2026}, Pantheon+~\cite{brout2022}, and Union3~\cite{rubin2025} SNIa data increases the preference for $(w_0,w_a)$ to deviate from $(-1,0)$. The posteriors for these three cases are shown in Figure~\ref{fig:w0wa}. The distances of the $(w_0,w_a)$ constraints from the $\Lambda$CDM expectation are $2.9\sigma$, $2.5\sigma$, and $3.3\sigma$, respectively. Two of the three SNe samples, when combined with $\baodesi$ and CMB, now show less than a $3\sigma$ preference for an evolving dark-energy equation of state.

We further examine these constraints when allowing the sum of the neutrino masses, $\sum m_{\nu}$, to vary. In the resulting $w_{0}w_{a}+\sum m_{\nu}$ model, the additional freedom in the late-time expansion and growth histories broadens the allowed region of parameter space relative to the case with fixed neutrino mass. Although the constraints are driven primarily by the combination of primary CMB and BAO data, CMB lensing can further tighten them by providing additional sensitivity to the amplitude of late-time structure growth and helping to partially break degeneracies between $\sum m_{\nu}$ and the dark-energy parameters.

\begin{figure}
\includegraphics[width=1.00\linewidth]{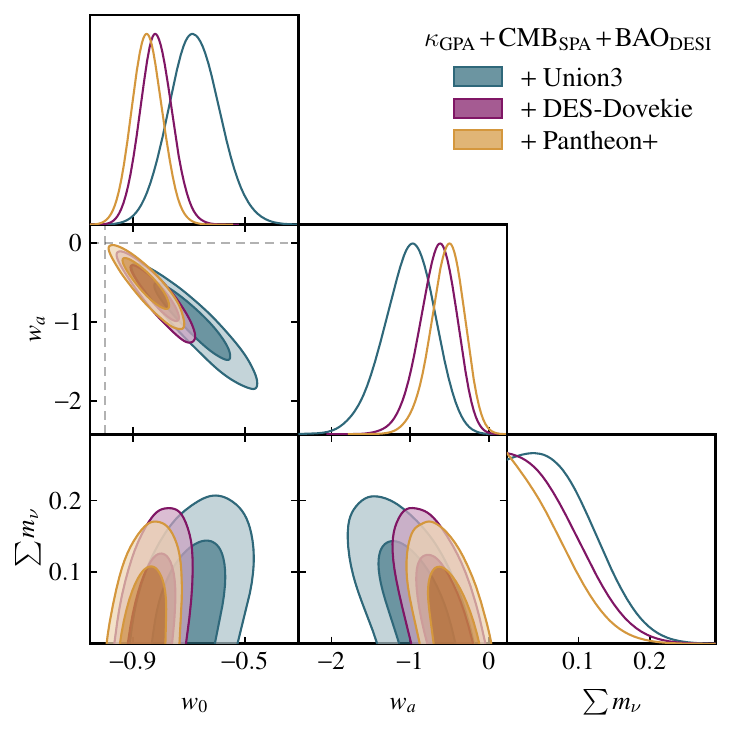} 
\caption{Same as Figure~\ref{fig:w0wa}, but with $\summnu$ additionally allowed to vary.}
\label{fig:w0wamnu}
\end{figure}

Figure~\ref{fig:w0wamnu} shows constraints from the combination of  $\kgpa$, $\cmbspa$, and $\baodesi$ with the DES-Dovekie, Pantheon+, and Union3 SNIa data. The inferred location and extent of the $(w_0,w_a)$ posterior remain dependent on the adopted SNIa data used. Allowing $\sum m_{\nu}$ to vary only subtly weakens the constraints, while the relative shifts in the preferred $(w_0,w_a)$ regions among the SNIa compilations remain largely unchanged.

\subsection{Comparison with optical lensing and other multi-probe results}
\label{sec:LSScomparison}
In this section, we compare our constraints on the matter density $\Omega_{\rm m}$ and the amplitude of matter fluctuations $S_{8}$ with those obtained by the Dark Energy Survey (DES; \cite{flaugher2015,des2016,des2018}), Kilo-Degree Survey (KiDS; \cite{kuijken2015}), the Hyper Suprime-Cam Strategic Program (HSC-SSP; \cite{hsc}), and the combined DECADE+DES\,Y3 dataset, referred to as DECam 13k \cite{anbajagane2025}.

Although the underlying physics that sources weak lensing is identical for galaxies and the CMB, the two cosmological probes differ fundamentally in the nature and redshift distribution of their background sources. For galaxy weak lensing, the observed distortions are measured from the shapes of galaxies that span a broad redshift range, typically extending up to $z \sim 1.5$, rather than from a single, well-defined source plane as in CMB lensing. Because the lensing kernel is most sensitive to structure located roughly halfway between the observer and the sources, galaxy weak lensing is most sensitive to large-scale structure at $z\sim0.5$. 
In addition, at a fixed angular scale, galaxy lensing probes smaller physical scales (higher $k$ modes) than CMB lensing. As a result, galaxy weak lensing provides a complementary view of structure formation, focusing on the late-time Universe and the mildly to strongly nonlinear regime, while CMB lensing probes structure over a broader range of redshifts and is relatively more sensitive to intermediate redshifts and less nonlinear scales.

For this comparison, we recompute the optical-survey constraints using the baseline analysis setup adopted in the DES\,Y3 + KiDS-1000 cosmic-shear joint analysis~\cite{deskids2023}: we use the nonlinear alignment (NLA) intrinsic alignment model, assume three massless neutrino species, adopt the \texttt{mead2020} nonlinear matter power spectrum model while varying the $\log(T_{\rm AGN})$ parameter and sample directly in $H_{0}$ instead of $\theta_{\rm MC}$. Consequently, our IA modeling differs from the baseline choices adopted in these analyses: HSC-Y3~\cite{li2023} and DECam\,13k~\cite{anbajagane2025} use the TATT model, while KiDS-Legacy~\cite{angus2025} adopts a mass-dependent IA model. We also omit the shear-ratio likelihood used in the DES\,Y3 analyses~\cite{amon2022,secco2022}, following the choice in~\cite{deskids2023}. All inferences are performed with the \texttt{Nautilus} sampler in \texttt{CosmoSIS}.

\begin{figure*}[t]
    \centering
    \includegraphics[width=\linewidth]{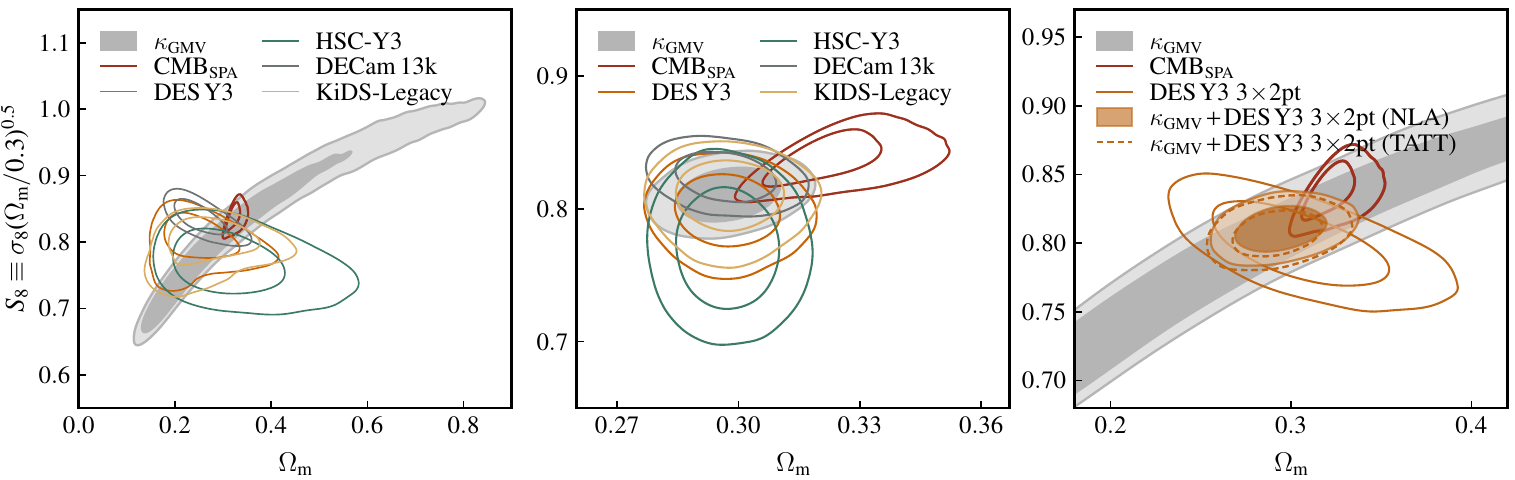}
    \caption{{\bf Left:} Constraints on $\Omega_{\rm m}$ and $S_8$ using lensing alone from DES\,Y3 (orange), HSC-Y3 (green), DECam\,13k (gray), KiDS-Legacy (gold), and our GMV results (filled gray), compared with constraints from $\cmbspa$ (wine red). {\bf Center:} Same as the left panel, but combining the cosmic shear measurements with DESI DR2. {\bf Right:} Constraints from DES\,Y3 3$\times$2pt and the joint combination of DES\,Y3 3$\times$2pt + $\kgmv{}$, compared with $\cmbspa$. These constraints are recomputed using the common analysis assumptions described in Section~\ref{sec:LSScomparison} and therefore may differ from the corresponding published baseline results. }
    \label{fig:lss_gmv_compare}
\end{figure*}

As discussed previously, the CMB lensing auto-spectrum most tightly constrains $\som$, while galaxy weak lensing probes a steeper degeneracy direction in $\sigma_8\Omega_{\rm m}^{\alpha}$ because of its sensitivity over a narrower redshift range and greater sensitivity to nonlinear scales. The amplitude of structure fluctuations in galaxy weak-lensing analyses is typically parameterized as
$S_{8}\equiv \sigma_{8}(\Omega_{\rm m}/0.3)^{0.5}$.
In this parameterization, galaxy weak lensing measurements have often favored lower values of $S_{8}$ than those inferred from primary CMB observations. More recent cosmic shear analyses, however, have generally found values of $S_{8}$ that are in closer agreement with primary CMB constraints. Cosmic shear constraints obtained by reanalyzing the published likelihoods within our common analysis setup are summarized in Table~\ref{table:s8}, with the corresponding posteriors shown in Figure~\ref{fig:lss_gmv_compare}. Using the same model assumptions and analysis settings, our CMB lensing measurement gives
\begin{equation}
S_{8}=0.822^{+0.068}_{-0.126}\hspace{0.3cm} (\kappa_{\rm GMV}). \nonumber
\end{equation}

We quantify the level of agreement between these measurements and the $\cmbspa$ result in the one-dimensional $S_{8}$ parameter space.\footnote{For this comparison, we use primary $\cmbspa$ constraints obtained with the settings matched to those of the large-scale structure analysis in Table~\ref{tab:theory_settings}, rather than the baseline settings used for the CMB lensing + primary CMB analysis.} The level of consistency varies across these measurements, ranging from very close agreement to $2.1\sigma$ for the cosmic-shear-only constraints. This variation may in part reflect differences in the treatment of several important systematics, including baryonic effects, nonlinear structure formation, photometric redshifts, and intrinsic alignment.

While cosmic shear directly constrains $S_8$, it only weakly constrains $\Omega_{\rm m}$. We therefore also consider the combination with $\baodesi{}$, which pins down $\Omega_{\rm m}$ and allows the consistency between the measurements to be assessed with this degeneracy substantially reduced. Since $\baodesi{}$ favors somewhat lower values of $\Omega_{\rm m}$ than the primary CMB, the resulting consistency reflects both the cosmic shear constraints and the preferred $\Omega_{\rm m}$ region of the $\baodesi$ data. The resulting constraints are summarized in Table~\ref{table:s8}. We find that the overall level of agreement remains similar, with consistency ranging from 0.8 to 2.1$\sigma$.

Finally, we consider the DES\,Y3 3$\times$2pt analysis, which combines measurements of galaxy clustering, cosmic shear, and galaxy-galaxy lensing~\citep{desy3_3x2}. Using our unified analysis settings, we obtain
\begin{align}
S_{8}=0.798^{+0.020}_{-0.020}
\hspace{0.5cm}
({\rm DES\,Y3}\ \textnormal{3$\times$2pt}).
\nonumber
\end{align}
Our primary interest is in its combination with our CMB lensing measurement, which provides complementary information in the $\Omega_{\rm m}$--$S_{8}$ plane and further tightens the constraints. Before forming the joint likelihood, we assess the covariance between the two datasets. Using an analytical Gaussian covariance estimate, we verify that the correlations between the 3$\times$2pt correlation function bins and the CMB lensing auto-spectrum bandpowers are mostly below the percent level, with the largest values at the few-percent level. Since these correlations are small, we neglect the corresponding off-diagonal covariance blocks and combine the two likelihoods. With this setup, we obtain
\begin{align}
S_{8}&=0.811^{+0.011}_{-0.011} \hspace{0.5cm} ({\rm DES\,Y3\ \textnormal{3$\times$2pt}+\kgmv}).\nonumber
\end{align}
This combination of DES\,Y3 3$\times$2pt and CMB lensing yields a $1.4\%$ constraint on $S_8$, with an uncertainty smaller than that obtained from $\cmbspa$ ($S_{8}=0.838^{+0.013}_{-0.013}$). This joint constraint is consistent with $\cmbspa$ at the $1.6\sigma$ level in $S_{8}$, indicating agreement between the late-time structure constraints from DES\,Y3 and CMB lensing and those inferred from primary CMB observations. As a robustness test, we also repeat the analysis using the TATT intrinsic alignment model and find no significant change in the resulting constraints, as shown in Figure~\ref{fig:lss_gmv_compare}. This result can be compared with the previous constraint from combining DES\,Y3 3$\times$2pt with the CMB lensing map from \cite{omori2023}, corresponding to the so-called 6$\times$2pt analysis,
\begin{equation}
S_{8}=0.792^{+0.012}_{-0.012} \hspace{0.5cm} ({\rm DES}\,{\rm Y3}\ \textnormal{3$\times$2pt}+\kappa_{\rm SPTSZ}+\kappa_{\rm Planck}),\nonumber
\end{equation}
which additionally included the cross-correlations of CMB lensing with galaxy density and galaxy weak lensing. The larger number of probe combinations helps break degeneracies between cosmological and systematic parameters, leading to tighter control of the nuisance parameters. Nevertheless, we find an $8.3\%$ reduction in the marginalized uncertainty relative to the 6$\times$2pt analysis, despite using fewer probe combinations. The size of this improvement underscores the gain in constraining power delivered by the new CMB lensing map.

\begin{table*}
\caption{ Table summarizing $S_{8}$ constraints from the CMB-lensing measurement presented in this work, its combination with BAO, cosmic shear measurements from optical weak lensing surveys and their combinations with BAO, and DES\,Y3 $3\!\times\!2{\rm pt}$ constraints with and without the CMB lensing map presented in this work. The distance from $\cmbspa$ is measured using the one-dimensional marginalized $S_8$ posteriors. }
\centering
\begin{tabular}{l l c c}
\toprule
Probe combination & Dataset & $S_{8}$ & $\Delta_{\cmbspa}$ \\
\midrule
Primary CMB & $\cmbspa$ & $0.838^{+0.013}_{-0.013}$ & -- \\

\midrule

CMB lensing & $\kgmv$ & $0.822^{+0.068}_{-0.126}$ & $0.7\sigma$ \\
CMB lensing + BAO & $\kgmv + \baodesi$ & $0.811^{+0.013}_{-0.013}$ & $1.4\sigma$ \\

\midrule

\multirow{4}{*}{Cosmic shear only}
  & DES\,Y3 & $0.801^{+0.027}_{-0.019}$ & $1.2\sigma$ \\
  & KiDS-Legacy & $0.797^{+0.030}_{-0.016}$ & $1.4\sigma$ \\
  & HSC-Y3 & $0.767^{+0.032}_{-0.032}$ & $2.1\sigma$ \\
  & DECam\,13k & $0.836^{+0.018}_{-0.017}$ & $<0.1\sigma$ \\

\midrule

\multirow{4}{*}{Cosmic shear + BAO}
  & DES\,Y3 + $\baodesi$ & $0.797^{+0.020}_{-0.017}$ & $1.8\sigma$ \\
  & KiDS-Legacy + $\baodesi$ & $0.808^{+0.020}_{-0.014}$ & $1.2\sigma$ \\
  & HSC-Y3 + $\baodesi$ & $0.770^{+0.030}_{-0.030}$ & $2.1\sigma$ \\
  & DECam\,13k + $\baodesi$ & $0.824^{+0.013}_{-0.012}$ & $0.8\sigma$ \\

\midrule

3$\times$2pt & DES\,Y3 & $0.798^{+0.020}_{-0.020}$ & $1.7\sigma$ \\

\midrule

3$\times$2pt + CMB lensing & DES\,Y3 + $\kgmv$ & $0.811^{+0.011}_{-0.011}$ & $1.6\sigma$ \\
\bottomrule
\end{tabular}
\label{table:s8}
\end{table*}

\subsection{Constraints on the growth of structure}
 
In the previous sections, we have demonstrated that current CMB lensing measurements are broadly consistent with the primary CMB prediction when summarized by an integrated amplitude parameter such as $\sigma_8\Omega_{\rm m}^{0.25}$. However, this does not exclude the possibility of scale-dependent departures from the standard $\Lambda$CDM matter power spectrum. This motivates us to go beyond a single overall amplitude parameter, since deviations on particular scales may arise from physics or systematics that do not act as a simple rescaling of the matter power spectrum. CMB lensing provides a complementary probe to cosmic shear because its sensitivity to structure peaks at higher redshift. It therefore allows us to test whether the scale-dependent suppression suggested by late-time weak lensing measurements~\cite{amon2022b,chen2023,preston2023} is already present at intermediate redshifts. Such a suppression would affect both the overall amplitude and the shape of the CMB lensing power spectrum, motivating a test that goes beyond a single amplitude parameter.
 
To investigate whether this suppression varies with scale without imposing a specific functional form, we adopt the binned reconstruction approach introduced by \cite{doux2026}, which infers a scale-dependent but redshift-independent modification to the matter power spectrum in bins of $k$ relative to a fixed reference cosmology and nonlinear model.

The lensing (both CMB and galaxy) signal can be written as 
\begin{equation}\label{eq:limber_cmbkappa}
C_{\ell}^{ab}=\int d\chi\frac{W_{a}(\chi)W_{b}(\chi)}{\chi^{2}}P\left(k=\frac{\ell+1/2}{\chi},z(\chi) \right),
\end{equation}
where $\chi$ is the radial comoving distance, $W_{a/b}$ are the kernel window functions:
\begin{equation}
W_{\rm CMB}(\chi)=\frac{3\Omega_{\rm m}H_{0}^{2} }{2c^2}\frac{\chi}{a(\chi)}\frac{\chi_{*}-\chi}{\chi_{*}}
\end{equation}
for CMB lensing, where $\chi_*$ denotes $\chi$ at the surface of last scattering, $a(\chi)$ is the scale factor, and 
\begin{equation}
W_{\kappa_g}^i(\chi)
=
\frac{3\,\Omega_{\rm m} H_0^2}{2c^2}\,
\frac{\chi}{a(\chi)}
\int_{\chi}^{\chi_{\rm H}} d\chi_s \, n_i(\chi_s)\,
\frac{\chi_s-\chi}{\chi_s} 
\end{equation}
for galaxy lensing, where $n_i(\chi_{s})$ is the (normalized) source-galaxy distribution in bin $i$, $\int_0^{\chi_{\rm H}} d\chi\, n_i(\chi)=1$, and $\chi_{\rm H}$ is the horizon distance.

Next, we allow for a scale-dependent modification to the matter power spectrum,
\begin{equation}
P(k,z)=[1+\alpha(k)]P_{\rm fid}(k,z),
\end{equation}
where $P_{\rm fid}(k,z)$ is the nonlinear matter power spectrum computed at the $\cmbspa$ best-fit $\Lambda$CDM cosmology using the \texttt{HMcode2020} halo-model prescription~\cite{mead2021}, and $\alpha(k)$ describes the fractional deviation from the fiducial spectrum.

We perform a change of variable $k=(\ell+0.5)/\chi$ and transform to $\ln k$, such that the original line-of-sight integral over $\chi$ can equivalently be written as an integral over $k$. Discretizing this integral into bins of $k$, we define
\begin{equation}
\mathbf{W}_{\ell,k_{i}}^{ab}
=
\Delta_{\ln\,k}
\frac{W_{a}(\chi_{\ell}^{i}) W_{b}(\chi_{\ell}^{i})}{\chi_{\ell}^{i}}
P_{\rm fid}(k_{i},z(\chi_{\ell}^{i})),
\end{equation}
where $\chi_{\ell}^{i}=(\ell+0.5)/k_{i}$.
Thus, $\mathbf{W}_{\ell,k_i}^{ab}$ is a two-dimensional projection matrix in angular scale $\ell$ and physical scale $k_i$. For each $(\ell,k_i)$ pair, the Limber relation fixes the corresponding comoving distance $\chi_{\ell}^{i}$, and hence the redshift at which the matter power spectrum is evaluated, while the lensing kernels determine the weight assigned to matter fluctuations at that distance.

The model lensing spectrum can then be written as
\begin{equation}
C_{\ell}^{ab}=\sum_{i}{\bf W}_{\ell,k_{i}}^{ab}\left(1+\alpha_{i}\right),
\label{eq:alpha_projection}
\end{equation}
where $\alpha_i\equiv\alpha(k_i)$, and the summation over $k_i$ is equivalent to the original integration along the line of sight. For a fixed reference cosmology and fixed lensing kernels, $\mathbf{W}_{\ell,k_i}^{ab}$ is therefore a precomputed  matrix that maps modifications in each $k$-bin into the observable lensing spectrum. During inference, a new set of $\alpha_{i}$ values is proposed, systematic effects are applied to the corresponding model spectrum, and the resulting model is compared with the measured bandpowers using the full data covariance. We implement this framework in \texttt{CosmoSIS} and sample the posterior using the \texttt{Nautilus} sampler.

\begin{figure*}
\includegraphics[width=1.00\linewidth]{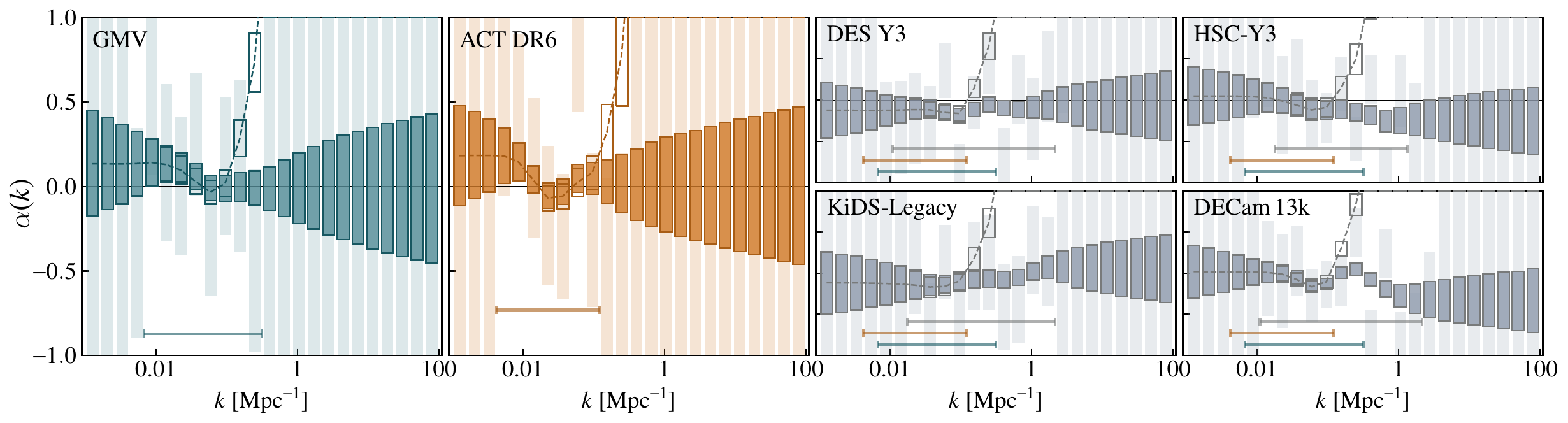} 
\caption{The light filled boxes show the raw, unsmoothed constraints on $\alpha(k) \equiv P(k)/P_{\rm nl}^{\rm fid}(k) - 1$, while the dark filled boxes show the corresponding smoothed constraints with $\sigma_{\rm s}=0.3$. The open boxes traced by dashed lines use the linear-theory reference, $\alpha(k) \equiv P(k)/P_{\rm lin}^{\rm fid}(k) - 1$. All constraints are measured from systematics-marginalized bandpowers with cosmology fixed to $\cmbspa$. The horizontal bars at the bottom indicate the range of modes over which each experiment has constraining power.
}
\label{fig:alpha_K}
\end{figure*}

Figure~\ref{fig:alpha_K} shows the constraints on $\alpha(k)$ using $n_{k}=24$ logarithmically spaced $k$-bins over the range $10^{-3}\,{\rm Mpc}^{-1} \leq k \leq 10^{2}\,{\rm Mpc}^{-1}$. We show results both with and without the smoothing prior $\sigma_{\rm s}=0.3$, defined as a Gaussian prior on the difference between adjacent bins, which is the default setting adopted in \cite{doux2026}. Since the smoothing prior introduces correlations between neighboring bins, we use the {\it unsmoothed} ($\sigma_{\rm s}=\infty$) posterior to determine which $k$-bins are independently informed by the data. The uniform $[-1,+1]$ prior on each bin has standard deviation $\sigma_{\rm prior}=1/\sqrt{3}\approx0.58$, and we retain bins whose posterior standard deviation is reduced below $0.53$ in the unsmoothed chain.\footnote{Each bin is assigned a weight $1-(\sigma_i/\sigma_{\rm prior})^2$, which quantifies how much its posterior variance is reduced relative to the prior. We rank the bins by this weight and retain the smallest set whose cumulative weight accounts for $90\%$ of the total across all bins. This requirement is satisfied for bins with $\sigma_i\lesssim0.53$, and we therefore adopt $\sigma_i<0.53$ as the threshold for defining constrained bins.} The $k$-ranges identified this way for $\kgmv$ and $\kact$ are shown as teal and orange bars at the bottom of each shear panel, with the gray bar indicating the corresponding $k$-range for the galaxy weak-lensing measurement shown in that panel.

Following \cite{doux2026}, we quantify consistency with $\alpha(k)=0$ using the {\it smoothed} ($\sigma_{\rm s}=0.3$) posterior, considering only the constrained bins. The $\kgmv$ measurement deviates from the baseline prediction by only $0.2\sigma$, while the deviation for ACT DR6 CMB lensing is $0.1\sigma$. For the four cosmic shear measurements from DES\,Y3, HSC-Y3, KiDS-Legacy, and DECam\,13k, the deviations are $0.3\sigma$, $0.3\sigma$, $0.1\sigma$, and $1.3\sigma$, respectively. These significances account for the substantial covariance between the reconstructed $\alpha_i$ bins and therefore cannot be inferred directly from the individual error bars shown in Figure~\ref{fig:alpha_K}. Thus, all probes are consistent with the predictions of the $\cmbspa$ best-fit cosmology, although DECam\,13k shows a marginally larger deviation, driven primarily by its statistical power.

A complementary question is whether the reconstructed $\alpha(k)$ constraints show a preference for nonlinear evolution in the matter power spectrum relative to a linear theory prediction. This test gives a qualitatively different result. For $\kgmv$, the linear theory is disfavored at $3.2\sigma$, indicating that the SPT-3G CMB lensing measurement is sensitive to nonlinear growth over the reconstructed range of $k$ modes. In contrast, ACT DR6 shows a weaker deviation of $0.2\sigma$ in the same test. The galaxy weak-lensing datasets show much stronger evidence for nonlinear structure, as expected from their greater sensitivity to lower redshifts and smaller physical scales, with all surveys exhibiting high-significance departures from the linear-theory prediction.

We next compare the reconstructed $\alpha(k)$ constraints over scales on which both CMB lensing and cosmic shear are sensitive, allowing a more direct comparison of the inferred matter clustering on matched physical scales. For each pair of datasets, we therefore restrict the comparison to bins that are constrained by both probes, and evaluate the consistency of the corresponding $\alpha(k)$ measurements over those bins. For the comparison between $\kgmv$ and individual galaxy weak lensing datasets, we find no significant scale-dependent mismatch over the jointly constrained range of $k$ modes. The differences are $1.3\sigma$ for DES\,Y3, $<0.1\sigma$ for HSC-Y3, $0.5\sigma$ for KiDS-Legacy, and $0.7\sigma$ for DECam\,13k. Comparisons between $\kact$ and the galaxy weak-lensing datasets also show no significant scale-dependent disagreement, with differences of $0.5\sigma$ for DES\,Y3, $0.8\sigma$ for HSC-Y3, $0.9\sigma$ for KiDS-Legacy, and $1.5\sigma$ for DECam\,13k. The direct comparison between $\kgmv$ and $\kact$ is similarly consistent, with a $1.3\sigma$ difference.

Overall, we find that the $\kgmv$ reconstruction is highly consistent with the common $\mathrm{CMB}_{\rm SPA}$ reference prediction, while several cosmic shear datasets show larger, though generally not statistically significant, deviations. Direct comparisons between CMB lensing and cosmic shear also show no significant disagreement over their jointly constrained $k$ range. However, because the two probes weight substantially different redshift ranges, we regard these direct comparisons primarily as consistency checks rather than as a means of localizing any difference in the underlying matter power spectrum. A more direct comparison of the scale and redshift dependence would require an extension to $\alpha(k,z)$.

\section{Summary and outlook}\label{sec:summary}
In this work, we present a new lensing map based on data from the $1500\,\sqdeg$ SPT-3G Main field collected between 2019 and 2020. We reconstruct the lensing map using the quadratic-estimator formalism, jointly combining the temperature and polarization fields while fully accounting for their correlations. The resulting lensing map is dominated by the polarization channel on large angular scales and is therefore less susceptible to foreground contamination on those scales, yielding cosmological constraints that are robust to residual biases. We validate the lensing auto-spectrum using simulations and a suite of consistency tests, finding no detectable bias at the precision of our validation after accounting for foreground and instrumental systematics.

We first measure the amplitude of the reconstructed lensing power spectrum relative to the $\Lambda$CDM lensing spectrum computed at the $\cmbspa$ best-fit cosmology. We find $A_{\rm recon}^{\theta_{\rm fix}}=1.015\pm0.021$, fully consistent with unity. This provides a direct consistency test between the intermediate-redshift matter distribution probed by CMB lensing, whose kernel peaks at around $z\sim2$, and the $\Lambda$CDM model inferred from primary CMB measurements at $z\sim1100$.

Our SPT-3G CMB lensing measurement alone constrains the structure-growth parameter to $\sigma_8\Omega_{\rm m}^{0.25}=0.6046\pm0.0096$. Combining this with ACT\,DR6 and \planck{} PR4 CMB lensing measurements improves the constraint to $0.6020\pm0.0084$, the most precise determination of this parameter from CMB lensing to date. Combining our SPT-3G CMB lensing measurement with the primary CMB further constrains the late-time cosmological parameters, yielding $\Omega_{\rm m}=0.3117 \pm 0.0059$, $\sigma_{8}=0.8124 \pm 0.0044$, and $H_{0} = 67.60 \pm 0.42$. These joint constraints also have important implications for beyond-$\Lambda$CDM cosmological models, particularly when compared with BAO measurements. Given the known differences between primary CMB+CMB lensing and BAO measurements, we measure the distance between $\kgpa+\cmbspa$ and $\baodesi$ in the $\Omega_{\rm m}$--$hr_{\rm d}$ plane. We find that the distance between the two datasets decreases from $2.8\sigma$ in C26 to $1.9\sigma$ in this work. As discussed in Appendix~\ref{app:tau_prior}, this reduction reflects both the updated CMB lensing measurement and the different treatment of the low-$\ell$ polarization information used to constrain $\tau$.

This shift toward greater consistency in $\Lambda$CDM can also shift the preferred values of additional parameters in extended models relative to previous studies. For example, the upper limit on the sum of neutrino masses is relaxed to $\sum m_\nu < 0.072\,{\rm eV}$ at 95\% C.L.; the preference for non-zero spatial curvature is reduced to $1.9\sigma$; and the distance of $(w_0,w_a)$ from $(-1,0)$ is reduced to $2.1\sigma$. These results demonstrate that CMB lensing measurements can play a key role in determining how primary CMB and BAO constraints project onto extended cosmological models, and future CMB lensing measurements will provide a more definitive test of whether a discrepancy with BAO is present.

Our CMB lensing measurement is highly complementary to optical large-scale-structure measurements, which constrain different combinations of $\sigma_8$ and $\Omega_{\rm m}$. Combining our baseline lensing measurement with DES\,Y3 $3\!\times\!2$pt, we obtain $S_8 \equiv \sigma_8 \sqrt{\Omega_{\rm m} / 0.3} = 0.811\pm0.011$, a 1.4\% constraint. With major large-scale-structure surveys such as {\it Euclid}~\cite{euclid}, Rubin-LSST~\cite{rubin}, and {\it Roman}~\cite{roman} now underway, joint analyses with CMB lensing will provide increasingly powerful constraints on the growth of structure in the near future.
  
Beyond the cosmological constraints from the auto-spectrum presented in this work, the reconstructed SPT-3G lensing map can be used for a broad range of scientific applications. In particular, the high signal-to-noise lensing map can be used to construct a high-fidelity template of lensing $B$ modes over the $1500\,\sqdeg$ region overlapping with the deep degree-scale $B$-mode observations of the BICEP series of experiments~\cite{bk18,nakato2026b}, enabling more effective delensing and tighter constraints on the amplitude of primordial gravitational-wave $B$ modes, a major science goal of the South Pole Observatory~\cite{natoli2026}. 

In addition, the lensing map will be used for cross-correlation studies with large-scale-structure surveys~\cite{ouellette2026}, as well as with other CMB secondary tracers such as the CIB and tSZ, advancing our understanding of structure growth and astrophysical processes. Cross-correlations with line-intensity mapping and 21\,cm surveys offer a particularly promising route to extending these studies to higher redshifts, where direct auto-spectrum measurements remain challenging because of foregrounds and instrumental systematics.

Observations of the SPT-3G Main field have continued well beyond the two years of data used in this work. The field was observed to a similar per-year depth in $2021$--$2023$ and 2025, and observations will continue until the deployment of the SPT-3G+ camera~\cite{natoli2026}. These maps will enable significant gains in lensing signal-to-noise when employing methods beyond the quadratic estimator~\cite{ge2025,carronlewis2017,millea2020}, allowing more efficient removal of lensing $B$ modes and thereby improving the sensitivity of searches for primordial gravitational waves.

Beyond the Main field analyzed in this work, SPT-3G has observed several additional fields to reduce lensing sample variance on large scales and to increase the overlap with external surveys. These include a set of three ``Summer" fields totaling $2600\,\sqdeg$, observed during parts of every observing season, and a $6000\,\sqdeg$ ``Wide" survey observed throughout 2024~\cite{prabhu2024}. Lensing analyses using observations of these additional fields are ongoing. As part of this effort, a first reconstruction using the Summer fields has demonstrated that SPT-3G lensing analyses can be successfully extended beyond the Main field despite differences in observing conditions~\cite{levy2026}. The lensing measurement obtained by combining the different SPT-3G survey fields, covering approximately $10000\,\sqdeg$ in total (Ext-10k survey), will push the lensing SNR toward 100 and significantly improve cosmological parameter constraints~\cite{vitrier2026,prabhu2024}.

\begin{acknowledgments}
YO thanks Eric Baxter for his scientific collaboration, mentorship, and friendship. His support and guidance were instrumental in making this work possible. We also remember Karim Benabed for his scientific contributions, mentorship, and dedication to the collaboration.

The South Pole Telescope program is supported by the National Science Foundation (NSF) through awards OPP-1852617 and OPP-2332483. Partial support is also provided by the Kavli Institute of Cosmological Physics at the University of Chicago. 
Argonne National Laboratory’s work was supported by the U.S. Department of Energy, Office of High Energy Physics, under contract DE-AC02-06CH11357. 
The UC Davis group acknowledges support from Michael and Ester Vaida. 
Work at the Fermi National Accelerator Laboratory (Fermilab), a U.S. Department of Energy, Office of Science, Office of High Energy Physics HEP User Facility, is managed by Fermi Forward Discovery Group, LLC, acting under Contract No. 89243024CSC000002.
The Melbourne authors acknowledge support from the Australian Research Council’s Discovery Project scheme (No. DP210102386). 
The Paris group has received funding from the European Research Council (ERC) under the European Union’s Horizon 2020 research and innovation program (grant agreement No 101001897), and funding from the Centre National d’Etudes Spatiales. 
The SLAC group is supported in part by the Department of Energy at SLAC National Accelerator Laboratory, under contract DE-AC02-76SF00515.
WLKW acknowledges support from an Early Career Research Award DE-SC0026376 of the Department of Energy.
We gratefully acknowledge the computing resources provided on Crossover, a high-performance computing cluster operated by the Laboratory Computing Resource Center at Argonne National Laboratory. This research used resources of the Argonne Leadership Computing Facility, which is a U.S. Department of Energy Office of Science User Facility operated under contract DE-AC02-06CH11357. This work was completed in part with resources provided by the University of Chicago’s Research Computing Center. Some of the computing for this project was performed on the Sherlock cluster. We would like to thank Stanford University and the Stanford Research Computing Center for providing computational resources and support that contributed to these research results. This work used the resources of the SLAC Shared Science Data Facility (S3DF) at SLAC National Accelerator Laboratory. S3DF is a shared High-Performance Computing facility, operated by SLAC, that supports the scientific and data-intensive computing needs of all experimental facilities and programs of the SLAC National Accelerator Laboratory. SLAC is operated by Stanford University for the U.S. Department of Energy’s Office of Science. This research was done using services provided by the OSG Consortium \cite{osg07,osg09,osg_ext1,osg_ext2} which is supported by the National Science Foundation awards 2030508 and 2323298. This work relied on the \texttt{NumPy} package for numerical computations~\citep{numpy}, the \texttt{SciPy} package for scientific computing~\citep{scipy}, the \texttt{JAX} package for automatic differentiation and GPU/TPU acceleration~\citep{jax18}, the \texttt{healpy} package for HEALPix-based spherical map operations~\citep{healpix,healpy}, the \texttt{astropy} package for astronomical calculations~\citep{astropy}, and the \texttt{Matplotlib} package for plotting~\citep{matplotlib}.
Cosmological calculations and parameter inference were performed using the \texttt{CAMB}~\citep{camb,camb2}, \texttt{Cobaya}~\citep{cobaya}, \texttt{CosmoSIS}~\citep{cosmosis}, and \texttt{nautilus}~\citep{nautilus} packages.
Posterior sampling analysis and plotting were performed using the \texttt{GetDist} package~\citep{Lewis_getdist}, and posterior consistency was assessed using the \texttt{tensiometer} package~\citep{raveri2021}..
\end{acknowledgments}

\clearpage
\onecolumngrid

\appendix

\section{Spectrum-level bias correlated with the mean-field estimate}\label{app:mfnoise}
In this appendix, we outline the derivation we use to arrive at the form of the spectrum-level bias term $N_L^{\rm MFcorr}$ in Equation~\eqref{eq:mfnoise}.
We write the raw reconstruction as:
\begin{equation}\label{eqn:A1}
\widehat{\phi}^{\rm raw}_{i} = \phi^{\rm true}_{i} + \phi_{\rm MF}^{\rm true} + n_{i} + \epsilon^{\phi}_{i},
\end{equation}
where $\phi^{\rm true}_{i}$ is the input lensing field, $\phi_{\rm MF}^{\rm true}$ is the true mean-field, $n_i$ is the reconstruction noise realization (whose spectrum includes $N_L^{(0)}$ and $N_L^{(1)}$), and $\epsilon^{\phi}_{i}$ is a term that sources $N_L^{\rm MFcorr}$ whose properties we describe next. 

We test the mean-field subtraction using two different spectrum constructions that should be equivalent in expectation. In the first, we construct  $\langle\widehat{\phi}\rangle_{j\neq i}$ using all the available simulation realizations except realization $i$, and form
\begin{equation}
C_L\left(\widehat{\phi}^{\rm raw}_{i},
\widehat{\phi}^{\rm raw}_{i}-\langle\widehat{\phi}\rangle_{j\neq i}\right).
\end{equation}
Although the mean field is subtracted only from one leg, this provides an unbiased estimate in expectation: the mean-field contribution is removed from the second leg, so there is no mean-field auto-spectrum contribution, while the finite-simulation noise in the leave-one-out mean-field estimate is independent of realization $i$ and therefore does not correlate with the raw reconstruction in the first leg. In the second approach, we split the simulations into two non-overlapping halves and construct independent mean-field estimates, $\langle\widehat{\phi}\rangle_{h1}$ and $\langle\widehat{\phi}\rangle_{h2}$. We then form
\begin{equation}
C_L\left(
\widehat{\phi}^{\rm raw}_{i}-\langle\widehat{\phi}\rangle_{h1},
\widehat{\phi}^{\rm raw}_{i}-\langle\widehat{\phi}\rangle_{h2}
\right).
\end{equation}
Because disjoint mean-field estimates are used in the two legs, the noise arising from the finite number of simulations used to estimate each mean field is uncorrelated between the two legs and therefore does not bias the cross-spectrum. The two approaches should therefore give the same result on average.

In the lensed simulations, however, we find a small but non-zero difference between the two constructions. We parameterize this discrepancy by introducing an additional component of the reconstructed field, $\epsilon_i^{\phi}$, whose cross-spectrum with the estimated mean field is non-zero:
\begin{equation}\label{eqn:A3}
C_{L}({\epsilon^{\phi}_{i}, \langle \widehat{\phi}\rangle_{j \neq i}}) \neq 0.
\end{equation}
The superscript $\phi$ indicates that this component is introduced to describe an effect that is present in the lensed simulations but absent in the unlensed simulations. By construction, we require $\epsilon^{\phi}$ to be uncorrelated with $n$, $\phi_{\rm MF}^{\rm true}$, and $\phi^{\rm true}$, such that $\epsilon^{\phi}$ represents a convenient decomposition of the observed residual rather than a unique physical identification of the underlying effect. 

In our standard pipeline, we use the cross-mean-field approach to estimate the lensing power spectrum. Including the additional $\epsilon^{\phi}$ term introduced above gives
\begin{align} \label{eqn:A4}
C_{L}^{\hat{\phi}_{h1}\hat{\phi}_{h2}} &= C_L( \widehat{\phi}^{\rm raw}_{i} - \langle \widehat{\phi}\rangle_{h1 \neq i} \, , \widehat{\phi}^{\rm raw}_{i}- \langle \widehat{\phi}\rangle_{h2 \neq i} ) \nonumber \\
&= C_L^{\phi\phi, {\rm true}} + N_L^{(0)} + N_L^{(1)} + C_{L}(\epsilon^{\phi}_{i},\epsilon^{\phi}_{i})  - C_{L}({\epsilon^{\phi}_{i}, \langle \widehat{\phi}\rangle_{h1 \neq i}}) - C_{L}({\epsilon^{\phi}_{i}, \langle \widehat{\phi}\rangle_{h2 \neq i}}).
\end{align}
The last three terms denote the extra spectrum-level bias we have to subtract.
This bias can be estimated by forming a cross-spectrum similar to Equation~\eqref{eqn:A4} but subtracting the input $\phi$ from both terms (Equation~\eqref{eq:mfnoise}):
\begin{align}
N_L^{\rm MFcorr} &\equiv \frac{1}{N_{\rm sim}}\sum_i C_L( \widehat{\phi}^{\rm raw}_{i} - \langle \widehat{\phi}\rangle_{h1 \neq i} - \phi^{\rm true}_i \, , \widehat{\phi}^{\rm raw}_{i}- \langle \widehat{\phi}\rangle_{h2 \neq i}- \phi^{\rm true}_i ) - N_L^{(0)} - N_L^{(1)} \\
& = \frac{1}{N_{\rm sim}}\sum_i C_{L}(\epsilon^{\phi}_{i},\epsilon^{\phi}_{i})   - C_{L}({\epsilon^{\phi}_{i}, \langle \widehat{\phi}\rangle_{h1 \neq i}}) - C_{L}({\epsilon^{\phi}_{i}, \langle \widehat{\phi}\rangle_{h2 \neq i}}).
\end{align}

While we do not have an exact form for what constitutes $\epsilon^{\phi}$, we can comment on what this excess power is {\it not} sourced from. 
One possibility is that the mean-field estimate $ \langle \widehat{\phi}\rangle$ is not an unbiased estimate of the true mean-field. This can arise because our noise realizations are correlated (at a few percent level) between different simulation indices. The noise realization correlations originate from the fact that they are drawn from sign-flip combinations of observed maps. This leads to both the mean-field estimate not converging to the true mean-field and the mean-field cross-spectrum using two halves of disjoint simulations having residuals over the true mean-field power. While this effect exists, we know that this is negligible because we use the mean-field estimates from lensed simulations (which uses these noise realizations) on the unlensed sims test and we do not observe this bias.
Another possibility is that the reconstruction noise $n_i$ correlates with the mean-field estimate $ \langle \widehat{\phi}\rangle_{j \neq i}$ because of instrumental noise correlations between realization $i$ and non-$i$ that enter the mean-field estimate. Again, this effect exists but is negligible because it would otherwise also have appeared in the unlensed set of simulations.

In summary, we identify this mean-field spectrum bias as arising from a component of the reconstructed $\phi$ field that correlates with the mean-field estimate. We compute the size of this term using simulations and subtract it from our data lensing spectrum estimate.

\section{Calibration of simulations}\label{sec:sim_calibration}
\begin{figure*}[t]
\centering

\begin{minipage}[t]{0.48\textwidth}\vspace{0pt}
\centering
\includegraphics[width=\linewidth]{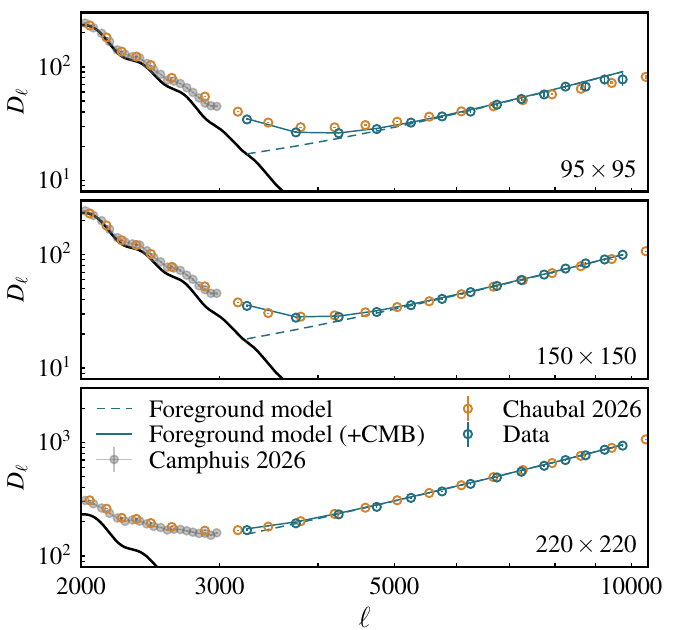}
\end{minipage}
\hspace{-0.0cm}
\begin{minipage}[t]{0.48\textwidth}\vspace{0pt}
\includegraphics[width=\linewidth]{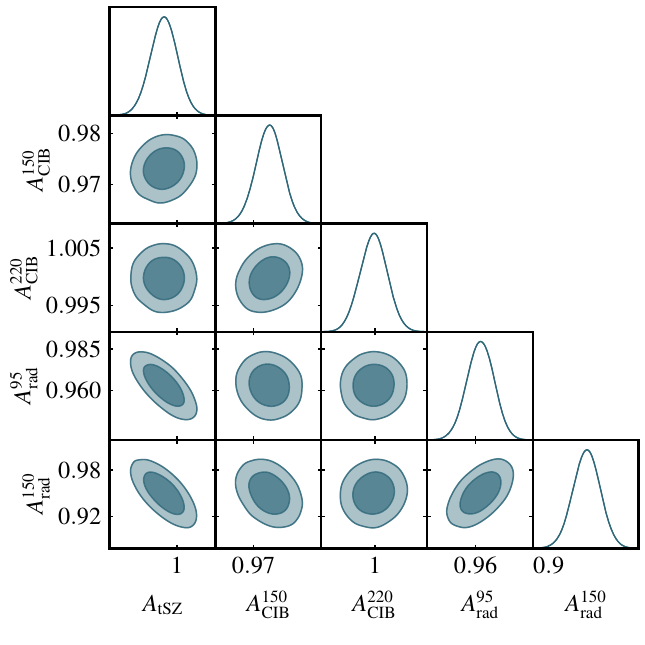}
\centering
\end{minipage}
\caption{{\bf Left:} Bandpowers used in C26 and the high-$\ell$ bandpowers measured for this work, shown as gray and teal points, respectively. The high-$\ell$ bandpowers use the same masking prescription as C26 and extend the measurement to the smaller angular scales required for this purpose. The best-fit foreground model is shown by the teal dashed line, while the total CMB+foreground model is shown by the solid teal line. Although only the auto-spectra are shown here, the cross-spectra are also included in the fit. For reference, the orange points show the measurements from \cite{chaubal2026}, which use the same maps and point-source masking threshold but do not mask galaxy clusters. {\bf Right:} Posteriors for the five foreground amplitude parameters $A_{\rm tSZ}$, $A_{\rm CIB}^{150}$, $A_{\rm CIB}^{220}$, $A_{\rm rad}^{95}$, and $A_{\rm rad}^{150}$.}
\label{fig:fg_fit_model}
\end{figure*}

Although the astrophysical components in the \textsc{Agora} simulation are calibrated against observational data, small discrepancies may remain in the output maps. We therefore apply small multiplicative calibration factors to the raw simulation products so that their high-$\ell$ ($3000 <\ell <10000$) power spectrum amplitudes better match the SPT-3G observations. The single-frequency temperature maps are modeled as
\begin{equation}
{M}_{\nu}
=
\left({M}^{\rm CMB}+{M}^{\rm kSZ}\right)
+ A^{\rm tSZ}{M}^{\rm tSZ}_{\nu}
+ A^{\rm CIB}_{\nu}{M}^{\rm CIB}_{\nu}
+ A^{\rm rad}_{\nu}{M}^{\rm rad}_{\nu}.
\end{equation}
For the tSZ contribution, we apply a single amplitude to the underlying Compton-$y$ map before converting it to temperature units at each frequency. The same amplitude parameter therefore applies to all frequency channels. In contrast, the CIB and radio-source amplitudes are allowed to vary independently in each frequency channel. However, we fix $A^{\rm CIB}_{95}$ and $A^{\rm rad}_{220}$ to 1 since their amplitudes are small and we have little constraining power on those amplitudes. In total, we fit five amplitude parameters. We perform this calibration in two stages:\\
\begin{itemize}
\item\textbf{Stage I}: the foreground amplitudes were calibrated before the production of the Gaussian simulations, using preliminary beam and calibration models. This stage was intended to provide approximate agreement between \agora{} and the SPT-3G observations at high-$\ell$ so that we could start producing simulations for this work.

\item\textbf{Stage II}: the calibration was repeated after the beam model was updated and the final map-calibration parameters were determined. This second stage is used primarily to define the prior range for the foreground-emulator parameters. In this fit, the CMB contribution to the SPT-3G spectra is fixed to the prediction from $\cmbspa$, and the high-$\ell$ spectra are refit using the updated beam and calibration models. This simple foreground amplitude-scaling approach does not necessarily provide a good fit at the precision of the measured spectra. Rather than interpreting the resulting statistical uncertainties as precise constraints on the foreground amplitudes, we conservatively inflate the Gaussian uncertainties derived from the measured spectra by a factor of $\mathord{\sim}10$, which brings the goodness of fit to $\chi^2/{\rm dof}\sim1$. The resulting broader parameter constraints are used to set the widths of the foreground-parameter priors adopted in the cosmological inference.
\end{itemize}
In the left panel of Figure~\ref{fig:fg_fit_model}, we compare the measured data bandpowers with the total foreground model obtained by applying the fitted amplitude parameters to the templates. 

\newpage
\normalsize
\section{Bandpower covariance matrix}\label{app:bpcm_cond}
\begin{figure}
\includegraphics[width=0.6\linewidth]{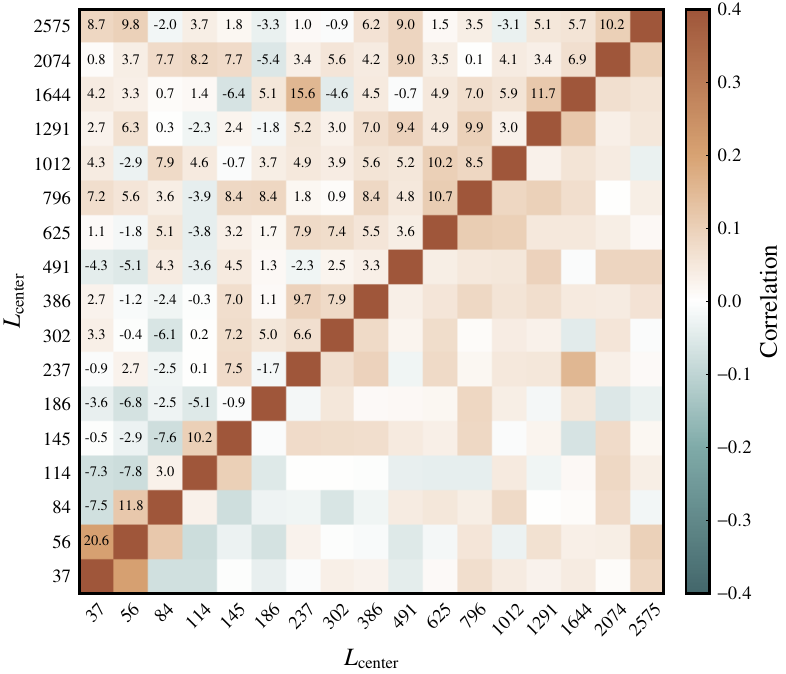} 
\caption{The correlation matrix of the bandpower covariance for $\kgmv$, constructed from 498 debiased simulation lensing bandpowers.
The $x$- and $y$-axis labels are the bin centers.
The values of the off-diagonal correlation are included in the grid, expressed as percentages.
Overall, the off-diagonal correlations are small with coherent positive correlations across the higher $L$ bins. 
}
\label{fig:bpcm_corr}
\end{figure}

In the baseline analysis, we construct the bandpower covariance matrix (BPCM) $ \mathbb{C}_{bb'}$ by computing
\begin{equation}
\mathbb{C}_{bb'} = \frac{1}{N-1} \sum^N_{i=1} ( \hat{C}^i_{b} -  \bar{\hat{C}}_{b} ) (  \hat{C}^i_{b'} -  \bar{\hat{C}}_{b'} ), 
\end{equation}
where $\hat{C}^i_{b}$ denotes the $N_{L}^{(0),{\rm SA}}$-debiased lensing bandpower at bin $b$ from simulation $i$ and $\bar{\hat{C}}_{b'}$ is the mean of the debiased bandpowers across all $N = 498$ simulation realizations. The $N_{L}^{(0),{\rm SA}}$ correction is mean-calibrated so that its average over simulations matches the corresponding simulation-based $N_{L}^{(0)}$ estimate.
Figure~\ref{fig:bpcm_corr} shows the bandpower correlation matrix. It shows mild ($\lesssim$20\%) correlations between bandpowers relatively far off from the diagonal, with coherent positive 
correlations among the high-$L$ bins.
The eigenvalue spectrum and the condition number of the covariance matrix suggest that there are no noisy modes that require regularization or removal. We therefore retain the full measured covariance structure.

To see the impact of these off-diagonal correlations on parameter constraints, we construct two BPCM variants in which elements far from the diagonal are smoothed or zeroed, replace the baseline BPCM with each variant, and run MCMC chains on simulated bandpowers. In the first variant, we zero all elements in the correlation matrix that are more than nine bins away from the diagonal, smooth elements more than four bins away from the diagonal using the average of their four nearest neighbors, and leave the remaining elements unchanged. In the second variant, we zero all elements that are more than four bins away from the diagonal. From each modified correlation matrix, we reconstruct the corresponding BPCM. 

We run MCMC chains sampling only cosmological parameters (fixing instrumental and foreground systematics)
on lensing bandpowers from one simulation. The chain mean and MAP values of $\sigma_8\Omega_{\rm m}^{0.25}$ across the baseline and the two variants are  within 0.15$\sigma$ of each other, where $\sigma$ denotes the $1\sigma$ uncertainty on $\sigma_8\Omega_{\rm m}^{0.25}$
on data in our baseline run (freeing calibration and foreground parameters). The uncertainties on $\sigma_8\Omega_{\rm m}^{0.25}$ across the three cases are within 3\% of each other.
These tests indicate that the inferred constraint on $\sigma_8\Omega_{\rm m}^{0.25}$ is robust to alternative treatments of the off-diagonal covariance.

\section{CMB marginalization in BPCM for lensing-only chains}\label{app:cmb_marg}
In this section, we quantify the increase in the diagonal elements of the total covariance matrix after adding the CMB-marginalization (CMBmarg) term to the base BPCM used in the lensing-only chains.
We discuss two approaches for constructing the CMBmarg term: our baseline cosmology-agnostic approach and an alternative cosmology-dependent prescription. 

\begin{figure}[t]
\centering

\begin{minipage}[t]{0.48\linewidth}
\centering
\includegraphics[width=\linewidth]{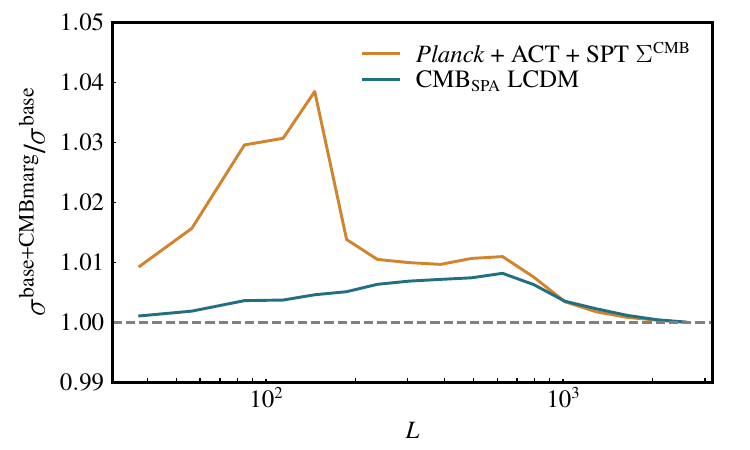}
\caption{
The ratio of the square root of the diagonal of the lensing bandpower covariance matrix with and without the CMBmarg term.
The lines denote the increase in the size of the error bars of the lensing bandpowers due to the uncertainties in the primary CMB measurements.
We adopt the cosmology agnostic choice of using the CMB {\it Lite} covariances from \planck{} PR3, ACT\,DR6, and SPT-3G D1 (orange) in this analysis.
We also show, in teal, an alternative prescription of constructing the CMBmarg term that is cosmology-dependent.
Our baseline choice increases the uncertainty by up to 4\% in the lensing-only chains.
}
\label{fig:cmbmarg_ratio}
\end{minipage}
\hfill
\begin{minipage}[t]{0.48\linewidth}
\centering
\includegraphics[width=\linewidth]{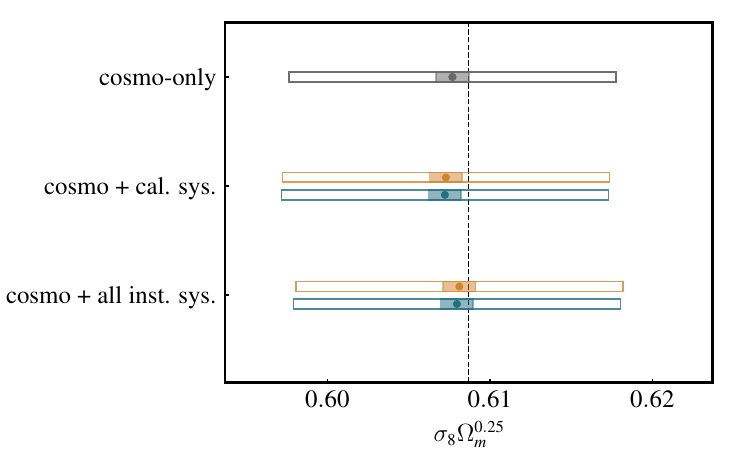}
\caption{
Inference pipeline validation at intermediate steps.
We show the mean, standard error $\sigma_m$ (filled), and measurement error $\sigma$ (open)
on $\sigma_8\Omega_{\rm m}^{0.25}$ for freeing (top) cosmology-only, (middle) cosmology and calibration parameters,
and (bottom) cosmology and all instrumental systematic parameters.
For the cases where instrumental systematic parameters are freed, we include the case where the systematics are modeled through the analytic instrumental-systematics approach (orange) and the emulator approach (teal).
In all cases, the mean $\som$ values are within $2\sigma_m$ of the input value, with no statistically significant bias detected.
}
\label{fig:pipe_val}
\end{minipage}

\end{figure}

Recall that for lensing-only chains, we add the CMBmarg term to the base BPCM 
\begin{equation}
	\mathbb{C}_{bb'} \rightarrow \mathbb{C}_{bb'} + \mathbb{C}^{\rm CMBmarg}_{bb'}.
\end{equation}
In the baseline case, we construct $\mathbb{C}^{\rm CMBmarg}_{bb'}$ by projecting the primary CMB bandpower uncertainties
and their correlations onto the lensing bandpower space using the $M^x$ matrices, where $x \in \{{TT, TE, EE}\}$, which encodes the
change in the lensing response due to changes in the primary CMB  spectra: 
\begin{equation}
	 \mathbb{C}^{\rm CMBmarg}_{bb'} = M^{x}_{b\ell_b} \,
	\Sigma^{\rm CMB; {\it xy}}_{\ell_b \ell'_b}
	M^{y}_{\ell'_b b'}, 
\end{equation}
where $\Sigma^{\rm CMB}$ denotes a primary CMB spectrum covariance and $x,y \in \{{TT, TE, EE}\}$. 
When the CMB covariance matrix is binned, this term can be approximated by Equation~\eqref{eqn:cmb_marg}, as is done for this analysis. 
Here $\Sigma^{\rm CMB}$ is cosmology agnostic: it comes from the {\it Lite} covariances from {\it Planck} PR3, ACT\,DR6, and SPT-3G D1, which were constructed requiring the CMB component to be the same across all frequencies without imposing a cosmological model~\cite{Prince2019, balkenhol2025cmblite}.
We show in Figure~\ref{fig:cmbmarg_ratio} the ratio of the square root of the diagonal of the total BPCM after adding the CMBmarg term to that of the baseline BPCM. Around $L\sim100$, the baseline CMBmarg term increases the lensing bandpower uncertainties by up to $4\%$. Above $L\sim300$, the increase is at most $1\%$.

An alternative prescription~\cite{madhavacheril2024} to construct the CMBmarg term is to estimate $\Sigma^{\rm CMB}$ from the scatter in spectra generated from posterior samples. Specifically, for each of the {\it Planck} PR3, ACT\,DR6, and SPT-3G D1 primary-CMB chains, we draw 5000 samples and use the corresponding $TT$, $TE$, and $EE$ spectra to form $\mathbb{C}^{\rm CMBmarg}_{bb'}$ by:
\begin{equation}
	 \mathbb{C}^{\rm CMBmarg}_{bb'} = {\rm cov}\left( \sum_{x} M^{x}_{b\ell_b} C_{\ell_b}^{x} \right),
\end{equation}
where $x$ runs through $TT$, $TE$, and $EE$. 
We neglect the $\Sigma^{\rm CMB; {\it xy}}$ blocks where $x \neq y$ in this formulation as they contribute negligibly. 
This prescription, rather than marginalizing over the uncertainty of the primary CMB bandpowers, marginalizes over the spread of the allowed
primary CMB spectra given $\Lambda$CDM parameters that fit all three datasets. 
This prescription therefore depends on the assumed cosmological model and is more restrictive than our baseline approach.
This is reflected in the minimal increase ($< 1\%$) in the square root of the BPCM diagonal when adding this term (see Figure~\ref{fig:cmbmarg_ratio}).
We choose the cosmology-agnostic approach in building the CMBmarg term in this analysis.

\section{Analytic modeling of instrumental systematics}\label{app:analytic_sys_model}
The instrumental systematics considered in this work include temperature and polarization calibration ($T_{\rm cal}, P_{\rm cal}$), 
temperature beam uncertainties ($\eta$), and polarized beam shape caused by the depolarization of sidelobes $(\beta_{\rm pol}^{\nu})$. 
These are effects that modify the shape of the measured primary CMB spectra.
Therefore, their impact on the measured lensing spectrum enters through the change of the response function $\mathcal{R}_L$
evaluated at different instrumental systematic parameter values.  

For a given set of sampled beam and calibration parameters, the primary CMB model spectra transform as
\begin{equation} \label{eqn:sys_cmb2pt}
	C_{\ell}^{XY} \rightarrow C_{\ell}^{XY} \mathcal{B}_{\ell}^{XY} / \mathcal{C}^{XY},
\end{equation}
where $XY \in \{TT, TE, EE\}$, $\mathcal{B}_{\ell}$ and $\mathcal{C}$ denote beam and calibration factors.
The beam factors are functions of ratios of the LC-weighted perturbed beam to the fiducial beam in temperature $b_{\ell}^{T}$ and in polarization $b_{\ell}^{E}$: 

\begin{align}\label{eqn:beams}
	b_{\ell}^{T} & =\frac{ \sum_{\nu} w^{T, \nu}_{\ell} (B_{\ell}^{T, \nu} + \sum_{ i=1,\ldots,4} \eta_i dB_{\ell, i}^{T, \nu}) } { \sum_{\nu} w^{T, \nu}_{\ell} B_{\ell}^{T, \nu}},  \\ 
	b_{\ell}^{E} & = \frac{ \sum_{\nu} w^{E, \nu}_{\ell} B_{\ell}^{P, \nu} (\beta_{\rm pol}^{\nu})} { \sum_{\nu} w^{E, \nu}_{\ell} B_{\ell}^{P, \nu} (\beta_{\rm pol, fid}^{\nu})}, 
\end{align}
where $B_{\ell}^{T, \nu}$ is the per-frequency temperature beam and 
\begin{equation}
B_{\ell}^{P, \nu} = \frac { B_{\ell}^{{\rm main}, \nu} +\beta_{\rm pol}^{\nu} (B_{\ell}^{T, \nu}-B_{\ell}^{{\rm main}, \nu}) } { B_{800}^{{\rm main}, \nu} +\beta_{\rm pol}^{\nu} (B_{800}^{T, \nu}-B_{800}^{{\rm main}, \nu})}.   
\end{equation}
Here, $\nu$ denotes the frequency band and $w^{\nu}_{\ell}$ denotes the LC weights
(which are precomputed and held fixed). The term $dB_{\ell,i}^{T, \nu}$ represents the $i$-th eigenmode of the
temperature beam uncertainty covariance matrix, and we vary the first four 
of these eigenmodes by sampling $\eta_{i}$. Note that $\eta_{i}$ carries no $\nu$
dependence, as the same value is applied per eigenmode which carries information of the beam uncertainties from all three frequency bands. 
The combined beam factors are then given by
\begin{equation}
    \mathcal{B}_{\ell}^{XY} = b_{\ell}^{X}\, b_{\ell}^{Y}.
\end{equation}
The calibration factors are
\begin{equation}
\begin{aligned}
\mathcal{C}^{TT} &= T_{\rm cal}^{2}, \\ 
\mathcal{C}^{TE} &= T_{\rm cal}^{2}  P_{\rm cal}, \\ 
\mathcal{C}^{EE} &= T_{\rm cal}^{2} P_{\rm cal}^{2} .\\ 
\end{aligned}
\end{equation}

In the baseline analysis, we emulate and marginalize over both the instrumental and foreground effects (Equation \eqref{eqn:modelspec_emu}). In the analytic instrumental-systematics approach, we instead pull the instrumental systematic parameters $\boldsymbol{\theta^s}$ out of the emulator term and include them through the change of the primary CMB model spectra (Equation~\eqref{eqn:sys_cmb2pt}), while the foreground parameters
$\boldsymbol{\theta^f}$ remain emulated:
\begin{align}\label{eqn:modelspec_full}
C_{b}^{\kappa \kappa,\mathrm{model}}(\boldsymbol{\Theta}) & = C_b^{\kappa \kappa}({\boldsymbol{\theta^c}}) \overbracket[0.8pt]{\frac{C_b^{\kappa \kappa}({\boldsymbol{\theta^c}_{\rm fid}},\boldsymbol{\theta^s}_{\rm fid} , \boldsymbol{\theta^f})}{C_b^{\kappa \kappa}({\boldsymbol{\theta^c}_{\rm fid}}, {\boldsymbol{\theta^s}_{\rm fid}, {\boldsymbol{\theta^f}=0 }} )}}^{\rm emulated}\nonumber \\
&+M^{\kappa\kappa}_{bL} \left(C_{L}^{\kappa\kappa}(\boldsymbol{\theta^c})-C_{L}^{\kappa\kappa}(\boldsymbol{\theta^c}_{\rm fid})\right)\nonumber\\
&+\sum_{x\in \lbrace TT,TE,EE \rbrace } M^{x}_{b\ell} \left(C_{\ell}^{x}(\boldsymbol{\theta^c}, \overbracket[0.8pt]{\boldsymbol{\theta^s}}^{\rm sys})-C_{\ell}^{x}(\boldsymbol{\theta^c}_{\rm fid})\right).
\end{align}
The change in primary CMB model spectra from shifts in systematic parameters propagates into the model lensing spectrum through the response. This approach has the advantage of allowing the impact of the instrumental systematics to be calculated analytically, without introducing errors from the emulator. However, it also has limitations in its current form, as discussed in Appendix~\ref{app:emu_sys}, which led us to baseline the emulator approach in this analysis. We compare the parameters from the two approaches in Appendix~\ref{app:pipe_vad}.

\section{Inference pipeline validation}\label{app:pipe_vad}
In this section, we show that our cosmological inference pipeline is unbiased at intermediate steps when not all of the parameters are varied. For the lensing-only case, we test three parameter subsets: (1) only cosmological parameters, (2) cosmological and calibration parameters, and (3) cosmological and all instrumental-systematic parameters. For each case, the mean recovered MAP value of $\sigma_8\Omega_{\rm m}^{0.25}$ across the simulation realizations is consistent with the input truth. For these tests, we use the set of simulations with Gaussian foregrounds.  When instrumental systematics are included, we test both the analytic instrumental-systematics approach
(Appendix~\ref{app:analytic_sys_model}) and the emulator approach (Section~\ref{sec:emulator_train}). Instrumental systematic parameters include temperature and polarization calibration, temperature beam uncertainties, and polarization beam modeling variations. For the baseline case, we fix the beam parameters because both temperature and polarization beam variations are included in the CMBmarg term and the nuisance-marginalized (except calibration) CMB bandpowers in lensing-only runs and in the model for the SPT-3G D1 $TT/TE/EE$ bandpowers in lensing+primary CMB runs. For the full case in which all parameters (cosmology, instrumental, and foreground) are freed, we test on the \textsc{Agora} set of sims, as detailed in Section~\ref{sec:agora_test}.

For each test, we run minimizers on 100 debiased simulated lensing spectra, using the same Cobaya settings for the {\tt CAMB} theory accuracy and sampled-parameter priors as in all other lensing-only simulation runs. The input lensing power spectrum for generating the lensed CMB skies was computed with higher-accuracy settings ({\tt l\_accuracy\_boost = 4} and {\tt accuracy\_boost = 4}) and using the default setting of \texttt{HMcode2016}~\cite{mead2016}. 
In order for the theory CAMB output in run time to match the input simulations spectrum without going to the extra high accuracies (which incurs significant computation time), for the simulation validation runs, we set {\tt  HMCode\_A\_baryon = 3.23} and {\tt HMCode\_eta\_baryon = 0.592} and {\tt mnu = 0.059} to match the simulation input lensing power spectrum to within 0.05\%.

We deem the pipeline to be unbiased if the mean of the 100 MAP $\sigma_8\Omega_{\rm m}^{0.25}$ values is within
2 $\sigma_m$ with $\sigma_m \equiv \frac{\sigma}{\sqrt{N_{\rm sims}}}$ and $\sigma$ from the mean of the 1$\sigma$
uncertainties on $\sigma_8\Omega_{\rm m}^{0.25}$ of the baseline chains (freeing cosmology, calibration, and foreground parameters) on data.

We show in Figure~\ref{fig:pipe_val} the mean, $\sigma_{\rm m}$, and $\sigma$ of $\sigma_8\Omega_{\rm m}^{0.25}$ for all three cases. For the cosmology-only case, the mean is $1\sigma_{\rm m}$ from the input truth. For the cosmology and calibration parameter case, the means are $1.4\sigma_{\rm m}$ and $1.5\sigma_{\rm m}$ from the input truth for the analytic instrumental-systematics and emulator approaches, respectively. For the cosmology and all instrumental-systematics parameter case, the corresponding offsets are $0.6\sigma_{\rm m}$ and $0.7\sigma_{\rm m}$, respectively. We therefore find no statistically significant bias in the inference pipeline at these intermediate steps.

\section{Characterization of systematic and foreground
marginalization}\label{app:emu_sys}

\begin{figure*}[t]
\centering
\begin{minipage}[t]{0.48\textwidth}
\centering
\includegraphics[width=\linewidth]{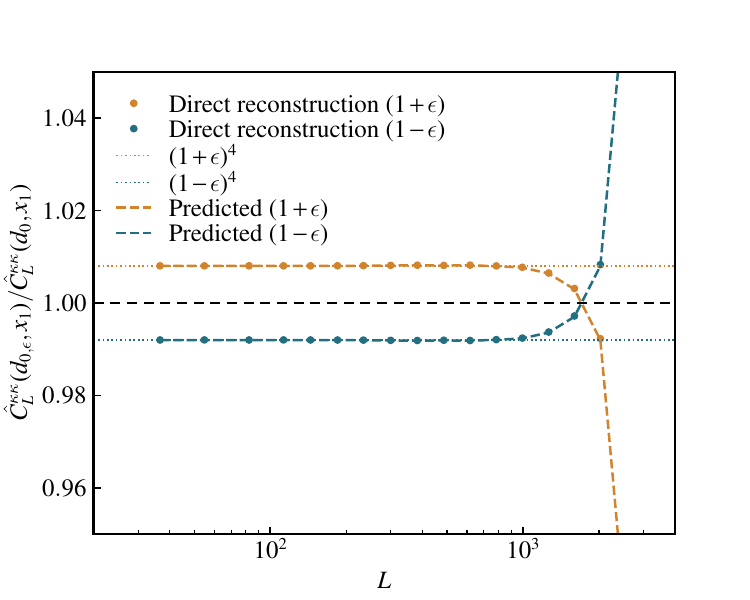}
\end{minipage}
\hfill
\begin{minipage}[t]{0.48\textwidth}
\centering
\includegraphics[width=\linewidth]{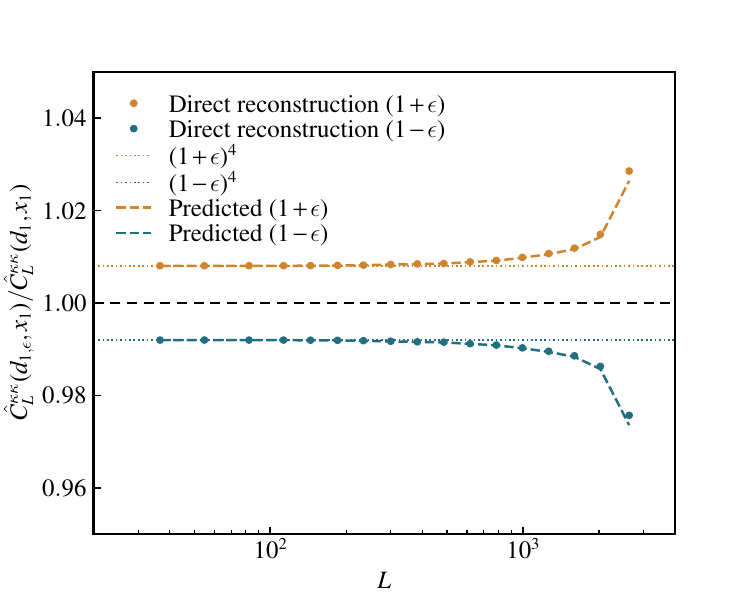}
\end{minipage}
\caption{ {\bf Left:} Ratio of perturbed to baseline debiased lensing spectra for a foreground-free data realization, $d_0$, with the calibration parameter varied by $\pm0.2\%$. {\bf Right:} Same as left, but with the foreground amplitude of the data realization set to the fiducial value, $d_1$, matching with the simulation ensemble $x_1$. In both panels, the simulations used to compute the lensing response and the $N_L^{(0),\rm RD}$ and $N_L^{(1)}$ biases have the fiducial foreground amplitudes, denoted by $x_1$. The ``Direct reconstruction'' points are obtained by multiplying the input $T/Q/U$ maps by $(1\pm\epsilon)$, rerunning the lensing reconstruction and bias subtraction, and taking the ratio of the resulting perturbed and unperturbed spectra. The ``Predicted'' curves are calculated from Equation~\eqref{eq:emul_bias} using only simulation products evaluated at $\epsilon=0$, without rerunning the reconstruction for the perturbed calibration. The $(1\pm\epsilon)^4$ lines show the analytic reference expected from the scaling of the raw four-point term alone. }
\label{fig:emul_validation_mv}
\end{figure*}

In this appendix, we investigate the high-$L$ turnover feature in the emulator correction as the calibration parameters are varied, as seen in Figure~\ref{fig:emul}.
We first show that this feature is not a numerical artifact of the emulator, but instead originates from the mismatch between the foreground amplitude of the reference point and the fiducial foreground amplitude included in the simulations used to compute the lensing response and the $N_L^{(0)}$, $N_L^{(1)}$ bias terms. We then derive an expression that predicts the resulting change in the reconstructed lensing spectrum using quantities evaluated from the unperturbed simulations, and use it to understand why the turnover is particularly pronounced for the foreground-free reference adopted in the emulator.

We characterize the effects of foreground contamination and instrumental systematics on the reconstructed lensing spectrum through a multiplicative correction factor applied to the theoretical lensing spectrum. This correction factor is measured from simulations at discrete points in the foreground and instrumental-systematic parameter space, and the emulator provides an efficient continuous interpolation between these points:
\begin{equation}
C_b^{\kappa \kappa}({\boldsymbol{\theta^c}}) \longrightarrow C_b^{\kappa \kappa}({\boldsymbol{\theta^c}}) \overbracket[0.8pt]{\frac{C_b^{\kappa \kappa}({\boldsymbol{\theta^c}_{\rm fid}}, \boldsymbol{\theta^f, \theta^s})}{C_b^{\kappa \kappa}({\boldsymbol{\theta^c}_{\rm fid}}, {\boldsymbol{\theta^s}_{\rm fid}, {\boldsymbol{\theta^f}=0 }} )}}^{\rm emulated},
\end{equation}
where $C_b^{\kappa \kappa}({\boldsymbol{\theta^c}})$ is the raw theoretical spectrum. The denominator of the multiplicative correction, which we call the reference point, is chosen such that when $C_b^{\kappa \kappa}({\boldsymbol{\theta^c}_{\rm fid}}, \boldsymbol{\theta^f, \theta^s})$ is evaluated at the fiducial systematics parameters and with no foregrounds, the multiplicative correction returns unity and the raw theoretical spectrum is recovered. The procedure for computing these spectra is similar to that used in Section~\ref{sec:agora_test}: we treat the single \textsc{Agora} realization as data and use Gaussian simulation realizations to compute the response function and noise bias terms. In the data realization, each individual foreground component is multiplied by the amplitude scaling defined by $\boldsymbol{\theta^f}$, and the beam and calibration defined by $\boldsymbol{\theta^s}$ are applied prior to running the lensing reconstruction. For the Gaussian simulations, the foreground amplitudes $\boldsymbol{\theta^f}_{\rm fid}$ and systematics parameters $\boldsymbol{\theta^s}_{\rm fid}$ are fixed to their fiducial values, matching the procedure used in the baseline analysis.

Let us denote the data map with fiducial systematics and no foregrounds as $d_{0}$, where the subscript represents the foreground amplitude. We also denote the Gaussian simulation ensemble that assume fiducial foreground amplitude as $x_{1}$. Expressed in terms of the maps used in the calculation, the denominator of the multiplicative factor can be written as
\begin{equation}
\hat{C}^{\kappa\kappa}_{L}(d_0,x_1)=C_L([d_0,d_0],[d_0,d_0])-N_L^{(0),{\rm RD}}(d_0,x_1)-N_L^{(1)}(x_1).
\end{equation}
Following the notation of Section~\ref{sec:N0N1}, we define the data-simulation contribution to the realization-dependent $N_L^{(0)}$ as
\begin{equation}\label{eq:rdn0_dx}
N_L^{(0),dx}(d_0,x_1)\equiv\left\langle C_L([d_0,x_{1,i}],[d_0,x_{1,j}])+C_L([x_{1,i},d_0],[x_{1,j},d_0])+C_L([d_0,x_{1,i}],[x_{1,i},d_0])+C_L([x_{1,i},d_0],[d_0,x_{1,i}])\right\rangle,
\end{equation}
where $x_{1,i}$ and $x_{1,j}$ denote different realizations drawn from the $x_1$ simulation ensemble. The realization-dependent $N_L^{(0)}$ can then be written as
\begin{equation}\label{eq:rdn0_dx_relation}
N_L^{(0),{\rm RD}}(d_0,x_1)=N_L^{(0),dx}(d_0,x_1)-N_L^{(0)}(x_1).
\end{equation}
Now consider a small multiplicative calibration perturbation, $d_{0,\epsilon}=(1+\epsilon)d_0$. Since the lensing estimator is quadratic in the input map, the reconstructed convergence map scales as $\hat{\kappa}[d_{0,\epsilon},d_{0,\epsilon}]=(1+\epsilon)^2\hat{\kappa}[d_0,d_0]$. Therefore the raw auto-spectrum contains four data legs and scales as $(1+\epsilon)^4$. In contrast, each term in $N_L^{(0),dx}$ contains two data legs and two simulation legs, and therefore scales only as $(1+\epsilon)^2$, while $N_{L}^{(0)}(x_{1})$ and $N_{L}^{(1)}(x_{1})$ are constructed entirely from simulations and therefore remain unchanged. The perturbed debiased spectrum is therefore
\begin{equation}\label{eq:bias2}
\hat{C}^{\kappa\kappa}_{L}(d_{0,\epsilon},x_1)=(1+\epsilon)^4 C_L([d_0,d_0],[d_0,d_0])-(1+\epsilon)^2N_L^{(0),dx}(d_0,x_1)+N_L^{(0)}(x_1)-N_L^{(1)}(x_1).
\end{equation}
Taylor expanding to first order in $\epsilon$ gives
\begin{align}
\Delta \hat{C}^{\kappa\kappa}_{L}&=4\epsilon\,C_L([d_0,d_0],[d_0,d_0])-2\epsilon\,N_L^{(0),dx}(d_0,x_1)+\mathcal{O}(\epsilon^2)\\
&=\epsilon\left[4\hat{C}^{\kappa\kappa}_{L}(d_0,x_1)+4N_L^{(1)}(x_1)+2\left(
N_L^{(0),{\rm RD}}(d_0,x_1)-N_L^{(0)}(x_1)\right)\right]+\mathcal{O}(\epsilon^2),
\end{align}
and hence
\begin{equation}\label{eq:emul_bias}
\frac{\hat{C}^{\kappa\kappa}_{L}(d_{0,\epsilon},x_1)}
{\hat{C}^{\kappa\kappa}_{L}(d_0,x_1)}
=
1+\epsilon\left[
4
+
4\frac{N_L^{(1)}(x_1)}{\hat{C}^{\kappa\kappa}_{L}(d_0,x_1)}
+
2\frac{N_L^{(0),{\rm RD}}(d_0,x_1)-N_L^{(0)}(x_1)}
{\hat{C}^{\kappa\kappa}_{L}(d_0,x_1)}
\right]
+
\mathcal{O}(\epsilon^2).
\end{equation}
Equation~\eqref{eq:emul_bias} makes the origin of the turnover clear. In addition to the expected scale-independent calibration scaling, the multiplicative correction contains a term proportional to the difference $N_L^{(0),{\rm RD}}(d_0,x_1)-N_L^{(0)}(x_1)$. This difference arises because the data-like realization $d_0$ is foreground free, while the Gaussian simulations $x_1$ used in the realization-dependent $N_L^{(0)}$ subtraction contain the fiducial foreground level. This residual becomes increasingly important at high $L$, where the disconnected bias is large compared with the debiased lensing signal, and produces the turnover seen in Figure~\ref{fig:emul_validation_mv}.

We also consider the scenario of using a data-like realization $d_1$ with the same fiducial foreground amplitude as the Gaussian simulations $x_1$. In this case, the data and simulation power are closely matched, so that $N_L^{(0),{\rm RD}}(d_1,x_1)\simeq N_L^{(0)}(x_1)$ and the term responsible for the strong turnover is largely removed. The multiplicative correction then reduces to
\begin{equation}
\frac{\hat{C}^{\kappa\kappa}_{L}(d_{1,\epsilon},x_1)}
{\hat{C}^{\kappa\kappa}_{L}(d_1,x_1)}
\simeq
1+4\epsilon
\left[
1+
\frac{N_L^{(1)}(x_1)}
{\hat{C}^{\kappa\kappa}_{L}(d_1,x_1)}
\right],
\end{equation}
which shows that even if the foreground powers in the data and simulations are matched, the multiplicative correction has some scale dependence. On scales where $N_L^{(1)}\ll \hat{C}_L^{\kappa\kappa}$, this further reduces to
\begin{equation}
\frac{\hat{C}^{\kappa\kappa}_{L}(d_{1,\epsilon},x_1)}
{\hat{C}^{\kappa\kappa}_{L}(d_1,x_1)}
\simeq
1+4\epsilon
\simeq
(1+\epsilon)^4,
\end{equation}
recovering the expected scale-independent calibration scaling to first order. This behavior is seen in the right panel of Figure~\ref{fig:emul_validation_mv}: matching the foreground amplitudes strongly suppresses the turnover, while the remaining scale dependence is consistent with the contribution from $N_L^{(1)}$. 

The $d_{0}$ configuration considered above is specific to the foreground-free reference point adopted in our emulator. In contrast, the $d_{1}$ case is closer to a standard quadratic-estimator analysis, in which simulations with an assumed foreground power are used to compute the response and noise-bias terms. Equation~\eqref{eq:emul_bias} shows that even in this case, varying the calibration can introduce a small scale dependence through the simulation-derived bias terms, most notably $N_L^{(1)}$. Any residual mismatch between the foreground power in the data and simulations introduces an additional contribution proportional to $N_L^{(0),{\rm RD}}-N_L^{(0)}$, which can further enhance this scale dependence. In practice, these effects are expected to be small given the tight calibration uncertainties and the close agreement between the foreground power in the data and simulations. Nevertheless, they may become increasingly important to model as future CMB lensing measurements reach higher precision. In our analysis, these effects are naturally captured by the emulator.

\section{Measuring distances between posteriors}\label{app:tensiometer}
We measure the distance between the posterior constraints from various datasets using the non-Gaussian parameter-shift estimator implemented in \texttt{tensiometer}. Given two independent posterior distributions,
$P_1(\theta)$ and $P_2(\theta)$, the method constructs the posterior distribution of parameter shifts~\citep{raveri2019},
\begin{equation}
\Delta \theta \equiv \theta_1 - \theta_2 ,
\end{equation}
with density
\begin{equation}
P_{\Delta}(\Delta\theta)=\int d\theta \,P_1(\theta)\,P_2(\theta-\Delta\theta) .
\end{equation}
The point $\Delta\theta=0$ corresponds to exact agreement between the
two constraints. The tension probability is then defined as the posterior
mass lying at higher density than the no-shift point,
\begin{equation}\label{eq:pshift}
P_{\rm shift}=\int_{P_{\Delta}(\Delta\theta) > P_{\Delta}(0)}d\Delta\theta\,P_{\Delta}(\Delta\theta).
\end{equation}
Thus, if $\Delta\theta=0$ lies in the tail of the parameter-shift
distribution, $P_{\rm shift}$ is large and the two datasets are in
tension. This shift probability is converted into an effective Gaussian significance by using $N_{\sigma}=\sqrt{2}{\rm erf}^{-1}(P_{\rm shift})$. 

In practice, samples of $\Delta\theta$ are obtained by differencing paired draws from two MCMC chains, and the density $P_{\Delta}$ is estimated from these samples using the machinery implemented in \texttt{tensiometer}: for one-dimensional shifts we use a kernel density estimate, while for multi-dimensional shifts $P_{\Delta}$ is modeled with normalizing flows trained on the difference samples \citep{raveri2021}. The integral in Equation~\eqref{eq:pshift} is then evaluated by drawing Monte Carlo samples from the difference distribution.

\section{Dependence of nonlinear scale modeling on $T_{\rm AGN}$}\label{app:tagn}

Our fiducial model adopts the \texttt{HMcode2020} (\texttt{mead2020}) prescription. To investigate the impact of baryonic feedback on our constraints, we switch to the \texttt{HMcode2020\_feedback} (\texttt{mead2020\_feedback}) variant and allow $T_{\rm AGN}$ to vary over the range $7.0$ to $8.5$. In this framework, $T_{\rm AGN}$ characterizes the strength of AGN feedback through an effective heating temperature, expressed in units of $\log_{10}(T/{\rm K})$. Larger values correspond to more energetic feedback that expels gas more efficiently from halo centers, and suppresses the nonlinear matter power spectrum at intermediate and small scales. This parameterization has been widely adopted in the literature as a phenomenological approach to marginalize over uncertainties on baryonic feedback effects without explicitly running hydrodynamical simulations. The model is calibrated against suites of hydrodynamical simulations with different feedback strengths, effectively compressing the impact of complex subgrid physics into a single parameter, $T_{\rm AGN}$. Sampling over this parameter therefore provides a physically motivated and computationally efficient way to propagate baryonic modeling uncertainty into weak lensing, galaxy clustering, and CMB lensing analyses. However, CMB lensing is intrinsically only weakly sensitive to $T_{\rm AGN}$, since much of its constraining power comes from relatively large scales and higher redshifts where baryonic-feedback effects are weaker.

We also test two alternative nonlinear prescriptions: the Takahashi revision of \texttt{Halofit} \cite{takahashi2012} and the older \texttt{HMcode2016} (\texttt{mead2016}), which was adopted in the ACT DR6 lensing analysis \cite{madhavacheril2024}. These tests probe the sensitivity to the choice of nonlinear matter-power-spectrum prescription, while the impact of baryonic feedback is tested separately using the \texttt{HMcode2020\_feedback} model described above. For the $\kgmv$ lensing-only analysis, the resulting constraints on $\som$ for the different nonlinear prescriptions and baryonic-feedback treatments are
\begin{equation}
\som=
\begin{cases}
0.6046\pm0.0096 \hspace{1cm} \texttt{HMcode2020}\ {\rm (fiducial)} \\[0.05cm]
0.6009\pm0.0096 \hspace{1cm} \texttt{Revised Halofit} \\[0.05cm]
0.6053\pm0.0094 \hspace{1cm} \texttt{HMcode2016} \\[0.05cm]
0.6064\pm0.0097 \hspace{1cm} \texttt{HMcode2020}\ {\rm with\   feedback}.\nonumber
\end{cases}
\end{equation}
We find that marginalizing over baryonic-feedback effects through $T_{\rm AGN}$ has a negligible impact on the lensing-only constraint on $\sigma_8\Omega_{\rm m}^{0.25}$, consistent with the weak sensitivity to $T_{\rm AGN}$ discussed above. We similarly find essentially no change when replacing \texttt{HMcode2020} with \texttt{HMcode2016}. In contrast, adopting the revised \texttt{Halofit} prescription shifts $\sigma_8\Omega_{\rm m}^{0.25}$ by approximately $0.5\sigma$, indicating a modest dependence on the nonlinear matter-power-spectrum prescription. A similar shift was noted in the joint analysis of DES\,Y3 and KiDS-1000~\cite{deskids2023}.

\section{Choice of $\tau$ prior}\label{app:tau_prior} 
\begin{figure}[h]
\includegraphics[width=0.48\linewidth]{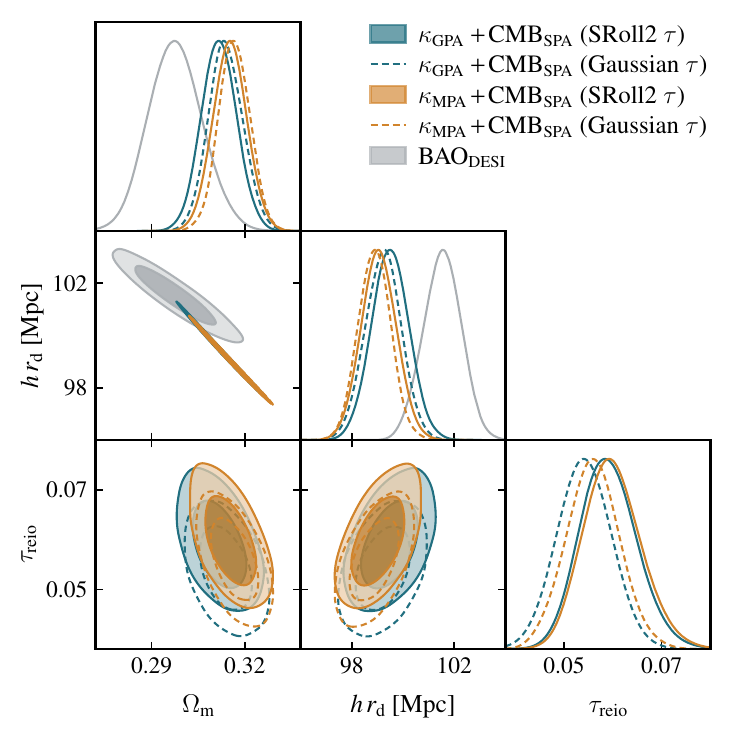} 
\caption{Constraints on $\Omega_{\rm m}$ and $hr_{\rm d}$ for the dataset combinations $\kgpa+\cmbspa$ and $\kmpa+\cmbspa$, using either a Gaussian prior on $\tau=\mathcal{N}(0.051,0.006)$ or the SRoll2 low-$\ell$ $EE$ likelihood, compared to those from $\baodesi$.}
\label{fig:Omegam_hrdrag_sroll2tau_Gausstau}
\end{figure}

In C26, the distance between the $\kmpa$+$\cmbspa$ and $\baodesi$ was evaluated in the $\Omega_{\rm m}$--$h r_{\rm d}$ parameter space, yielding a distance of $2.8\sigma$. That analysis adopted a Gaussian prior $\tau=\mathcal{N}(0.051,0.006)$, taken from the \planck{} NPIPE large-scale polarization constraint \citep{planck2020LFIHFItau}. In our baseline analysis, we instead use a broad top-hat prior on $\tau$ together with the SRoll2 low-$\ell$ polarization likelihood, so that the optical depth is constrained directly by the data combination used. This choice has a modest but non-negligible impact on the inferred distance between datasets. For $\kgpa+\cmbspa$, the consistency with $\baodesi$ changes from $2.3\sigma$ when using the Gaussian $\tau$ prior to $1.9\sigma$ when using SRoll2. 
For $\kmpa+\cmbspa$, the consistency with $\baodesi$ changes from $2.8\sigma$ when using the Gaussian $\tau$ prior to $2.5\sigma$ when using SRoll2. We compute the distance between the two datasets using both normalizing flows in \texttt{tensiometer} and Gaussian distances, and verify that they agree to within 0.05$\sigma$. This comparison shows that the quoted consistency metrics are not purely properties of the high-significance lensing and BAO measurements, but can also depend on the treatment of the optical-depth information in the primary CMB likelihood. This behavior is expected because $\tau$ is partially degenerate with the scalar amplitude through the primary CMB constraint on $A_{\rm s}e^{-2\tau}$, and therefore can propagate into the inferred growth amplitude and its degeneracy with $\Omega_{\rm m}$ and $hr_{\rm d}$. These results are shown in Figure~\ref{fig:Omegam_hrdrag_sroll2tau_Gausstau}.

\clearpage
\bibliography{apssamp}

\end{document}